\documentclass[prd,10pt,aps]{revtex4-2}
\usepackage{graphicx,epstopdf}
\usepackage[caption=false]{subfig}
\usepackage{appendix}
\usepackage[T1]{fontenc}
\usepackage{lmodern}
\usepackage[dvipsnames,x11names]{xcolor}
\usepackage[psdextra,colorlinks=true,linkcolor=Magenta!80!black,
citecolor=Magenta!70!black,urlcolor=Magenta!80!black]{hyperref}
\usepackage[sort&compress]{natbib}
\usepackage{morefloats}
\usepackage[pdf]{pstricks}
\usepackage{amsmath}
\usepackage{amssymb}
\usepackage{amsfonts}
\usepackage{rotating}
\usepackage{cancel}
\usepackage{mathtools}
\usepackage{bbm}
\usepackage{dsfont}
\usepackage{bbold}
\usepackage{multirow}
\usepackage{ulem}
\usepackage{physics}
\usepackage{orcidlink}
\usepackage{colortbl}
\usepackage{slashed}
\usepackage{booktabs}
 \usepackage{xcolor}
 \usepackage{colortbl}
 \usepackage{array}

\def\pslash{p\!\!\!\slash }
\def\p_1slash{p1\!\!\!\slash }
\def\p_2slash{p2\!\!\!\slash }
\def\qslash{q\!\!\!\slash }

\def\eslash{\varepsilon\!\!\!\slash }

\definecolor{headerblue}{RGB}{28, 66, 110}
\definecolor{rowgray}{RGB}{245, 247, 250}

\begin{document}

\title{Deciphering the nature of $P^{\Sigma}_{\psi s}$ pentaquarks in the light of their electromagnetic multipole moments}

\author{Ula\c{s}~\"{O}zdem\orcidlink{0000-0002-1907-2894}}%
\email[]{ulasozdem@aydin.edu.tr}
\affiliation{ Health Services Vocational School of Higher Education, Istanbul Aydin University, Sefakoy-Kucukcekmece, 34295 Istanbul, T\"{u}rkiye}

\begin{abstract}
We calculate the electromagnetic multipole moments of the $\Sigma$-type strange hidden-charm pentaquarks $P^{\Sigma}_{\psi s}$ --- which form an isospin triplet with three distinct charge states ($\Sigma^+$, $\Sigma^0$, $\Sigma^-$) offering independently measurable electromagnetic properties --- using QCD light-cone sum rules, employing six interpolating currents for the spin-$\frac{1}{2}$ channel and seven for the spin-$\frac{3}{2}$ channel, each built from diquark--diquark--antiquark operators with scalar or axial-vector diquark content. We compute the magnetic dipole moment $\mu$ for all channels and, for spin-$\frac{3}{2}$, the electric quadrupole $\mathcal{Q}$ and magnetic octupole $\mathcal{O}$ moments, the latter two for the first time for this system, and provide the first systematic quark-flavor decomposition of all three multipole moments for $P^{\Sigma}_{\psi s}$. The results show a systematic dependence on the diquark spin. Scalar diquark currents yield charm-sector-dominated, flavor-insensitive moments ($\mu \in [-1.92,\,-1.21]\,\mu_N$ for spin-$\frac{1}{2}$ and $|\mu| \lesssim 1.2\,\mu_N$ for spin-$\frac{3}{2}$), consistent with the heavy-quark spin symmetry limit. Axial-vector diquark currents produce larger, flavor-sensitive moments with sign reversals governed by the $e_u/e_d = -2$ charge ratio. For $\mathcal{Q}$, scalar-diquark currents give oblate deformations ($Q_0 \approx -2.0\times 10^{-2}~\mathrm{fm}^2$) dominated by the charm sector, while two-axial-vector-diquark currents predict prolate values up to $Q_0 = +8.0\times 10^{-2}~\mathrm{fm}^2$, with a sign reversal for $[su][uc]\bar{c}$ in two currents. Currents with scalar antiquark coupling yield a topology-independent octupole value $\mathcal{O} \approx -0.25\times 10^{-3}~\mathrm{fm}^3$, which may serve as a lattice QCD benchmark. Comparison with constituent quark model predictions identifies four qualitative discriminants: $|\mu| \gtrsim 3\,\mu_N$ in the spin-$\frac{1}{2}$ sector; the sign of $\mu$ for the $[su][uc]\bar{c}$ state in the spin-$\frac{3}{2}$ sector; a non-zero $\mathcal{Q}$, which vanishes in the $S$-wave molecular approximation; and the sign correlation between $\mathcal{Q}$ and $\mathcal{O}$, which probes the $1/m_q$ mass weighting of the magnetization distribution.
\end{abstract}


\maketitle

\section{Introduction}\label{sec:introduction}

The existence of hadronic states beyond the conventional quark--antiquark mesons and three-quark baryons has been a central question of QCD for decades. The observation of the $X(3872)$ by Belle in 2003~\cite{Belle:2003nnu} opened a new era: this state, whose mass, width, and decay pattern resist interpretation within the quark model,
is now widely accepted as the first confirmed tetraquark candidate and has catalyzed an extensive experimental and theoretical program on exotic hadrons. The theoretical landscape of proposed internal structures spans compact multiquark configurations, hadronic molecules, and kinematic effects, and a consensus on the internal organization of most candidates remains elusive~\cite{Esposito:2014rxa, Esposito:2016noz, Olsen:2017bmm, Lebed:2016hpi, Nielsen:2009uh, Brambilla:2019esw, Agaev:2020zad, Chen:2016qju, Ali:2017jda, Guo:2017jvc, Liu:2019zoy, Yang:2020atz, Dong:2021juy, Dong:2021bvy, Chen:2022asf, Meng:2022ozq}.

Hidden-charm pentaquarks entered experimental reach in 2015 when LHCb observed two candidates, $P_{\psi}^{N}(4380)$ and $P_{\psi}^{N}(4450)$, in the $J/\psi p$ invariant mass spectrum from $\Lambda^0_b \to J/\psi p K^-$~\cite{LHCb:2015yax}. A subsequent higher-statistics analysis resolved the $P_{\psi}^{N}(4450)$ into two structures, $P_{\psi}^{N}(4440)$ and $P_{\psi}^{N}(4457)$, and revealed a third state, $P_{\psi}^{N}(4312)$~\cite{LHCb:2019kea}. These candidates, with minimal quark content $uudc\bar{c}$, established hidden-charm pentaquark spectroscopy as an active field. The search for strange analogs followed: LHCb reported evidence for $P^{\Lambda}_{\psi s}(4459)$ in $\Xi^-_b \to J/\psi\Lambda K^-$ decays~\cite{LHCb:2020jpq} and subsequently observed $P^{\Lambda}_{\psi s}(4338)$ in the same channel~\cite{LHCb:2022ogu}. These $udsc\bar{c}$ candidates are the first established strange hidden-charm pentaquarks. More recently, Belle reported a $P^{\Lambda}_{\psi s}$-like signal at $(4471.7 \pm 4.8 \pm 0.6)$~MeV
with significance $3.3\sigma$ including systematic uncertainties~\cite{Belle:2025pey}. 
Despite this experimental progress, the nature of these states remains unresolved. Their spin-parity quantum numbers are not experimentally established, and the question of whether they are compact multiquark configurations, hadronic molecules, or threshold effects is open. Both the non-strange and strange sectors of the $P_{\psi}^N$ family have been extensively analyzed theoretically, including mass spectra, decay widths, and production mechanisms~\cite{Esposito:2014rxa, Esposito:2016noz, Olsen:2017bmm, Lebed:2016hpi, Nielsen:2009uh, Brambilla:2019esw, Agaev:2020zad, Chen:2016qju, Ali:2017jda, Guo:2017jvc, Liu:2019zoy, Yang:2020atz, Dong:2021juy, Dong:2021bvy, Chen:2022asf, Meng:2022ozq}. By contrast, the $\Sigma$-type strange hidden-charm pentaquarks, $P^{\Sigma}_{\psi s}$ ($S = -1$, $I = 1$), have received considerably less attention. Unlike the $\Lambda$-type states, which are isospin singlets, the $\Sigma$-type states form an isospin triplet ($\Sigma^+$, $\Sigma^0$, $\Sigma^-$) with quark content $\{uus, uds, dds\}c\bar{c}$, offering three charge states whose electromagnetic properties can, in principle, be independently characterized. No experimental candidate for $P^{\Sigma}_{\psi s}$ has been confirmed to date, but the state is expected in the same mass region as $P^{\Lambda}_{\psi s}$ and is actively searched for in $\Sigma_b$ and $\Xi_b$ baryon decays \cite{LHCb:2018roe}. 

Resolving the internal structure of these states requires observables that probe hadron geometry beyond mass and width. Electromagnetic multipole moments are particularly well suited for this purpose. The magnetic dipole moment encodes the net alignment of quark spins and circulating currents, and in a composite system it depends not only
on the constituent charges and spins but also on the spatial wave function, making it sensitive to the internal correlation structure. The electric quadrupole moment  measures the deviation of the charge distribution from spherical symmetry and vanishes identically for any $S$-wave two-body bound state in the point-constituent approximation, since the orbital angular momentum between the constituents is zero and each hadron is treated as spherically symmetric. A non-zero electric quadrupole moment therefore constitutes model-independent evidence for an internal structure that cannot be reduced to a simple $S$-wave molecule: it signals either orbital excitation, non-spherical constituent deformation, or a compact multiquark configuration in which the quarks are organized below the hadronic scale. The magnetic octupole moment  probes the higher-order anisotropy of the magnetization distribution and provides complementary information on the relative orientation of spin and charge degrees of freedom. Together, these three observables constrain the internal structure from different and partially independent directions, making their simultaneous prediction substantially more discriminating than any single observable alone.

The electromagnetic properties of exotic hadrons have been studied using a variety of theoretical approaches, including the light-cone QCD sum rule (LCSR) method, which combines the OPE in the presence of an external electromagnetic field with hadronic dispersion relations, as well as the constituent quark model, the chiral quark model, and effective field theory frameworks. Applied to hidden-charm pentaquarks, these approaches have yielded predictions for the magnetic moments of both non-strange and strange states~\cite{Ozdem:2025jda,  Ozdem:2024rch,  Ozdem:2024rqx,  Ozdem:2023htj,  Ozdem:2022kei, Wang:2016dzu,  Ortiz-Pacheco:2018ccl, Xu:2020flp, Ozdem:2018qeh,  Ozdem:2021ugy, Li:2021ryu, Gao:2021hmv, Guo:2023fih, Wang:2022nqs, Wang:2022tib, Ozdem:2024jty, Li:2024wxr, Li:2024jlq,  Mutuk:2024ltc, Mutuk:2024jxf, Mutuk:2024ach, Ozdem:2024usw, Ozdem:2025fks, Zhu:2025abk, Ozdem:2026gmn, Ozdem:2025ion,Mutuk:2026zxp}. Across all these approaches, predictions have been predominantly restricted to the magnetic dipole moment, with only a subset of studies extending to higher multipoles. For the $\Sigma$-type strange hidden-charm sector, magnetic dipole moments of $P^{\Sigma}_{\psi s}$ states have been computed within the constituent quark model~\cite{Li:2024wxr, Li:2025ddx}, but the electric quadrupole and magnetic octupole moments have not been addressed in the literature within any framework. The present work fills this gap and, in addition, provides the first LCSR-based analysis of the electromagnetic multipole moments of $P^{\Sigma}_{\psi s}$. The LCSR approach is particularly well suited to this system because it operates directly at the quark and gluon level, incorporates non-perturbative QCD dynamics through photon distribution amplitudes, and naturally accommodates the diquark operator structures used to construct the interpolating currents.

A key methodological issue in LCSR~\cite{Chernyak:1990ag, Braun:1988qv, Balitsky:1989ry}  calculations of multiquark electromagnetic properties is the choice of interpolating current. The interpolating current does not uniquely represent the physical state: different currents with the same quantum numbers but different Dirac and color structures couple to the same state with different overlap amplitudes and probe different aspects of the internal quark correlations. In studies of conventional hadrons this ambiguity is less consequential because the space of acceptable currents is small; for pentaquarks, the much larger operator space means that predictions can vary substantially across currents. This sensitivity is not a weakness of the approach but rather a feature: by systematically varying the current structure, one maps out which aspects of the electromagnetic response are robust and which are strongly model dependent. A calculation that employs a single current and quotes a single prediction without exploring this dependence provides incomplete information. In the present work we address this directly by employing six independent interpolating currents for the spin-$1/2$ channel and seven for the spin-$3/2$ channel, each constructed from diquark--diquark--antiquark operators with distinct Dirac structures. The diquark basis is motivated by the expected importance of diquark correlations in the hidden-charm sector: one-gluon exchange in the color-antitriplet channel is attractive, making tightly correlated diquark pairs energetically favorable and providing a natural building block for compact multiquark states~\cite{Wang:2010sh, Kleiv:2013dta}. By varying the scalar ($J^P = 0^+$) versus axial-vector ($J^P = 1^+$) diquark content of the operator, we probe the two qualitatively different internal spin configurations that the diquark picture allows, and the resulting variation of the predicted moments provides a direct measure of the sensitivity of electromagnetic observables to the internal spin-flavor correlations. In this paper we present predictions for the magnetic dipole, electric quadrupole, and magnetic octupole moments of $P^{\Sigma}_{\psi s}$ with $J^P = 1/2^-$ and $3/2^-$, including a quark-level decomposition that reveals the physical origin of each contribution. We compare our results with molecular model predictions and identify four experimentally accessible discriminants between the compact diquark and molecular pictures. For the spin-$3/2$ channel, the electric quadrupole and magnetic octupole moments are computed here for the first time for this system.

The paper is organized as follows. Section~\ref{sec:formalism} introduces the LCSR framework, constructs the correlation functions, defines the interpolating currents, and derives the sum rules for the magnetic dipole, electric quadrupole, and magnetic octupole moments. Section~\ref{sec:numerical} provides the numerical implementation: input parameters, working regions for $M^2$ and $s_0$. Section~\ref{sec:results} presents and discusses the results, including the quark-level decomposition, a comparison with molecular model predictions, and a discussion of experimental discriminants.

\begin{widetext}
 
\section{Formalism} \label{sec:formalism}

 \subsection{General framework of LCSR and interpolating currents for $P^{\Sigma}_{\psi s}$ states} \label{subsec:general}

The calculation of electromagnetic multipole moments within the LCSR 
framework can be conceptually initiated using three-point correlation 
functions, where the photon is explicitly represented by an 
electromagnetic current operator. For the spin-$\frac{1}{2}$ and 
spin-$\frac{3}{2}$ channels, such three-point correlators take the form
\begin{align}
\mathcal T_{\alpha}(p,q) &= i^2 \int d^4x \, d^4y \, e^{ip \cdot x + iq \cdot y} \,
\langle 0 | \mathcal{T} \{ J(x) \, J_{\alpha}^{\gamma}(y) \, 
\bar{J}(0) \} | 0 \rangle,
\label{eq:3pt_12}\\[4pt]
\mathcal T_{\alpha\mu\nu}(p,q) &= i^2 \int d^4x \, d^4y \, 
e^{ip \cdot x + iq \cdot y} \,
\langle 0 | \mathcal{T} \{ J_\mu(x) \, J_{\alpha}^{\gamma}(y) \, 
\bar{J}_\nu(0) \} | 0 \rangle,
\label{eq:3pt_32}
\end{align}
respectively, where $J(x)$ is the interpolating current for the 
spin-$\frac{1}{2}$ pentaquark, $J_\mu(x)$ for the spin-$\frac{3}{2}$ 
pentaquark, and $J_{\alpha}^{\gamma}(y)$ is the electromagnetic current. While conceptually transparent, this three-point formulation becomes technically cumbersome when separating short-distance perturbative contributions from long-distance non-perturbative effects, particularly for soft photons relevant to static electromagnetic properties such as multipole moments.

To overcome this difficulty, we employ the external background electromagnetic field method~\cite{Ball:2002ps, Novikov:1983gd, Ioffe:1983ju}. In this formalism, the photon is treated not as a quantum current operator but as a classical weak background field, 
which allows us to reduce the three-point correlators to two-point correlators evaluated in the presence of this external field:
\begin{align}
\mathcal T(p,q) &= i \int d^4x \, e^{ip \cdot x} \,
\langle 0 | \mathcal{T} \{ J(x) \, \bar{J}(0) \} | 
0 \rangle_F,
\label{eq:corr_12}\\[4pt]
\mathcal T_{\mu\nu}(p,q) &= i \int d^4x \, e^{ip \cdot x} \,
\langle 0 | \mathcal{T} \{ J_\mu(x) \, \bar{J}_\nu(0) \} | 
0 \rangle_F,
\label{eq:corr_32}
\end{align}
where the subscript $F$ indicates evaluation in the electromagnetic background field characterized by the field strength tensor
\begin{align}
F_{\alpha\beta} = i \bigl( \varepsilon_\alpha q_\beta 
- \varepsilon_\beta q_\alpha \bigr) e^{-iq \cdot x},
\label{eq:Fmunu}
\end{align}
with $\varepsilon_\alpha$ and $q_\beta$ the photon polarization vector and four-momentum, respectively. It should be noted that $\mathcal T(p,q)$ stands for $\varepsilon^\alpha \mathcal T_\alpha(p,q)$, and similarly $\mathcal T_{\mu\nu}(p,q)$ for 
$\varepsilon^\alpha \mathcal T_{\alpha\mu\nu}(p,q)$.

Since the background field is treated as infinitesimally weak, each correlator admits a systematic expansion in powers of the field strength~\cite{Ball:2002ps, Novikov:1983gd, Ioffe:1983ju}:
\begin{align}
\mathcal T(p,q) &= \mathcal T^{(0)}(p,q) + \mathcal T^{(1)}(p,q) + \cdots,
\label{eq:expand_12}\\[4pt]
\mathcal T_{\mu\nu}(p,q) &= \mathcal T_{\mu\nu}^{(0)}(p,q) 
+ \mathcal T_{\mu\nu}^{(1)}(p,q) + \cdots,
\label{eq:expand_32}
\end{align}
where the leading terms $\mathcal T^{(0)}(p,q)$ and $\mathcal T_{\mu\nu}^{(0)}(p,q)$ 
correspond to the field-free two-point correlators relevant for mass sum rules, while the linear-response terms $\mathcal T^{(1)}(p,q)$ and $\mathcal T_{\mu\nu}^{(1)}(p,q)$ contain precisely the electromagnetic multipole moments to be extracted. The higher-order terms do not contribute at the order retained in this work. The background field formalism provides a gauge-invariant separation of soft and hard photon contributions and systematically organizes the calculation via photon distribution amplitudes (DAs).

Each correlator is computed in two representations. On the hadronic side, it is expressed in terms of physical hadron matrix elements by inserting a complete set of states; this representation encodes the electromagnetic form factors of interest. On the QCD side, the same correlator is evaluated by substituting the interpolating currents, contracting quark fields via Wick's theorem, and computing the result using the operator product expansion (OPE) near the light cone, with non-perturbative effects parametrized by photon DAs. Equating the two representations via quark-hadron duality and applying double Borel transformation yields the sum rules from which the electromagnetic multipole moments are extracted.

The QCD side of both correlation functions is evaluated by substituting the currents into Eqs.~(\ref{eq:corr_12})--(\ref{eq:corr_32}), contracting quark fields via Wick's theorem, and inserting the quark propagators. The light-quark propagator, including the leading gluon correction, is~\cite{Balitsky:1987bk}
\begin{equation}
S_q^{ab}(x) = \frac{\delta^{ab}}{2\pi^2 x^2} \left( \frac{i\not\!x}{x^2} - \frac{m_q}{2} \right)
- \frac{ig_s}{16\pi^2 x^2}\int_0^1 dv\,G^{ab}_{\mu\nu}(vx)
\bigl[\bar{v}\not\!x\,\sigma^{\mu\nu}+v\,\sigma^{\mu\nu}\not\!x\bigr],
\label{eq:Sq}
\end{equation}
where $g_s$ is the strong coupling constant and $G^{ab}_{\mu\nu} \equiv G^A_{\mu\nu}\,t^A_{ab}$ is the gluon field strength tensor in the fundamental representation, with $t^A$ the SU(3) color generators. The charm-quark 
propagator is~\cite{Belyaev:1985wza}
\begin{align}
S_c^{ab}(x) &= \frac{m_c^2\delta^{ab}}{4\pi^2}
\Biggl[\frac{K_1(m_c\sqrt{-x^2})}{\sqrt{-x^2}}
+i\frac{\not\!x\,K_2(m_c\sqrt{-x^2})}{-x^2}\Biggr]
-i\frac{g_s m_c}{16\pi^2}\int_0^1 dv\,G^{ab}_{\mu\nu}(vx)
\Biggl[\frac{K_1(m_c\sqrt{-x^2})}{\sqrt{-x^2}}
\bigl(\sigma^{\mu\nu}\not\!x+\not\!x\,\sigma^{\mu\nu}\bigr)
\nonumber\\
&\quad
+2\sigma^{\mu\nu}K_0(m_c\sqrt{-x^2})\Biggr],
\label{eq:Sc}
\end{align}
where $K_n(z)$ are modified Bessel functions of the second kind. The leading term of both the light- and heavy-quark propagators represents the free (perturbative) part, whereas the subsequent terms encode the non-perturbative interactions with the gluon background.

Having specified the quark propagators that enter the correlation functions, we now turn to the coupling of the external photon to the quark lines. This coupling proceeds 
through two distinct mechanisms that capture complementary kinematic regimes of the photon--quark interaction.

In the hard regime, the photon couples directly and locally to a quark line via a perturbative electromagnetic vertex. Each quark propagator in the correlator receives in turn a single-photon insertion,
\begin{equation}
S^{\rm free}(x) \;\longrightarrow\;
\int d^4z\;S^{\rm free}(x-z)\;\not\!{A}(z)\;S^{\rm free}(z),
\label{eq:pert}
\end{equation}
where $A_\alpha (z) = -\frac{1}{2} F_{\alpha\beta} z^\beta = -\frac{1}{2} (\varepsilon_\alpha q_\beta - \varepsilon_\beta q_\alpha) z^\beta$. The resulting
contributions enter the OPE at leading order in $\alpha_s$ and are
proportional to the electric charge of the quark line that absorbs
the photon.

In the soft regime, the photon interacts coherently with the non-perturbative QCD vacuum and is described by its DAs. A light-quark propagator is replaced by the bilocal vacuum-to-photon matrix element
\begin{equation}
S_{\alpha\beta}^{ab}(x) \;\longrightarrow\;
-\frac{1}{4}\bigl[\bar{q}^a(x)\,\Gamma_i\,q^b(0)\bigr]
(\Gamma_i)_{\alpha\beta},
\qquad
\Gamma_i \in \bigl\{1,\;\gamma_5,\;\gamma_\rho,\;
i\gamma_5\gamma_\rho,\;\sigma_{\rho\sigma}/2\bigr\}.
\label{eq:nonpert}
\end{equation}
The resulting matrix elements
\begin{equation}
\langle\gamma(q)|\bar{q}(x)\,\Gamma_i\,q(0)|0\rangle
\qquad\text{and}\qquad
\langle\gamma(q)|\bar{q}(x)\,\Gamma_i\,G_{\mu\nu}(vx)\,q(0)|0\rangle
\label{eq:photon_me}
\end{equation}
parametrize the two-particle and three-particle Fock components of the photon wave function, respectively. The two-particle matrix elements give rise to the twist-2 DA $\varphi_\gamma(u)$, the twist-3 DAs $\psi^{(a,v)}(u)$ and $\mathbb{A}(u)$, and the twist-4 DAs $h^{(\gamma,G)}(u)$; the three-particle matrix elements generate the DAs $\mathcal{S}(\alpha_i)$, $\mathcal{A}(\alpha_i)$ ,$\mathcal{V}(\alpha_i)$, and
$\mathcal{T}(\alpha_i)$, where $\alpha_i = (\alpha_{\bar{q}},\alpha_q,\alpha_g)$ denote the momentum fractions of the antiquark, quark, and gluon constituents of the photon. All
DAs are retained up to twist-4 and parametrized following~\cite{Ball:2002ps}. 
The soft-photon replacement is applied only to light-quark propagators. For the charm quark, $m_c \gg \Lambda_{\rm QCD}$ guarantees that the interaction is perturbative at all relevant scales, so the soft mechanism does not contribute to the charm line.

\subsubsection{Interpolating currents for $P^{\Sigma}_{\psi s}$ states} \label{subsubsec:currents}

The interpolating currents for $P^{\Sigma}_{\psi s}$ states are constructed from diquark--diquark--antiquark operators. The diquark basis is motivated by the attractive one-gluon-exchange interaction in the color-antitriplet channel, which favors tightly correlated quark pairs as building blocks of compact multiquark states~\cite{Wang:2010sh, Kleiv:2013dta}. By varying the Dirac structure of the diquarks between scalar ($J^P=0^+$) and axial-vector ($J^P=1^+$) types, we probe the two qualitatively distinct spin configurations available within this picture.

For the $J^P=\frac{1}{2}^-$ channel we employ six currents:
\begin{align}
J_1(x) &= \mathcal{A}\, [s^T_d(x)\Gamma_1 q_e(x)]\,[q^T_f(x)\Gamma_1 c_g(x)]\,
\Gamma_3\bar{c}^T_c(x),
\label{eq:J1_12}\\[3pt]
J_2(x) &= \mathcal{A}\, [s^T_d(x)\Gamma_1 q_e(x)]\,[q^T_f(x)\Gamma_2 c_g(x)]\,
\Gamma_4\bar{c}^T_c(x),
\label{eq:J2_12}\\[3pt]
J_3(x) &= \mathcal{A}\, [q^T_d(x)\Gamma_2 q_e(x)]\,[s^T_f(x)\Gamma_1 c_g(x)]\,
\Gamma_4\bar{c}^T_c(x),
\label{eq:J3_12}\\[3pt]
J_4(x) &= \mathcal{A}\, [q^T_d(x)\Gamma_2 q_e(x)]\,[s^T_f(x)\Gamma_2 c_g(x)]\,
\Gamma_3\bar{c}^T_c(x),
\label{eq:J4_12}\\[3pt]
J_5(x) &= \frac{\mathcal{A}}{\sqrt{2}}\Bigl\{
[q^T_d(x)\Gamma_2 q_e(x)]\,[s^T_f(x)\Gamma_2 c_g(x)]
-[q^T_d(x)\Gamma_2 s_e(x)]\,[q^T_f(x)\Gamma_2 c_g(x)]
\Bigr\}\Gamma_3\bar{c}^T_c(x),
\label{eq:J5_12}\\[3pt]
J_6(x) &= \frac{\mathcal{A}}{\sqrt{2}}\Bigl\{
[q^T_d(x)\Gamma_2 q_e(x)]\,[s^T_f(x)\Gamma_1 c_g(x)]
-[q^T_d(x)\Gamma_2 s_e(x)]\,[q^T_f(x)\Gamma_1 c_g(x)]
\Bigr\}\Gamma_4\bar{c}^T_c(x).
\label{eq:J6_12}
\end{align}
For the $J^P=\frac{3}{2}^-$ channel we employ seven currents:
\begin{align}
J^1_\mu(x) &= \mathcal{A}\,
[s^T_d(x)\Gamma_1 q_e(x)]\,[q^T_f(x)\Gamma_2 c_g(x)]\,
\Gamma_3\bar{c}^T_c(x),
\label{eq:J1_32}\\[3pt]
J^2_\mu(x) &= \mathcal{A}\,
[q^T_d(x)\Gamma_2 q_e(x)]\,[s^T_f(x)\Gamma_1 c_g(x)]\,
\Gamma_3\bar{c}^T_c(x),
\label{eq:J2_32}\\[3pt]
J^3_\mu(x) &= \mathcal{A}\,
[q^T_d(x)\Gamma_2 q_e(x)]\,[s^T_f(x)\Gamma_5 c_g(x)]\,
\Gamma_6\bar{c}^T_c(x),
\label{eq:J3_32}\\[3pt]
J^4_\mu(x) &= \mathcal{A}\,
[q^T_d(x)\Gamma_5 q_e(x)]\,[s^T_f(x)\Gamma_2 c_g(x)]\,
\Gamma_6\bar{c}^T_c(x),
\label{eq:J4_32}\\[3pt]
J^5_\mu(x) &= \frac{\mathcal{A}}{\sqrt{2}}\Bigl\{
[q^T_d(x)\Gamma_2 q_e(x)]\,[s^T_f(x)\Gamma_1 c_g(x)]
-[q^T_d(x)\Gamma_2 s_e(x)]\,[q^T_f(x)\Gamma_1 c_g(x)]
\Bigr\}\Gamma_3\bar{c}^T_c(x),
\label{eq:J5_32}\\[3pt]
J^6_\mu(x) &= \frac{\mathcal{A}}{\sqrt{2}}\Bigl\{
[q^T_d(x)\Gamma_2 q_e(x)]\,[s^T_f(x)\Gamma_5 c_g(x)]
-[q^T_d(x)\Gamma_2 s_e(x)]\,[q^T_f(x)\Gamma_5 c_g(x)]
\Bigr\}\Gamma_6\bar{c}^T_c(x),
\label{eq:J6_32}\\[3pt]
J^7_\mu(x) &= \frac{\mathcal{A}}{\sqrt{2}}\Bigl\{
[q^T_d(x)\Gamma_5 q_e(x)]\,[s^T_f(x)\Gamma_2 c_g(x)]
-[q^T_d(x)\Gamma_5 s_e(x)]\,[q^T_f(x)\Gamma_2 c_g(x)]
\Bigr\}\Gamma_6\bar{c}^T_c(x).
\label{eq:J7_32}
\end{align}
The color factor $\mathcal{A}=\varepsilon_{abc}\varepsilon_{ade}\varepsilon_{bfg}$
antisymmetrizes over color indices $a,\ldots,g$, ensuring that each diquark pair transforms as a color antitriplet $\bar{\mathbf{3}}_c$ and the five-quark operator is a color singlet. The symbol $q$ denotes a light quark ($u$ or $d$) and the superscript $T$ denotes Dirac transposition.

The Dirac matrices $\Gamma_i$ encode the spin-parity content of the diquarks and the coupling to the charm sector:
\begin{itemize}
\item $\Gamma_1 = C\gamma_5$: scalar diquark ($J^P=0^+$), antisymmetric in spin space, requires a flavor-antisymmetric quark pair in the color-antitriplet channel. Carries no free Lorentz index.

\item $\Gamma_2 = C\gamma_\mu$: axial-vector diquark ($J^P=1^+$), symmetric in spin space, requires a flavor-symmetric pair for same-flavor quarks. In the spin-$\frac{1}{2}$ currents its index is contracted internally; in the spin-$\frac{3}{2}$ currents it provides the free external Lorentz index $\mu$ of the current.

\item $\Gamma_3 = C$: scalar coupling of the diquark system to the charm sector, introduces no Lorentz index. Appears in $J_1(x)$, $J_4(x)$, $J_5(x)$ (spin-$\frac{1}{2}$) and $J^1_\mu(x)$, $J^2_\mu(x)$, $J^5_\mu(x)$ (spin-$\frac{3}{2}$).

\item $\Gamma_4 = \gamma_5\gamma^\mu C$: axial-vector coupling to the charm sector; exclusive to the spin-$\frac{1}{2}$ currents $J_2(x)$, $J_3(x)$, $J_6(x)$. Its index $\mu$ contracts the index of the accompanying $\Gamma_2$ diquark, yielding no free external index and projecting the operator onto $J^P=\frac{1}{2}^-$.

\item $\Gamma_5 = C\gamma_\alpha$: same Dirac structure as $\Gamma_2$ but carries an independent internal Lorentz index $\alpha\neq\mu$. Appears only in the spin-$\frac{3}{2}$ currents $J^3_\mu(x)$, $J^4_\mu(x)$, $J^6_\mu(x)$, $J^7_\mu(x)$, always coexisting with a $\Gamma_2$ diquark that provides the external index $\mu$. The separation of $\mu$ (external) and $\alpha$ (internal) is essential for the spin-$\frac{3}{2}$ projection.

\item $\Gamma_6 = \gamma_5\gamma^\alpha C$: axial-vector coupling carrying the internal index $\alpha$. Contracts the $\Gamma_5$ index in the spin-$\frac{3}{2}$ currents, leaving $\mu$ from $\Gamma_2$ as the sole free index. Its role is directly analogous to $\Gamma_4$ in the spin-$\frac{1}{2}$ sector: both provide axial-vector couplings that
contract an internal vector index, the distinction being that $\Gamma_4$ contracts what would be the external index (thereby reducing spin to $\frac{1}{2}$), while $\Gamma_6$ contracts a separate internal index (leaving $\mu$ free and projecting onto spin $\frac{3}{2}$).
\end{itemize}

The currents $J_5(x)$, $J_6(x)$, $J^5_\mu(x)$, $J^6_\mu(x)$, $J^7_\mu(x)$ are 
antisymmetrized under light-quark exchange and project onto a definite isospin channel. Their diquark content and coupling structure are summarized in Table~\ref{tab:currents}.

\begin{table}[htbp]
\centering
\caption{Diquark content of the interpolating currents. S~=~scalar
($J^P=0^+$), A~=~axial-vector ($J^P=1^+$). The isospin column indicates whether the current is a
direct product (D) or an antisymmetric superposition (AS) under
light-quark exchange.}
\label{tab:currents}
\renewcommand{\arraystretch}{1.25}
\setlength{\tabcolsep}{6pt}
\begin{tabular}{lllllc}
\toprule
\toprule
Current & 1st diquark & 2nd diquark & Antiquark & Isospin & $J^P$ \\
\midrule
\midrule
$J_1(x)$ & $[s^T\Gamma_1 q]$ S & $[q^T\Gamma_1 c]$ S & $\Gamma_3$ & D  & $\frac{1}{2}^-$ \\
$J_2(x)$ & $[s^T\Gamma_1 q]$ S & $[q^T\Gamma_2 c]$ A & $\Gamma_4$ & D  & $\frac{1}{2}^-$ \\
$J_3(x)$ & $[q^T\Gamma_2 q]$ A & $[s^T\Gamma_1 c]$ S & $\Gamma_4$ & D  & $\frac{1}{2}^-$ \\
$J_4(x)$ & $[q^T\Gamma_2 q]$ A & $[s^T\Gamma_2 c]$ A & $\Gamma_3$ & D  & $\frac{1}{2}^-$ \\
$J_5(x)$ & $[q^T\Gamma_2 q]$ A & $[s^T\Gamma_2 c]$ A & $\Gamma_3$ & AS & $\frac{1}{2}^-$ \\
$J_6(x)$ & $[q^T\Gamma_2 q]$ A & $[s^T\Gamma_1 c]$ S & $\Gamma_4$ & AS & $\frac{1}{2}^-$ \\
\midrule
\midrule
$J^1_\mu(x)$ & $[s^T\Gamma_1 q]$ S & $[q^T\Gamma_2 c]$ A & $\Gamma_3$ & D  & $\frac{3}{2}^-$ \\
$J^2_\mu(x)$ & $[q^T\Gamma_2 q]$ A & $[s^T\Gamma_1 c]$ S & $\Gamma_3$ & D  & $\frac{3}{2}^-$ \\
$J^3_\mu(x)$ & $[q^T\Gamma_2 q]$ A & $[s^T\Gamma_5 c]$ A & $\Gamma_6$ & D  & $\frac{3}{2}^-$ \\
$J^4_\mu(x)$ & $[q^T\Gamma_5 q]$ A & $[s^T\Gamma_2 c]$ A & $\Gamma_6$ & D  & $\frac{3}{2}^-$ \\
$J^5_\mu(x)$ & $[q^T\Gamma_2 q]$ A & $[s^T\Gamma_1 c]$ S & $\Gamma_3$ & AS & $\frac{3}{2}^-$ \\
$J^6_\mu(x)$ & $[q^T\Gamma_2 q]$ A & $[s^T\Gamma_5 c]$ A & $\Gamma_6$ & AS & $\frac{3}{2}^-$ \\
$J^7_\mu(x)$ & $[q^T\Gamma_5 q]$ A & $[s^T\Gamma_2 c]$ A & $\Gamma_6$ & AS & $\frac{3}{2}^-$ \\
\bottomrule
\bottomrule
\end{tabular}
\end{table}

\subsection{Spin-$\boldsymbol{\frac{1}{2}}$ sector}
\label{subsec:spin12_formalism}

\subsubsection{Hadronic representation}

Inserting a complete set of spin-$\frac{1}{2}$ intermediate states into
Eq.~(\ref{eq:corr_12}) gives
\begin{equation}
\mathcal T^{\rm had}(p,q) =
\frac{\langle 0|J|P^{\Sigma}_{\psi s}(p)\rangle\,
\langle P^{\Sigma}_{\psi s}(p)|
P^{\Sigma}_{\psi s}(p{+}q)\rangle_F\,
\langle P^{\Sigma}_{\psi s}(p{+}q)|\bar J|0\rangle}
{[p^2-m^2][(p+q)^2-m^2]}
+\cdots,
\label{eq:had_12}
\end{equation}
where $m$ is the pentaquark mass and the ellipsis denotes contributions from higher resonances and the continuum. The coupling of $J(x)$ to the physical state is parametrized by the pole residue $\lambda_P$:
\begin{equation}
\langle 0|J(0)|P^{\Sigma}_{\psi s}(p,s)\rangle
= \lambda_P\,\gamma_5\,u(p,s),
\label{eq:residue_12}
\end{equation}
with $u(p,s)$ the Dirac spinor. The electromagnetic vertex of the spin-$\frac{1}{2}$ pentaquarks is decomposed in the standard form~\cite{Leinweber:1990dv}:
\begin{equation}
\langle P^{\Sigma}_{\psi s}(p)|
P^{\Sigma}_{\psi s}(p{+}q)\rangle_F
= \varepsilon^\mu\,\bar u(p)\Bigl[
f_1(q^2)\gamma_\mu
+\frac{f_2(q^2)}{2m}\,i\sigma_{\mu\nu}q^\nu
\Bigr]u(p{+}q),
\label{eq:vertex_12}
\end{equation}
where $f_1(q^2)$ and $f_2(q^2)$ are the Dirac and Pauli form factors.

Substituting Eqs.~(\ref{eq:residue_12}) and (\ref{eq:vertex_12}) into Eq.~(\ref{eq:had_12}), using the spin sum for Dirac spinors, and projecting onto the Lorentz structure $\pslash \eslash \qslash$, the hadronic side becomes
\begin{equation}
\mathcal T^{\rm had}(p,q)
=\frac{\lambda_P^2\,[f_1(q^2)+f_2(q^2)]}
{[p^2-m^2][(p+q)^2-m^2]}+\cdots.
\label{eq:had_12_proj}
\end{equation}
The magnetic dipole moment is defined at $q^2=0$:
\begin{equation}
\mu_{P^{\Sigma}_{\psi s}}
= \frac{e}{2m}\bigl[f_1(0)+f_2(0)\bigr].
\label{eq:mu_def_12}
\end{equation}

\subsubsection{QCD representation}

After Wick contraction, the QCD side of the spin-$\frac{1}{2}$ correlator for the current $J_1(x)$ takes the explicit form
\begin{align}
\mathcal T^{\rm QCD}_1(p,q) ={}& {-i}\,
\mathcal A \mathcal A^\prime 
\int d^4x\,e^{ip\cdot x}
\langle 0|\,
\mathrm{Tr}\bigl[\Gamma_1 \, S_s^{ee'}(x) \Gamma_1^\prime \, S_q^{dd'T}(x)\bigr]
\nonumber\\
&\times
\mathrm{Tr}\bigl[\Gamma_1 \, S_c^{gg'}(x) \Gamma_1^\prime \, S_s^{ff'T}(x)\bigr]\,
\Gamma_3 \, S_c^{c'cT}(-x) \, \Gamma_3^\prime
\,|0\rangle_F,
\label{eq:qcd_J1_12}
\end{align}
where the color indices on the propagators are matched to those of the Levi-Civita tensors. The remaining five spin-$\frac{1}{2}$ currents lead to analogous expressions with the $\Gamma_1$ matrices replaced by the Dirac structures listed in Table~\ref{tab:currents}.

Inserting the propagators Eqs.~(\ref{eq:Sq})--(\ref{eq:Sc}), implementing the photon couplings via Eqs.~(\ref{eq:pert})--(\ref{eq:nonpert}), and performing the Fourier
transformation, the momentum-space OPE result for the coefficient of
the $\pslash \eslash \qslash$ structure is denoted $\mathcal{R}_k(M^2,s_0)$ after  double Borel transformation with respect to $-p^2$ and $-(p+q)^2$ and continuum subtraction.

\subsubsection{Sum rule and extraction of the magnetic dipole moment}

Equating the hadronic and QCD representations at the level of the
$\pslash \eslash \qslash$ structure, applying quark-hadron
duality for the continuum, and performing the double Borel transformation yields the master sum rule for each spin-$\frac{1}{2}$
current:
\begin{equation}
\mu_{P^{\Sigma}_{\psi s}}(J_k)\,\lambda_P^2(J_k)\,
e^{-m^2(J_k)/M^2}
= \mathcal{R}_k(M^2,s_0),
\qquad k=1,\ldots,6.
\label{eq:SR_12}
\end{equation}
Here $M^2$ is the Borel mass parameter, $s_0$ is the continuum
threshold, and $\mathcal{R}_k(M^2,s_0)$ is the Borel-transformed,
continuum-subtracted QCD-side function encoding the contributions of
all condensates and photon DAs retained in the OPE. The magnetic dipole
moment is then extracted as
\begin{equation}
\mu_{P^{\Sigma}_{\psi s}}(J_k)
= \frac{\mathcal{R}_k(M^2,s_0)}{\lambda_P^2(J_k)}\,
e^{m^2(J_k)/M^2}.
\label{eq:mu_extract_12}
\end{equation}
The pole residues $\lambda_P(J_k)$ and masses $m(J_k)$ are taken from
the spectroscopic sum-rule analysis of~\cite{Wang:2026dqi}, which
employs the same interpolating currents. The explicit form of
$\mathcal{R}_1(M^2,s_0)$ for the current $J_1(x)$ is given in
Appendix~\ref{appa}; the remaining functions are obtained by the same
procedure.

The double Borel transformation is performed with respect to the 
variables $-p^2$ and $-(p+q)^2$ simultaneously. On the hadronic 
side, the two-propagator structure transforms as
\begin{align}
\mathcal{B}\left\{ 
\frac{1}{[p^2-m_i^2]\,[(p+q)^2-m_f^2]} 
\right\} 
\longrightarrow 
e^{-m_i^2/M_1^2 - m_f^2/M_2^2},
\end{align}
where $M_1^2$ and $M_2^2$ are the Borel parameters associated 
with the initial and final pentaquark states, respectively. 
On the QCD side, the momentum-space denominators arising from 
the OPE transform according to
\begin{align}
\label{eq:Borel_QCD}
\mathcal{B}\left\{ 
\frac{1}{(m^2 - \bar{u}\,p^2 - u(p+q)^2)^{\alpha}} 
\right\} 
\longrightarrow 
(M^2)^{2-\alpha}\,\delta(u - u_0)\,e^{-m^2/M^2},
\end{align}
where the combined Borel parameter $M^2$ and the momentum 
fraction $u_0$ are defined through
\begin{align}
M^2 = \frac{M_1^2\,M_2^2}{M_1^2 + M_2^2}, 
\qquad 
u_0 = \frac{M_1^2}{M_1^2 + M_2^2}.
\end{align}
Since the initial and final states are the same 
$P^{\Sigma}_{\psi s}$ pentaquark, the two propagator 
denominators in the hadronic representation 
[Eqs.~(\ref{eq:had_12}) and (\ref{eq:had_32})] are 
identical at the pole, and there is no kinematical reason 
to weight one channel more than the other. The symmetric 
choice $M_1^2 = M_2^2 \equiv 2M^2$ is therefore the natural 
one, and it yields $u_0 = M_1^2/(M_1^2+M_2^2) = \tfrac{1}{2}$.  
Under this choice the double Borel transformation reduces 
to a single-variable transformation parametrized by $M^2$, 
and the delta function $\delta(u-u_0)$ arising from the 
Borel transformation of the QCD-side denominators localizes 
the photon momentum fraction at the symmetric point. The 
suppression of higher resonances and continuum states in 
both channels is governed by the Borel exponentials 
$\exp(-m_i^2/M_1^2)$ and $\exp(-m_f^2/M_2^2)$, which for 
the symmetric choice and $m_i = m_f = m$ reduce to 
$\exp(-m^2/M^2)$. Any asymmetric choice $M_1^2 \neq M_2^2$ 
would weight the two channels unequally and could either 
under-suppress excited states in the channel with larger 
$M_i^2$ or distort the symmetric structure of the initial 
and final state contributions. The symmetric choice 
preserves the kinematical equivalence of the two channels 
and is the standard prescription in QCD sum-rule analyses 
of static hadron properties~\cite{Ball:2002ps, Ioffe:2005ym}.

\subsection{Spin-$\boldsymbol{\frac{3}{2}}$ sector} \label{subsec:spin32_formalism}

\subsubsection{Hadronic representation}

Inserting a complete set of spin-$\frac{3}{2}$ intermediate states into
Eq.~(\ref{eq:corr_32}) gives
\begin{equation}
\mathcal T^{\rm had}_{\mu\nu}(p,q) =
\frac{\langle 0|J_\mu|P^{\Sigma^\ast}_{\psi s}(p)\rangle\,
\langle P^{\Sigma^\ast}_{\psi s}(p)|
P^{\Sigma^\ast}_{\psi s}(p{+}q)\rangle_F\,
\langle P^{\Sigma^\ast}_{\psi s}(p{+}q)|\bar J_\nu|0\rangle}
{[p^2-{m^\ast}^2][(p+q)^2-{m^\ast}^2]}
+\cdots,
\label{eq:had_32}
\end{equation}
where $m^\ast$ is the spin-$\frac{3}{2}$ pentaquark mass and the
ellipsis denotes contributions from higher resonances and the
continuum. The current couples to the physical state through
\begin{equation}
\langle 0|J_\mu(0)|P^{\Sigma^\ast}_{\psi s}(p,s)\rangle
= \lambda_{P^\ast}\,u_\mu(p,s),
\label{eq:residue_32}
\end{equation}
with $u_\mu(p,s)$ the Rarita--Schwinger spinor satisfying
$\gamma^\mu u_\mu=0$, $p^\mu u_\mu=0$, and the normalization condition
$\bar u_\mu u^\mu = -2m^\ast$.

The most general electromagnetic vertex consistent with Lorentz
invariance and parity conservation
reads~\cite{Nozawa:1990gt,Pascalutsa:2006up,Ramalho:2009vc}:
\begin{equation}
\langle P^{\Sigma^\ast}_{\psi s}(p_2)|
P^{\Sigma^\ast}_{\psi s}(p_1)\rangle_F
= -e\,\bar u_\alpha(p_2)\,\Gamma^{\alpha\beta}\,u_\beta(p_1),
\label{eq:vertex_32}
\end{equation}
with
\begin{equation}
\Gamma^{\alpha\beta} =
F_1(q^2)\,g^{\alpha\beta}\eslash
-\frac{1}{2m^\ast}\Bigl[
F_2(q^2)\,g^{\alpha\beta}\eslash \qslash
+F_4(q^2)\,\frac{q^\alpha q^\beta \eslash \qslash}{4{m^\ast}^2}
\Bigr]
+\frac{F_3(q^2)}{4{m^\ast}^2}\,q^\alpha q^\beta \eslash.
\label{eq:Gamma32}
\end{equation}
The form factors $F_1,\ldots,F_4$ are extracted in the static limit $q^2\to 0$ and related to the multipole moments through the standard multipole decomposition~\cite{Nozawa:1990gt,Pascalutsa:2006up,Ramalho:2009vc}.
The electromagnetic multipole form factor is defined as
\begin{align}
G_M(q^2) &= \left[F_1(q^2)+F_2(q^2)\right]
\big(1+\frac{4}{5}\kappa\big)
-\frac{2}{5} \kappa\left(1+\kappa\right)\left[F_3(q^2)+F_4(q^2)\right],
\label{eq:GM_q2}\\[4pt]
G_{E2}(q^2) &= F_1(q^2)
-\kappa \, F_2(q^2)
-\frac{1}{2} \left(1+\kappa\right)\left[F_3(q^2)
-\kappa \, F_4(q^2)\right],
\label{eq:GE2_q2}\\[4pt]
G_{M3}(q^2) &= \left[F_1(q^2)+F_2(q^2)\right]
-\frac{1}{2} \left(1+\kappa\right) \left[F_3(q^2)+F_4(q^2)\right]
,
\label{eq:GM3_q2}
\end{align}
with $\kappa=-q^2/4{m^\ast}^2$ being the kinematic factor. 

In the static limit $q^2\to 0$ ($\kappa\to 0$), this reduces to
$G_M(0)=F_1(0)+F_2(0)$, so that the magnetic dipole moment is
\begin{equation}
\mu_{P^{\Sigma^\ast}_{\psi s}}
= \frac{e}{2m^\ast}\,G_M(0)
= \frac{e}{2m^\ast}\bigl[F_1(0)+F_2(0)\bigr].
\label{eq:mu_def_32}
\end{equation}
Similarly, the electric quadrupole and magnetic octupole moments are
related to the form factors at $q^2=0$ by~\cite{Nozawa:1990gt,Pascalutsa:2006up,Ramalho:2009vc}:
\begin{align}
\mathcal{Q}_{P^{\Sigma^\ast}_{\psi s}}
&= \frac{e}{(2m^\ast)^2}\,G_{E2}(0)
= \frac{e}{(2m^\ast)^2}
\Bigl[F_1(0) - \tfrac{1}{2}F_3(0)\Bigr],
\label{eq:Q_def}\\[4pt]
\mathcal{O}_{P^{\Sigma^\ast}_{\psi s}}
&= \frac{e}{(2m^\ast)^3}\,G_{M3}(0)
= \frac{e}{(2m^\ast)^3}
\Bigl[F_1(0)+F_2(0)
-\tfrac{1}{2}\bigl(F_3(0)+F_4(0)\bigr)\Bigr].
\label{eq:O_def}
\end{align}

Substituting Eqs.~(\ref{eq:residue_32}) and (\ref{eq:Gamma32}) into
Eq.~(\ref{eq:had_32}), using the spin-$\frac{3}{2}$ projector
\begin{equation}
\sum_s u_\mu(p,s)\bar u_\nu(p,s)
= -(\not\!p+m^\ast)\Bigl[
g_{\mu\nu}-\tfrac{1}{3}\gamma_\mu\gamma_\nu
-\tfrac{p_\mu\gamma_\nu-p_\nu\gamma_\mu}{3m^\ast}
-\tfrac{2p_\mu p_\nu}{3{m^\ast}^2}
\Bigr],
\label{eq:RS_projector}
\end{equation}
and isolating the independent Lorentz structures, the hadronic side
takes the explicit form
\begin{align}
\mathcal T^{\rm had}_{\mu\nu}(p,q) &={}
\frac{\lambda_{P^\ast}^2}
{[p^2-{m^\ast}^2][(p+q)^2-{m^\ast}^2]}
\Bigl[
F_1(q^2)\,g_{\mu\nu}\pslash \eslash \qslash
-m^\ast F_2(q^2)\,g_{\mu\nu}\eslash \qslash
+\frac{F_3(q^2)}{2m^\ast}\,q_\mu q_\nu\ \eslash \qslash
\nonumber\\
&
+\frac{F_4(q^2)}{4{m^\ast}^3}\, (\varepsilon \cdot p)
q_\mu q_\nu \pslash \qslash
+\cdots
\Bigr] + \cdots,
\label{eq:had_32_final}
\end{align}
where the ellipsis inside the brackets denotes Lorentz structures
not relevant for the multipole moments, and the outer ellipsis denotes
contributions from higher resonances and the continuum.

Before proceeding to the QCD side, two important technical aspects
must be addressed. First, the interpolating current $J_\mu$ for
spin-$\frac{3}{2}$ states can also couple to spin-$\frac{1}{2}$
pentaquark states. This contamination can be parametrized as
\begin{equation}
\langle 0 | J_{\mu}(0) | B(p, s=1/2) \rangle = (A p_{\mu} + B \gamma_{\mu}) u(p, s=1/2),
\label{eq:spin12_coupling}
\end{equation}
where $A$ and $B$ are constants, and $u(p,s=1/2)$ is the 
Dirac spinor. The presence of these spin-$\tfrac{1}{2}$ 
admixtures in $J_\mu$ implies that the QCD-side correlator 
in Eq.~(\ref{eq:corr_32}), and consequently its Lorentz 
decomposition in Eq.~(\ref{eq:had_32_final}), contains 
not only the desired spin-$\tfrac{3}{2}$ pole but also 
spin-$\tfrac{1}{2}$ contaminations whose Lorentz structure 
follows from substituting Eq.~(\ref{eq:spin12_coupling}) 
into Eqs.~(\ref{eq:corr_32}) and Eq.~(\ref{eq:had_32}). 
Carrying out this substitution, the spin-$\tfrac{1}{2}$ 
contributions to the correlator are linear combinations of 
the four structures
\begin{equation*}
\gamma_\mu (\cdots) \,, 
\qquad 
(\cdots) \gamma_\nu \,, 
\qquad 
p_\mu (\cdots) \,, 
\qquad 
(\cdots) p_\nu \,,
\label{eq:spin12_contamination_structures}
\end{equation*}
where the parentheses denote the residual Dirac structure 
arising from the spin sum and the propagator denominators. 
In other words, every spin-$\tfrac{1}{2}$ contamination 
either carries a free $\gamma_\mu$ matrix on its left, a 
free $\gamma_\nu$ matrix on its right, or a free $p_\mu$ 
or $p_\nu$ four-momentum at the corresponding open Lorentz 
index. By contrast, the genuine spin-$\tfrac{3}{2}$ 
contribution arising from Eq.~(\ref{eq:had_32_final}) 
contains no such structures: the Rarita--Schwinger 
projector~Eq.~(\ref{eq:RS_projector}) is constructed precisely 
to be orthogonal to $\gamma^\mu$ and $p^\mu$ on the indices 
$\mu$ and $\nu$. 
To exploit this orthogonality, we follow the prescription 
of~\cite{Belyaev:1982cd, Belyaev:1993ss} and rearrange the 
Dirac structure of every term in the OPE expansion of the 
correlator into the canonical ordering
\begin{equation*}
\gamma_\mu \pslash \eslash \qslash \gamma_\nu \,,
\label{eq:canonical_ordering}
\end{equation*}
using the standard anticommutation relation 
$\{\gamma_\alpha,\gamma_\beta\}=2 g_{\alpha\beta}$. This 
reordering moves all $\gamma_\mu$ factors to the leftmost 
position and all $\gamma_\nu$ factors to the rightmost 
position in each term, while generating additional terms 
proportional to $g_{\mu\nu}$, $p_\mu$, and $p_\nu$ from 
the anticommutators. Once every term is brought into this 
canonical form, the spin-$\tfrac{1}{2}$ contaminations 
identified in 
Eq.~(\ref{eq:spin12_coupling}) are made 
manifest: they are precisely the terms that retain a 
$\gamma_\mu$ at the leftmost position, a $\gamma_\nu$ at 
the rightmost position, or a free $p_\mu$ or $p_\nu$ factor 
after the rearrangement. These contaminating terms are then 
discarded. This procedure ensures that the extracted form 
factors correspond purely to the spin-$\tfrac{3}{2}$ state.  
Second, the four Lorentz structures displayed in
Eq.~(\ref{eq:had_32_final}) form a complete set of linearly
independent structures that project onto the four form factors
$F_1$ through $F_4$. Specifically, 
$g_{\mu\nu}\pslash\eslash\qslash$ isolates $F_1$,
which enters $\mu$, $\mathcal{Q}$, and $\mathcal{O}$;
$g_{\mu\nu}\eslash\qslash$ isolates $F_2$,
which enters $\mu$ and $\mathcal{O}$;
$q_\mu q_\nu\eslash\qslash$ isolates $F_3$,
which enters $\mathcal{Q}$ and $\mathcal{O}$;
and $(\varepsilon\cdot p)\,q_\mu q_\nu\pslash\qslash$ isolates $F_4$,
which enters $\mathcal{O}$ only.
These four structures are sufficient to determine all form 
factors without redundancy.

\subsubsection{QCD representation}

After Wick contraction, the QCD side for the current $J^1_\mu(x)$ is
\begin{align}
\mathcal T^{\rm QCD}_{\mu\nu,1}(p,q) ={}& {-i}\,
\mathcal A \mathcal A^\prime
\int d^4x\,e^{ip\cdot x}
\langle 0|\,
\mathrm{Tr}\bigl[ \Gamma_2 \, S_c^{gg'}(x)\Gamma_2^\prime \, S_s^{ff'T}(x)\bigr]
\nonumber\\
&\times
\mathrm{Tr}\bigl[\Gamma_1 \, S_s^{ee'}(x) \Gamma_1^\prime\, S_q^{dd'T}(x)\bigr]\,
\Gamma_3 \,S_c^{c'cT}(-x)\Gamma_3^\prime
\,|0\rangle_F.
\label{eq:qcd_J1_32}
\end{align}
The remaining six spin-$\frac{3}{2}$ currents yield analogous
contractions with the $\Gamma_i$ matrices replaced according to Table~\ref{tab:currents}.

After inserting the propagators, implementing the photon couplings, and
Fourier transforming, the OPE coefficients of the four Lorentz
structures listed above are extracted. Denoting the Borel-transformed,
continuum-subtracted QCD-side functions as $\mathcal{L}^{(1)}_\ell$,
$\mathcal{L}^{(2)}_\ell$, $\mathcal{L}^{(3)}_\ell$, $\mathcal{L}^{(4)}_\ell$
for the $g_{\mu\nu}\pslash \eslash \qslash$,
$g_{\mu\nu} \eslash \qslash$,
$q_\mu q_\nu \eslash \qslash$, and
$(\varepsilon\cdot p) q_\mu q_\nu \pslash \qslash$ structures
respectively, the sum rules take the forms given in
Sec.~\ref{subsubsec:SR_32}.

\subsubsection{Sum rules and extraction of the multipole moments}
\label{subsubsec:SR_32}

Equating the hadronic and QCD representations structure by structure,
applying quark-hadron duality, and performing the double Borel transformation with respect to
$-p^2$ and $-(p+q)^2$ simultaneously for both propagator denominators yields four
sum rules per current, one for each form factor:
\begin{align}
F_1(0;J^\ell_\mu)\,
\lambda_{P^\ast}^2(J^\ell_\mu)\,
e^{-{m^\ast}^2(J^\ell_\mu)/M^2}
&= \mathcal{L}^{(1)}_\ell(M^2,s_0),
\label{eq:SR_F1_32}\\[4pt]
F_2(0;J^\ell_\mu)\,
\lambda_{P^\ast}^2(J^\ell_\mu)\,
e^{-{m^\ast}^2(J^\ell_\mu)/M^2}
&= \mathcal{L}^{(2)}_\ell(M^2,s_0),
\label{eq:SR_F2_32}\\[4pt]
F_3(0;J^\ell_\mu)\,
\lambda_{P^\ast}^2(J^\ell_\mu)\,
e^{-{m^\ast}^2(J^\ell_\mu)/M^2}
&= \mathcal{L}^{(3)}_\ell(M^2,s_0),
\label{eq:SR_F3_32}\\[4pt]
F_4(0;J^\ell_\mu)\,
\lambda_{P^\ast}^2(J^\ell_\mu)\,
e^{-{m^\ast}^2(J^\ell_\mu)/M^2}
&= \mathcal{L}^{(4)}_\ell(M^2,s_0),
\qquad \ell=1,\ldots,7.
\label{eq:SR_F4_32}
\end{align}
The multipole moments are then obtained by combining the extracted form
factors according to Eqs.~(\ref{eq:mu_def_32})--(\ref{eq:O_def}):
\begin{align}
\mu_{P^{\Sigma^\ast}_{\psi s}}(J^\ell_\mu)
&= \frac{e}{2m^\ast(J^\ell_\mu)}\,
\frac{e^{{m^\ast}^2(J^\ell_\mu)/M^2}}
{\lambda_{P^\ast}^2(J^\ell_\mu)}\, \Bigr[ \mathcal{L}^{(1)}_\ell-  \frac{1}{m^*(J^\ell_\mu)}\, \mathcal{L}^{(2)}_\ell\Bigl],
\label{eq:mu_extract_32}\\[4pt]
\mathcal{Q}_{P^{\Sigma^\ast}_{\psi s}}(J^\ell_\mu)
&= \frac{e}{[2m^\ast(J^\ell_\mu)]^2}\,
\frac{e^{{m^\ast}^2(J^\ell_\mu)/M^2}}
{\lambda_{P^\ast}^2(J^\ell_\mu)}\,
 \Bigr [ \mathcal{L}^{(1)}_\ell- m^*(J^\ell_\mu)\mathcal{L}^{(3)}_\ell \Bigl],
\label{eq:Q_extract_32}\\[4pt]
\mathcal{O}_{P^{\Sigma^\ast}_{\psi s}}(J^\ell_\mu)
&= \frac{e}{[2m^\ast(J^\ell_\mu)]^3}\,
\frac{e^{{m^\ast}^2(J^\ell_\mu)/M^2}} {\lambda_{P^\ast}^2(J^\ell_\mu)}\,\Bigr[  \mathcal{L}^{(1)}_\ell-\frac{1}{m^*(J^\ell_\mu)}\, \mathcal{L}^{(2)}_\ell  
-m^*(J^\ell_\mu)\, \mathcal{L}^{(3)}_\ell-2 m^{*3}(J^\ell_\mu)\mathcal{L}^{(4)}_\ell \Bigl]
.
\label{eq:O_extract_32}
\end{align}
The pole residues $\lambda_{P^\ast}(J^\ell_\mu)$ and masses
$m^\ast(J^\ell_\mu)$ are taken from \cite{Wang:2026dqi}.
The explicit analytical expressions for the magnetic dipole moments 
corresponding to the current $J^1_\mu(x)$ are collected in
Appendix~\ref{appa}; the remaining functions follow from the same
procedure with the Dirac structures of Table~\ref{tab:currents}.

\end{widetext}

\section{Numerical analysis}\label{sec:numerical}

This section presents the numerical implementation of the LCSR 
for the electromagnetic multipole moments of the $P^{\Sigma}_{\psi s}$
pentaquarks. We specify the input parameters, determine working regions
for the Borel mass $M^2$ and continuum threshold $s_0$ by applying
standard stability criteria, and estimate the uncertainties on the
extracted observables.

\subsection{Input parameters}
\label{subsec:inputs}

The calculation requires quark masses and non-perturbative QCD condensate
parameters as inputs. For the quark masses we use~\cite{ParticleDataGroup:2024cfk}
\begin{align}
m_s   &= 93.5  \pm 0.8~\text{MeV},  \\
m_c   &= 1.273 \pm 0.0046~\text{GeV},
\end{align}
with $m_u = m_d = 0$ in the isospin-symmetric chiral limit. For the non-perturbative condensate parameters we
adopt~\cite{Ioffe:2005ym, Narison:2018nbv}
\begin{align}
\langle \bar{q}q \rangle &= -(0.24 \pm 0.01)^3~\text{GeV}^3, \\
\langle \bar{s}s \rangle &= (0.8 \pm 0.1)\,\langle \bar{q}q \rangle, \\
\langle g_s^2 G^2 \rangle &= 0.48 \pm 0.14~\text{GeV}^4.
\end{align}
The ratio $\langle\bar{s}s\rangle/\langle\bar{q}q\rangle = 0.8\pm 0.1$
encodes the $SU(3)$ flavor breaking of the chiral condensate.  
The external electromagnetic field couples to the quarks through photon 
DAs, for which we adopt the parametrization of \cite{Ball:2002ps} up to twist-4. The magnetic susceptibility of the quark condensate,
\begin{equation}
\chi = -2.85 \pm 0.5~\text{GeV}^{-2},
\end{equation}
controls the leading-twist coupling of the quark condensate to the photon
DAs~\cite{Rohrwild:2007yt}. The nonperturbative constant
$f_{3\gamma} = -0.0039~\text{GeV}^2$~\cite{Ball:2002ps} enters at
dimension-2 in the OPE. The masses and pole residues of the
$P^{\Sigma}_{\psi s}$ states are taken from our previous spectroscopic
analysis~\cite{Wang:2026dqi} using the same interpolating currents.

\subsection{OPE structure and working regions}
\label{subsec:ope_stability}
 
The LCSR analysis involves two auxiliary parameters that must be
determined before the multipole moments can be extracted: the Borel mass
$M^2$ and the continuum threshold $s_0$. Neither is a physical
observable, and the extracted predictions should be stable under their
variation within a well-defined working region. Two conditions, one on
the OPE and one on the hadronic representation, are imposed simultaneously
to constrain this region from opposite sides in $M^2$.
 
The OPE for the correlation function in an external electromagnetic field
includes contributions from operator dimensions 1 through 7:
\begin{itemize}
  \item Dimension-1: $\langle\bar{q}q\rangle\chi$ and
    $\langle\bar{s}s\rangle\chi$;
  \item Dimension-2: $f_{3\gamma}$;
  \item Dimension-3: $\langle\bar{q}q\rangle$ and $\langle\bar{s}s\rangle$;
  \item Dimension-5: $\langle g_s^2 G^2\rangle\langle\bar{q}q\rangle\chi$
    and $\langle g_s^2 G^2\rangle\langle\bar{s}s\rangle\chi$;
  \item Dimension-6: $\langle g_s^2 G^2\rangle f_{3\gamma}$;
  \item Dimension-7: $\langle g_s^2 G^2\rangle\langle\bar{q}q\rangle$
    and $\langle g_s^2 G^2\rangle\langle\bar{s}s\rangle$.
\end{itemize}
As $M^2$ decreases, the Borel suppression of higher-dimensional terms
weakens and the OPE becomes increasingly dominated by the highest
retained operators. To ensure that the truncation at dimension-7 is
justified and does not distort the result, we require the OPE convergence
ratio
\begin{equation}
\mathrm{CVG}(M^2,s_0) \equiv
\frac{\left|\Pi^{\mathrm{Dim7}}(M^2,s_0)\right|}
     {\left|\Pi(M^2,s_0)\right|} < 1\%,
\label{eq:cvg}
\end{equation}
where $\Pi(M^2,s_0)$ denotes the relevant QCD sum-rule function after Borel transformation and continuum subtraction: $\mathcal{R}_{k}(M^2,s_0)$
for the spin-$\frac{1}{2}$ currents and $\mathcal{L}_\ell^{(i)}(M^2,s_0)$ for the
spin-$\frac{3}{2}$ currents (see Sec.~\ref{sec:formalism}), and
$\Pi^{\mathrm{Dim7}}$ is the dimension-7 contribution to the same
function. This condition sets a lower bound on $M^2$: for a
given $s_0$, only values of $M^2$ above a current-dependent threshold
satisfy CVG $<1\%$.
 
As $M^2$ increases, the Borel exponential suppresses the ground-state
contribution relative to the continuum, reducing the sensitivity of the
sum rule to the state of interest. We therefore impose a lower bound on
the pole contribution
\begin{equation}
\mathrm{PC}(M^2, s_0)
\equiv \frac{\Pi(M^2, s_0)}{\Pi(M^2, \infty)} \geq 40\%,
\label{eq:pc}
\end{equation}
where $\Pi(M^2,\infty)$ is the same function evaluated without continuum
subtraction. This condition sets an upper bound on $M^2$: beyond
a current-dependent value, the ground-state pole no longer dominates the
spectral integral and the extraction becomes unreliable.
 
\begin{figure}[htbp]
\centering
\includegraphics[width=0.35\textwidth]{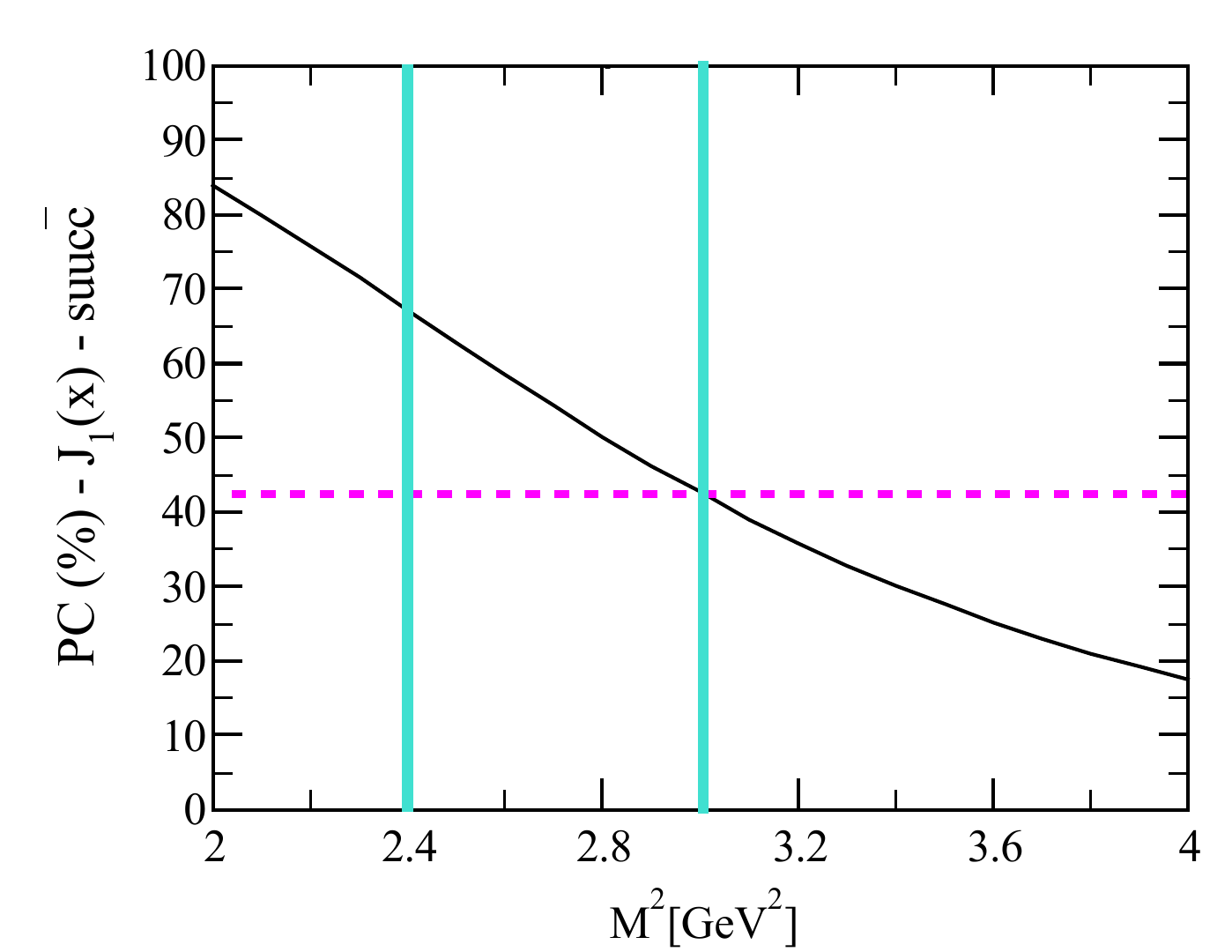} \qquad \qquad
\includegraphics[width=0.35\textwidth]{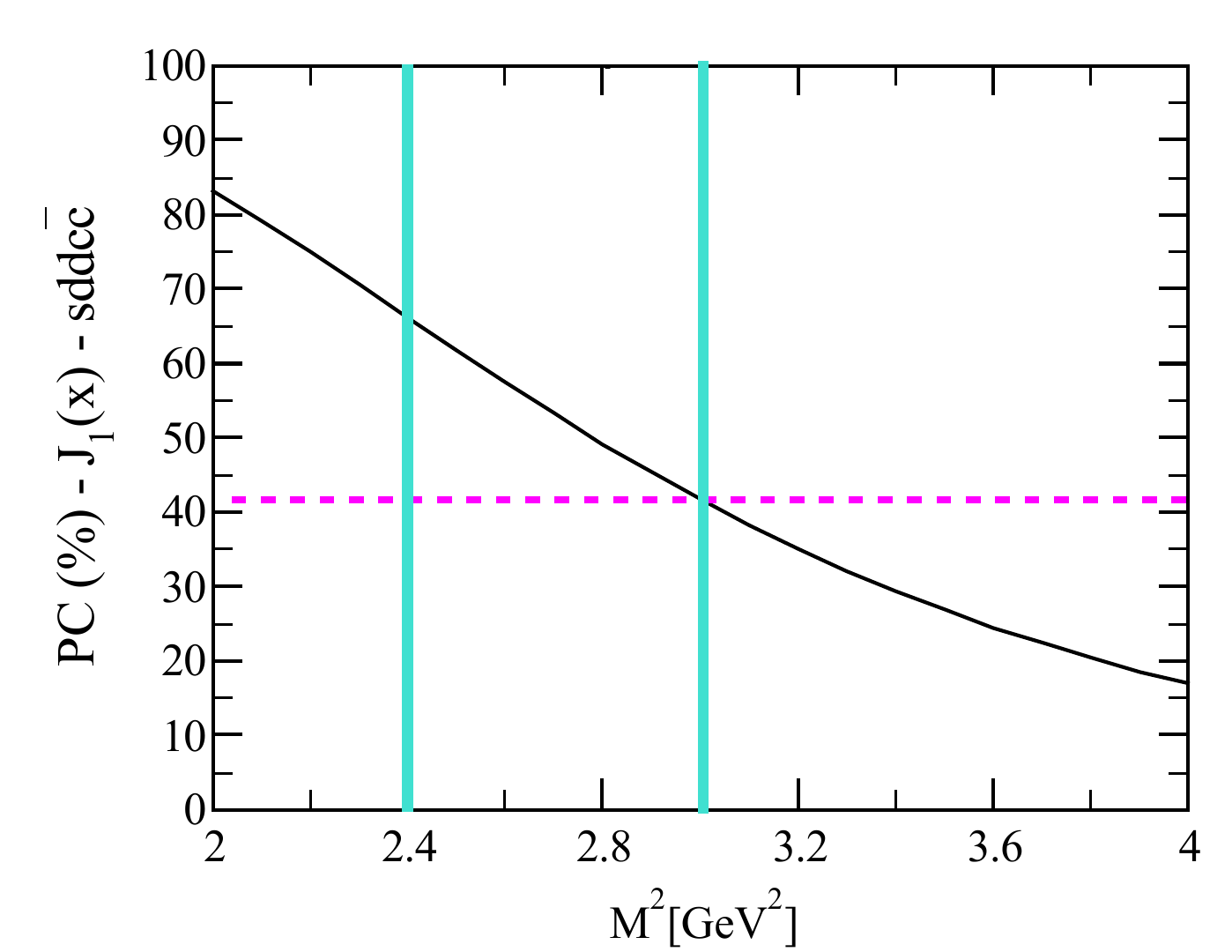}\\[15pt]
\includegraphics[width=0.35\textwidth]{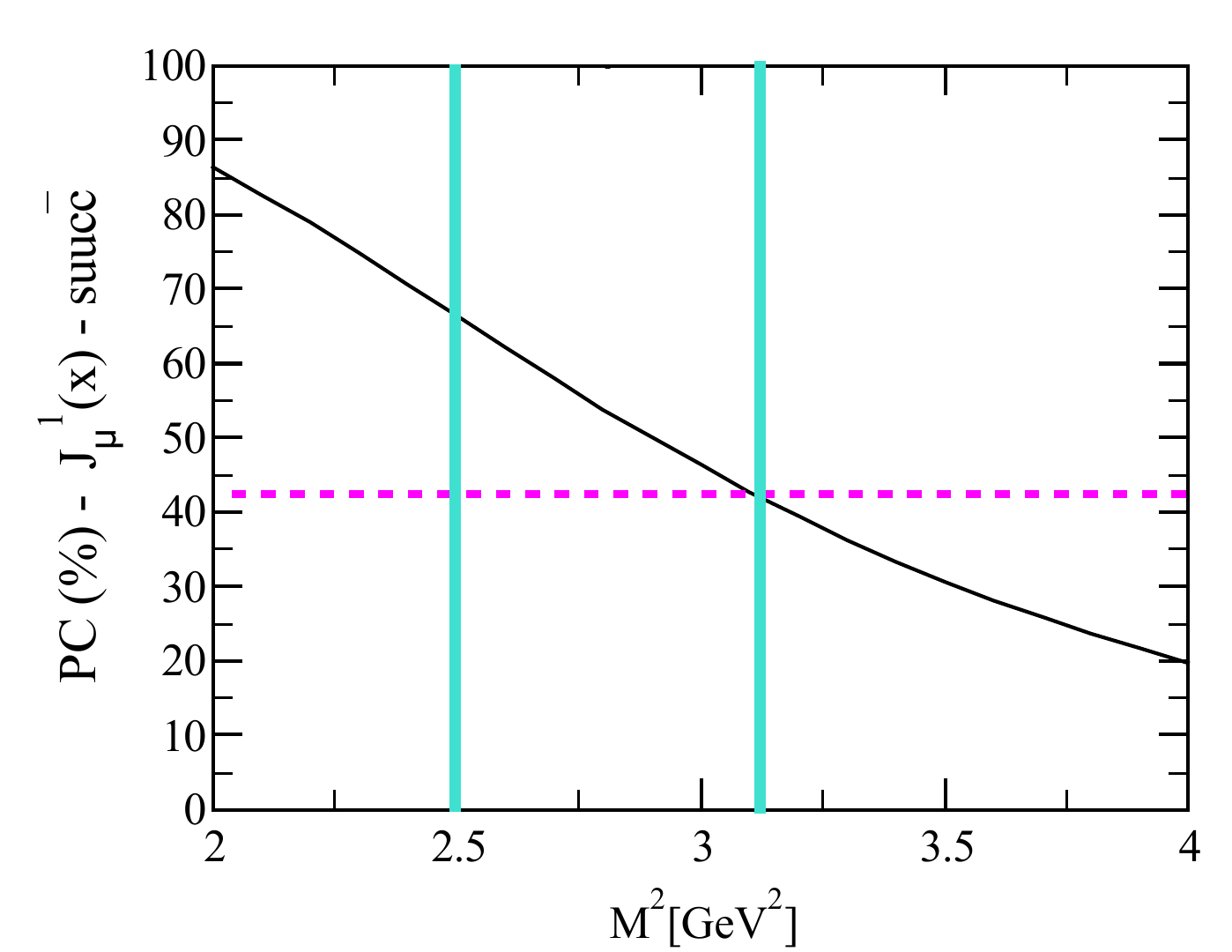} \qquad \qquad
\includegraphics[width=0.35\textwidth]{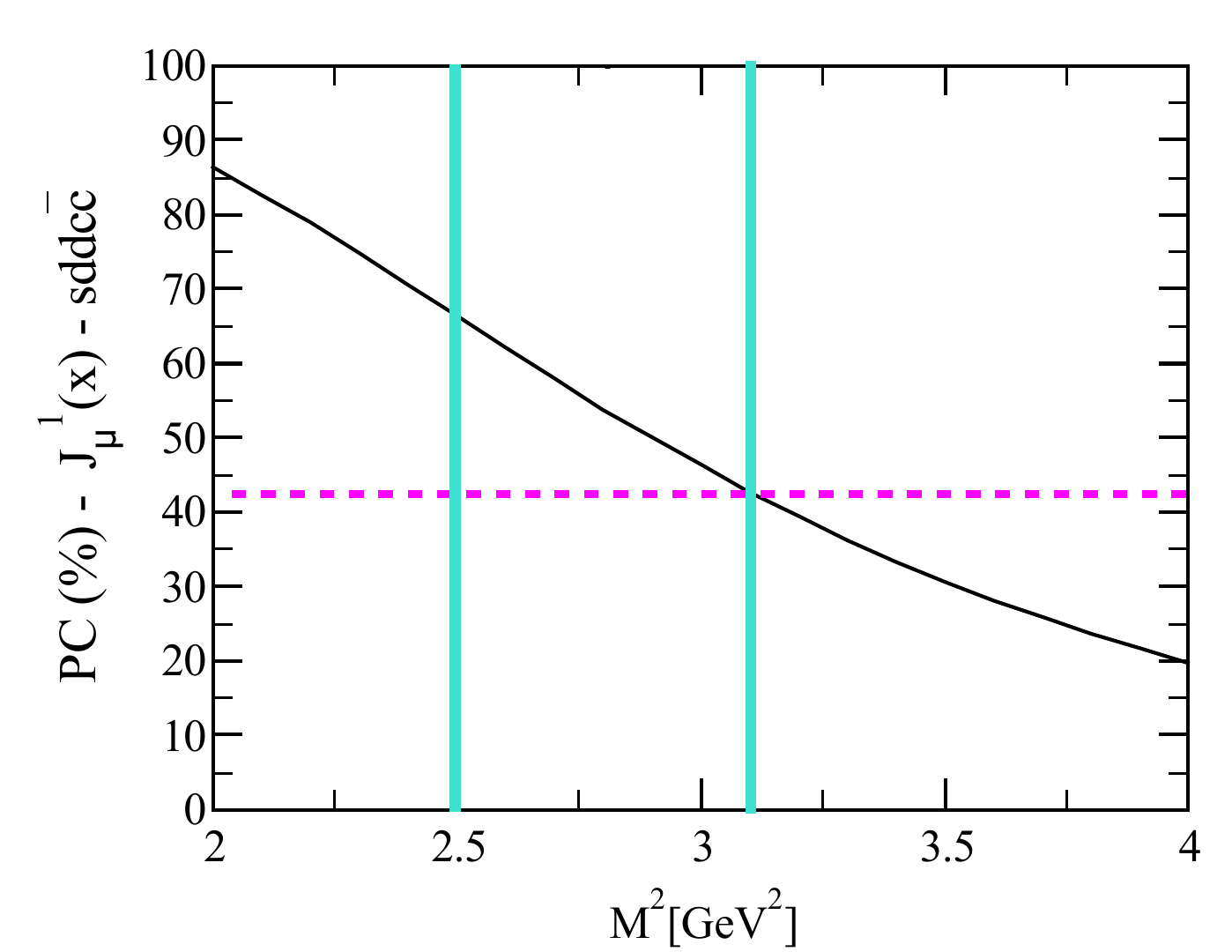}\\[15pt]
\includegraphics[width=0.35\textwidth]{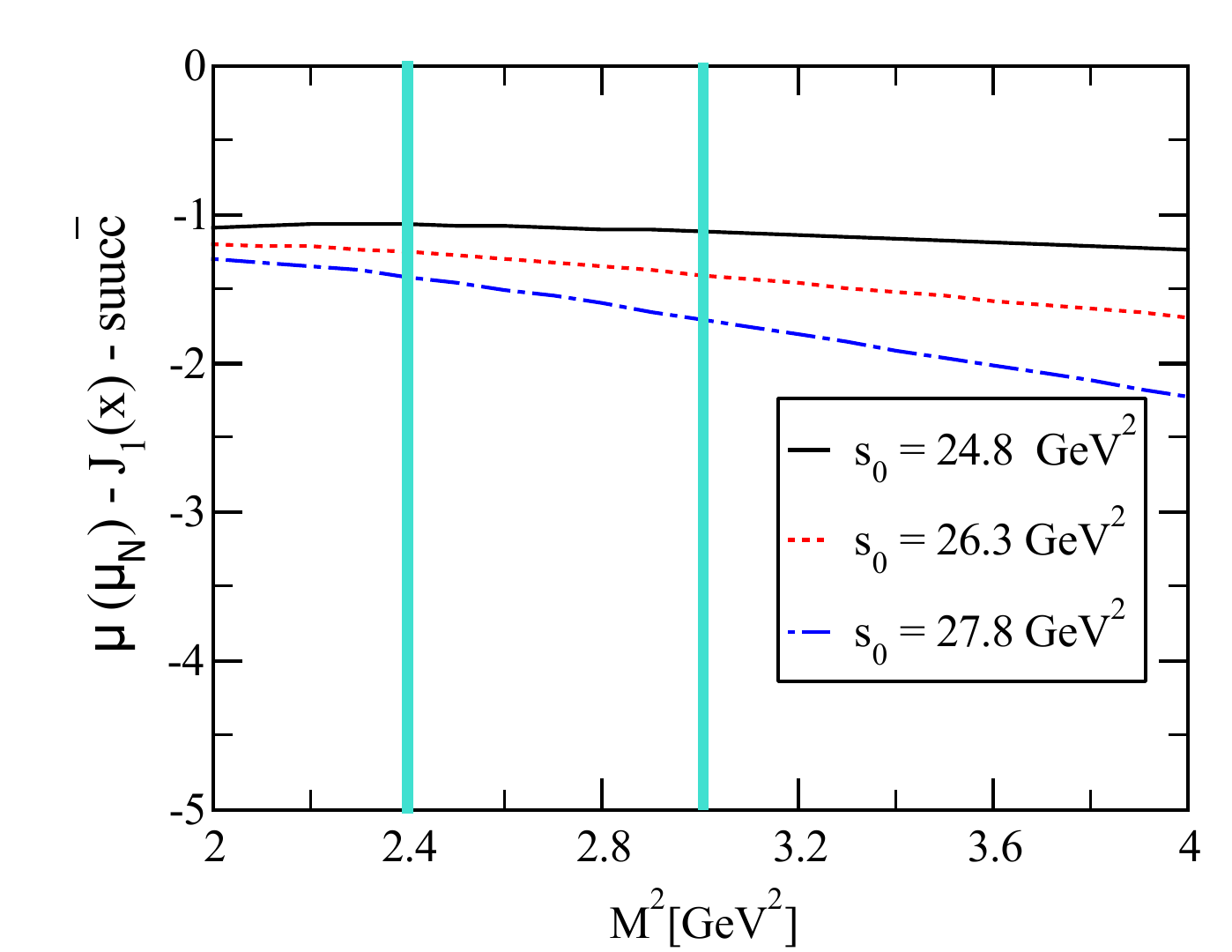} \qquad \qquad
\includegraphics[width=0.35\textwidth]{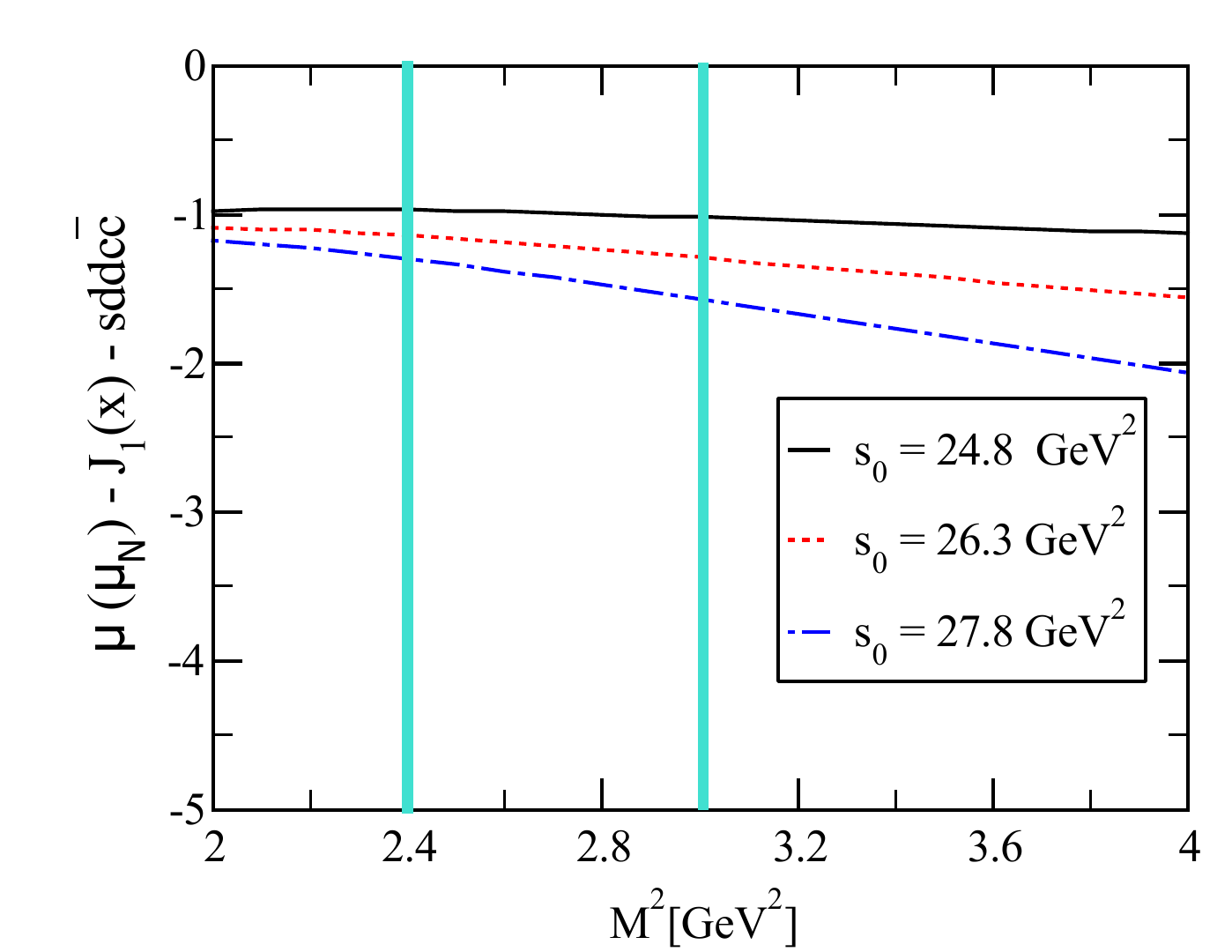}\\[15pt]
\includegraphics[width=0.35\textwidth]{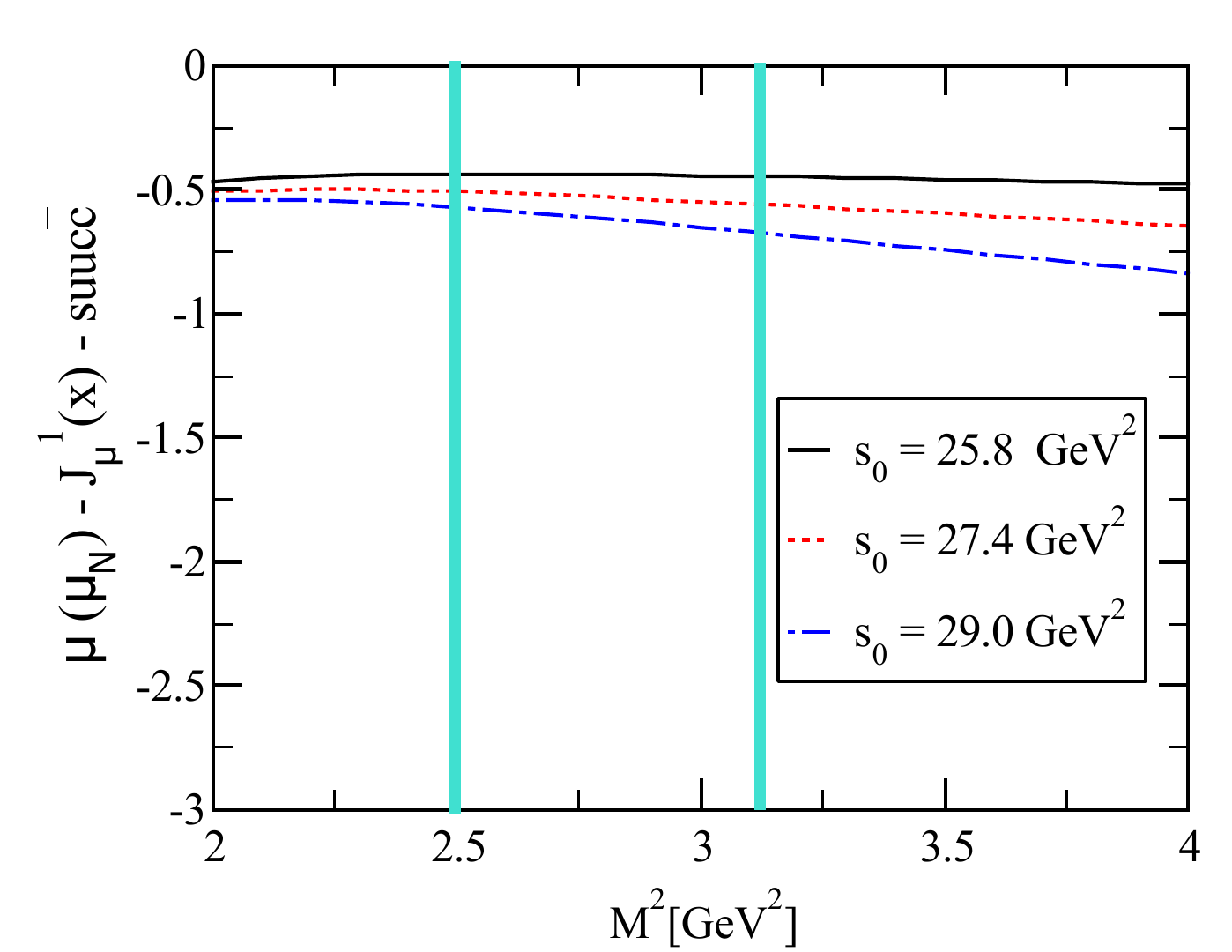} \qquad \qquad
\includegraphics[width=0.35\textwidth]{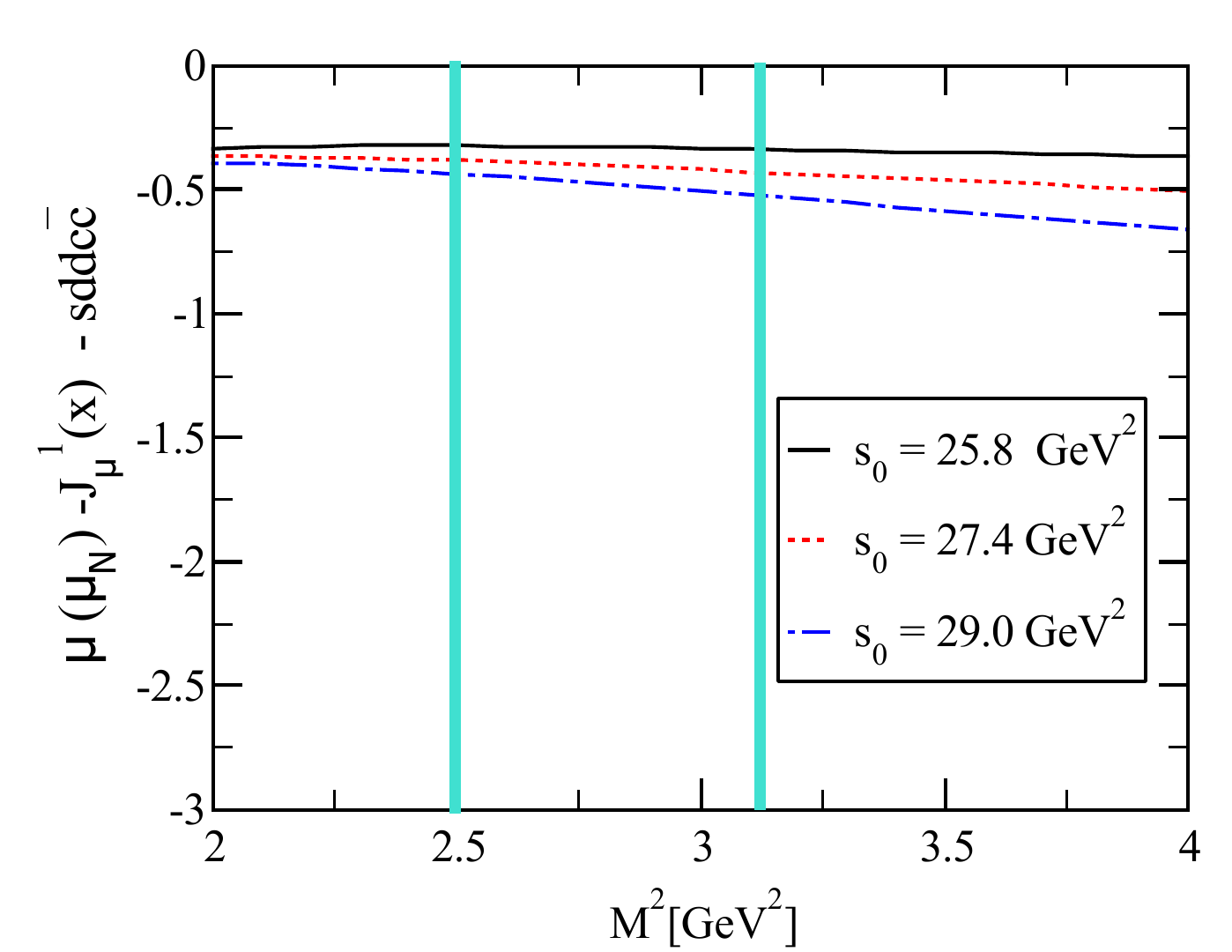}
\caption{LCSR analysis for the currents $J_1(x)$ (rows 1 and 3) and $J_\mu^1(x)$ (rows 2 and 4) in the $[su][uc]\bar{c}$ (left column) and $[sd][dc]\bar{c}$ (right column) configurations. Rows 1–2: pole contribution PC as a function of $M^2$ at the central $s_0$; the horizontal dashed line marks the PC value at the 
upper boundary of the Borel window, the shaded band indicates the working region. Rows 3–4: magnetic dipole moment $\mu$ (in $\mu_N$) as a function of $M^2$ for three values of $s_0$; the stability within the working region confirms the reliability of the extraction.}
\label{fig:analysis_J1}
\end{figure}

The working region for $M^2$ is thus the interval where both conditions
are simultaneously satisfied: CVG $<1\%$ from below and PC $\geq 40\%$
from above. Within this interval the extracted multipole moments vary by
at most $\sim 8\%$ with $M^2$ at fixed $s_0$, which we include in the
uncertainty budget.  
The continuum threshold parameter $s_0$ marks the beginning of excited-state and continuum contributions within the hadronic representation of the correlation function. From a phenomenological standpoint, $s_0$ is typically expected to lie slightly above the square of the ground-state mass. Following standard conventions in QCD sum-rule analyses, we adopt the range 
$(m_H + 0.5)^2 \, \text{GeV}^2 \leq s_0 \leq (m_H + 0.8)^2 \, \text{GeV}^2,$ 
where \( m_H \) denotes the mass of the hadronic state of interest. This selection has been shown to yield stable predictions in analogous studies and is therefore adopted throughout the present work. The resulting
$s_0$ values cluster in the range $23.5$--$29.1~\mathrm{GeV}^2$, as
listed in Table~\ref{tab:working_regions}.  
The CVG values obtained lie between $0.10\%$ and $0.69\%$ across all
currents, well within the $1\%$ bound. At the lower end of the Borel
window the PC values range from $66.6\%$ ($J^6_\mu(x)$) to $73.0\%$
($J^3_\mu(x)$); at the upper end from $40.1\%$ ($J^3_\mu(x)$) to $45.0\%$
($J_3(x)$). Currents involving more light-quark loops --- $J_3(x)$, $J_6(x)$,
$J^4_\mu(x)$, $J^6_\mu(x)$, $J^7_\mu(x)$ --- tend to produce larger CVG values
because they generate additional condensate structures at higher operator
dimensions; this variation should be kept in mind when interpreting the
results.  
Figure~\ref{fig:analysis_J1} illustrates the working-region
determination for the currents $J_1(x)$ and $J^1_\mu(x)$ as representative
examples. The upper panels display the PC as a
function of $M^2$ at the central value of $s_0$; the horizontal
dashed line marks the PC value at the upper boundary of the Borel
window for the corresponding current, and the shaded band identifies
the interval in $M^2$ where the PC condition (PC $\geq 40\%$) is
simultaneously satisfied with CVG $< 1\%$. The lower panels show the
magnetic dipole moment $\mu$ as a function of $M^2$ for three values
of $s_0$ spanning the working range. The extracted $\mu$ exhibits a
weak dependence on $M^2$ within the shaded region, while the
variation with $s_0$ is more pronounced but remains under control,
as reflected in the quoted uncertainty from $s_0$ variation.
\begin{table}[htbp]
\centering
\caption{Working regions in the $(M^2, s_0)$ plane determined by the
simultaneous requirements of OPE convergence (CVG $< 1\%$) and
pole dominance (PC $\geq 40\%$) for each interpolating current.
The CVG and PC values are evaluated at the lower and upper
boundaries of the Borel window.}
\label{tab:working_regions}
\renewcommand{\arraystretch}{1.35}
\setlength{\tabcolsep}{8pt}
\begin{tabular}{lcccc}
\toprule
Current
  & $s_0~(\mathrm{GeV}^2)$
  & $M^2~(\mathrm{GeV}^2)$
  & CVG (\%)
  & PC (\%) \\
\midrule
\multicolumn{5}{l}{\textit{Spin-$\frac{1}{2}$ sector}} \\[4pt]
$J_1(x)$  & $[24.8,\;27.8]$ & $[2.4,\;3.0]$ & $0.12$ & $[67.3,\;42.6]$ \\
$J_2(x)$  & $[25.9,\;29.1]$ & $[2.5,\;3.2]$ & $0.10$ & $[67.9,\;41.0]$ \\
$J_3(x)$  & $[24.7,\;27.7]$ & $[2.3,\;2.9]$ & $0.11$ & $[70.3,\;45.0]$ \\
$J_4(x)$  & $[25.0,\;28.0]$ & $[2.4,\;3.0]$ & $0.12$ & $[68.0,\;43.4]$ \\
$J_5(x)$  & $[23.5,\;26.5]$ & $[2.2,\;2.8]$ & $0.38$ & $[69.5,\;43.1]$ \\
$J_6(x)$  & $[23.5,\;26.5]$ & $[2.2,\;2.8]$ & $0.10$ & $[69.2,\;42.9]$ \\
\midrule
\multicolumn{5}{l}{\textit{Spin-$\frac{3}{2}$ sector}} \\[4pt]
$J_\mu^1(x)$  & $[25.8,\;29.0]$ & $[2.5,\;3.1]$ & $0.13$ & $[68.7,\;42.6]$ \\
$J_\mu^2(x)$  & $[24.8,\;27.8]$ & $[2.4,\;3.0]$ & $0.69$ & $[68.1,\;43.4]$ \\
$J_\mu^3(x)$  & $[25.2,\;28.2]$ & $[2.4,\;3.2]$ & $0.61$ & $[73.0,\;40.1]$ \\
$J_\mu^4(x)$  & $[25.2,\;28.2]$ & $[2.4,\;3.1]$ & $0.27$ & $[70.0,\;42.6]$ \\
$J_\mu^5(x)$  & $[23.8,\;26.8]$ & $[2.3,\;2.9]$ & $0.22$ & $[67.6,\;42.1]$ \\
$J_\mu^6(x)$  & $[24.7,\;27.7]$ & $[2.5,\;3.1]$ & $0.11$ & $[66.6,\;43.3]$ \\
$J_\mu^7(x)$  & $[24.7,\;27.7]$ & $[2.2,\;2.9]$ & $0.49$ & $[70.3,\;41.3]$ \\
\bottomrule
\end{tabular}
\end{table}

\section{Results and discussion}\label{sec:results}
Throughout this section, results are presented for all three 
charge states of the $P^{\Sigma}_{\psi s}$ isospin triplet: 
the $[sd][dc]\bar{c}$ ($\Sigma^-$-like), $[sd][uc]\bar{c}$ 
($\Sigma^0$-like), and $[su][uc]\bar{c}$ ($\Sigma^+$-like) 
configurations. The electromagnetic multipole moments of the 
$P^{\Sigma}_{\psi s}$ pentaquarks are computed using the 
interpolating currents introduced in Sec.~\ref{sec:formalism}. The numerical results are presented in Tables~\ref{tab:multipole_results_spin12}--\ref{tab:octupole_spin32}. 
The Borel parameter $M^2$, continuum threshold $s_0$, CVG, and PC for each current have been established in Sec.~\ref{sec:numerical} and are collected in Table~\ref{tab:working_regions} for reference.

The robustness of the LCSR predictions for the electromagnetic multipole moments of $P_{\psi s}^{\Sigma}$ states is assessed through a systematic sensitivity analysis. The total theoretical uncertainty is estimated to be in the range of $20$--$26\%$, obtained by varying each input parameter within its working range and combining the resulting deviations in quadrature, which provides a conservative estimate. The dominant source of uncertainty originates from the continuum threshold $s_0$ (at the level of $\sim 25\%$), reflecting the intrinsic sensitivity of the sum-rule approach to the modeling of higher-state contributions. Additional significant contributions arise from the spectroscopic  parameters, namely the pole residue $\lambda$ ($\sim 20\%$) and  the pentaquark mass $m$ ($\sim 15$--$20\%$). The remaining inputs, including the photon DAs, quark masses ($m_s, m_c$), and QCD condensates, contribute at the $5$--$10\%$ level. We have also verified that the extracted multipole moments exhibit good stability with respect to the Borel parameter and a moderate dependence on $s_0$ within the chosen working window. Importantly, these uncertainties do not affect the qualitative physical conclusions, such as the predicted signs of the moments or the structural differentiation among different diquark--diquark--antiquark configurations.
%
\subsection{Spin-$1/2$ sector}
\label{subsec:spin12}

For the $J^{P} = \frac{1}{2}^{-}$ channels only the magnetic dipole moment
$\mu$ is defined. The results are listed in Table~\ref{tab:multipole_results_spin12}
and the quark-level decomposition in Table~\ref{tab:quark_magnetic_spin12}. 
The magnetic dipole moments span a wide range across the six currents,
from $-7.40 \pm 1.92\,\mu_N$ ($J_3(x)$, $[su][uc]\bar{c}$) to
$+4.27 \pm 1.11\,\mu_N$ ($J_6(x)$, $[sd][dc]\bar{c}$), and exhibit a
systematic pattern that is entirely governed by the spin structure of the
interpolating operator rather than by numerical accident. 
\begin{table}[htbp]
\centering
\caption{Magnetic dipole moments $\mu$ (in $\mu_N$) of the 
$P^{\Sigma}_{\psi s}$ pentaquarks in the spin-$\tfrac{1}{2}$ 
sector, computed from the LCSR for each interpolating current 
and flavor configuration.}
\label{tab:multipole_results_spin12}
\renewcommand{\arraystretch}{1.25}
\setlength{\tabcolsep}{12pt}
\begin{tabular}{llc}
\toprule
Current & Configuration & $\mu_{P^{\Sigma}_{\psi s}}$ \\
\midrule
\multirow{3}{*}{$J_1(x)$}
  & $[sd][dc]\bar{c}$ & $-1.21 \pm 0.31$ \\
  & $[sd][uc]\bar{c}$ & $-1.27 \pm 0.33$ \\
  & $[su][uc]\bar{c}$ & $-1.32 \pm 0.35$ \\[4pt]
\multirow{3}{*}{$J_2(x)$}
  & $[sd][dc]\bar{c}$ & $\phantom{-}3.50 \pm 0.91$ \\
  & $[sd][uc]\bar{c}$ & $-0.10 \pm 0.03$ \\
  & $[su][uc]\bar{c}$ & $-3.70 \pm 0.96$ \\[4pt]
\multirow{3}{*}{$J_3(x)$}
  & $[sd][dc]\bar{c}$ & $\phantom{-}3.76 \pm 0.98$ \\
  & $[sd][uc]\bar{c}$ & $-1.82 \pm 0.47$ \\
  & $[su][uc]\bar{c}$ & $-7.40 \pm 1.92$ \\[4pt]
\multirow{3}{*}{$J_4(x)$}
  & $[sd][dc]\bar{c}$ & $-1.77 \pm 0.46$ \\
  & $[sd][uc]\bar{c}$ & $-1.85 \pm 0.48$ \\
  & $[su][uc]\bar{c}$ & $-1.92 \pm 0.49$ \\[4pt]
\multirow{3}{*}{$J_5(x)$}
  & $[sd][dc]\bar{c}$ & $-1.40 \pm 0.36$ \\
  & $[sd][uc]\bar{c}$ & $-1.47 \pm 0.38$ \\
  & $[su][uc]\bar{c}$ & $-1.54 \pm 0.40$ \\[4pt]
\multirow{3}{*}{$J_6(x)$}
  & $[sd][dc]\bar{c}$ & $\phantom{-}4.27 \pm 1.11$ \\
  & $[sd][uc]\bar{c}$ & $-0.84 \pm 0.22$ \\
  & $[su][uc]\bar{c}$ & $-5.94 \pm 1.54$ \\
\bottomrule
\end{tabular}
\end{table}

\paragraph{Scalar-diquark currents ($J_1(x)$, $J_4(x)$, $J_5(x)$):}
In these currents the light-quark diquarks are scalar,
$[q^T C\gamma_5\, q']$, pairing two quarks in a spin-singlet ($J^P = 0^+$)
configuration. The key consequence for electromagnetic observables is the
following: the spin-singlet diquark carries zero net angular momentum and
therefore contributes nothing to the magnetic moment through its spin. The
entire magnetic response of the system is then routed through the
charm-sector coupling, encoded at the operator level by the $\Gamma_3 = C$
or $\Gamma_4 = \gamma_5 \gamma^\mu C$ structure connecting the diquark
system to the charm propagator. This is directly visible in
Table~\ref{tab:quark_magnetic_spin12}: the charm contribution $\mu_c$ accounts for $95$--$99\%$ of the total in all three currents, while light-quark contributions are at the sub-percent level ($\lesssim 0.07\,\mu_N$).

The resulting moments, $\mu \in [-1.92,\,-1.21]\,\mu_N$, are nearly
flavor-independent: replacing $d$ by $u$ (or $s$ by $u$) changes the total
by at most $0.15\,\mu_N$. This flavor insensitivity is a direct consequence
of the spin-singlet suppression of light-quark contributions and constitutes
a characteristic electromagnetic signature of the scalar-diquark structure.
It is worth noting that this pattern is precisely what heavy-quark spin
symmetry predicts in the $m_c \to \infty$ limit: the light degrees of
freedom decouple from the heavy sector, and the electromagnetic properties
of the system are determined by the charm sector alone. The scalar-diquark
currents, by projecting the light quarks into spin-singlet pairs, realize
this decoupling at finite $m_c$ and at the operator level.

The superposition current $J_5(x)$, which is antisymmetric under light-quark
exchange and projects onto a definite isospin channel, yields
$\mu = -1.40 \pm 0.36$ and $-1.54 \pm 0.40\,\mu_N$, values indistinguishable
from those of $J_1(x)$ and $J_4(x)$ within uncertainties. This insensitivity
demonstrates a deeper point: the isospin projection --- that is, the flavor
antisymmetry of the operator --- has no effect on the magnetic moment when
the light quarks are in spin-singlet diquarks. The magnetic response is
blind to the isospin quantum number in this regime, which is a non-trivial
prediction that could in principle be tested if the magnetic moments of
different isospin components of the $P^{\Sigma}_{\psi s}$ multiplet become
experimentally accessible.

\paragraph{Axial-vector-diquark currents ($J_2(x)$, $J_3(x)$, $J_6(x)$):}
In these currents the diquarks involving light quarks are axial-vector,
$[q^T C\gamma_\mu\, q']$, forming spin-triplets ($J^P = 1^+$). A
spin-triplet diquark carries net angular momentum, and its coupling to the
external electromagnetic field is no longer suppressed. The light quarks
now participate actively in the magnetic response, and the resulting moments
are large, operator-dependent, and highly sensitive to the light-quark flavor.

The quark-level decomposition reveals the mechanism in detail. In $J_3(x)$
with $[su][uc]\bar{c}$, the $u$-quark contribution reaches
$\mu_u = -7.44\,\mu_N$ while $\mu_c \approx 0.00\,\mu_N$. The near-zero
charm contribution in $J_3(x)$ is not accidental: in this current the
heavy-light diquark is scalar, $[s^T C\gamma_5\, c]$, which suppresses
the charm spin coupling just as it suppresses light-quark contributions in
$J_1(x)$ and $J_4(x)$. The entire magnetic response is therefore carried by the
light axial-vector diquark $[q^T C\gamma_\mu\, q]$, producing a moment whose
magnitude is set entirely by the $u$-quark magnetic moment and its effective
spin polarization within the current. This is the limiting case of complete
light-quark dominance, complementary to the complete charm dominance observed
in $J_1(x)$ and $J_4(x)$.

In $J_2(x)$ and $J_6(x)$, both the charm sector and the light quarks contribute
at comparable levels: for $J_6(x)$ with $[sd][dc]\bar{c}$,
$\mu_d = +3.40\,\mu_N$ and $\mu_s = +0.87\,\mu_N$ alongside
$\mu_c \approx 0.00\,\mu_N$. The positive sign of the $d$-quark contribution
and the positive total moment in this configuration are direct consequences
of the $d$-quark charge  entering with a specific sign
determined by the spin polarization enforced by the axial-vector-diquark coupling.

A systematic consistency check across all currents and both spin sectors
is provided by the exact preservation of the 2:1 charge ratio between
$u$- and $d$-quark contributions. In every case where both flavor
configurations are present: $J_2(x)$: $(\mu_d, \mu_u) = (+2.40,-4.80)\,\mu_N$;
$J_3(x)$: $(+3.72,-7.44)\,\mu_N$; $J^4_\mu(x)$: $(-0.36,+0.72)\,\mu_N$;
$J^4_\mu(x)$ quadrupole: $(\mathcal{Q}_d, \mathcal{Q}_u) = (+3.05,-6.10)
\times 10^{-2}~\mathrm{fm}^2$. The ratio $\mu_u/\mu_d = e_u/e_d = -2$
is preserved to numerical precision at every multipole order. This is
a non-trivial internal consistency test: in the QCD sum-rule framework
the $u$- and $d$-quark contributions enter through distinct propagator
structures, and the exact recovery of the charge ratio confirms that the
isospin symmetry breaking encoded in the quark charges is correctly
propagated through the OPE at all orders retained in the calculation.

The sign reversal of the total moment between $[sd][dc]\bar{c}$ and
$[su][uc]\bar{c}$ configurations, observed in $J_2(x)$, $J_3(x)$, and $J_6(x)$,
follows directly from this charge ratio: when the light-quark contribution
dominates, replacing $d$  by $u$  reverses the sign of the dominant term and hence the total. In $J_1(x)$, $J_4(x)$, and $J_5(x)$, where light-quark contributions are at the $\lesssim 0.07\,\mu_N$ level, no sign change occurs. The sign reversal is therefore a diagnostic of light-quark dominance and, conversely, its absence is a diagnostic of the spin-singlet suppression.

The superposition current $J_6(x)$ exhibits constructive interference: its
moment ($4.27\,\mu_N$ for $[sd][dc]\bar{c}$) exceeds that of the
un-projected parent $J_2(x)$ ($3.50\,\mu_N$) by $22\%$. In the axial-vector-coupling sector, flavor and spin degrees of freedom are entangled in the operator
structure: the antisymmetric flavor combination of $J_6(x)$ selects a subset
of the spin-flavor space in which the $d$-quark polarization is amplified
relative to the symmetric case. This is in sharp contrast with the behavior
of $J_5(x)$ in the scalar sector, where the antisymmetrization has no effect.
The presence or absence of this interference effect is therefore itself a
discriminant between the scalar and axial-vector diquark regimes at the level of
the superposition currents.

\paragraph{Discriminative power and physical interpretation:}
The two groups of currents predict magnetic moments that differ by factors
of $3$--$6$ and, in several cases, have opposite signs. For the
$[sd][dc]\bar{c}$ configuration, the difference between $J_2(x)$ and $J_1(x)$
is $4.71 \pm 0.96\,\mu_N$, exceeding the combined uncertainty by a factor
of $\approx 5$. This separation is large enough that even a measurement
with $\sim 50\%$ precision would distinguish between the two scenarios.

The physical picture that emerges is the following. The six currents explored here probe two qualitatively different internal configurations of the $P^{\Sigma}_{\psi s}$ pentaquark: one in which the light quarks are dynamically inert (spin-singlet diquarks, charm-dominated moments, flavor-insensitive) and one in which the light quarks actively carry spin (axial-vector diquarks, large and flavor-sensitive moments, sign reversals between charge states). These are not two values of a continuous parameter but two qualitatively distinct internal structures. A future measurement of the
magnetic dipole moment will therefore not merely constrain a parameter within a model but will select between two physically distinct pictures of how the five quarks are organized inside the pentaquark.
\begin{table}[htbp]
\centering
\caption{Decomposition of the magnetic dipole moments of the
$P^{\Sigma}_{\psi s}$ pentaquarks into quark-flavor contributions
(in $\mu_N$) for the spin-$\tfrac{1}{2}$ sector, evaluated for 
the $[sd][dc]\bar{c}$, $[sd][uc]\bar{c}$, and $[su][uc]\bar{c}$ 
configurations across all interpolating currents.}
\label{tab:quark_magnetic_spin12}
\renewcommand{\arraystretch}{1.25}
\setlength{\tabcolsep}{9pt}
\begin{tabular}{llccccc}
\toprule
& & \multicolumn{4}{c}{Quark Contributions} & Total \\
\cmidrule(lr){3-6}
Current & Configuration
  & $\mu_u$ & $\mu_d$ & $\mu_s$ & $\mu_c$ & $\mu_{\mathrm{tot}}$ \\
\midrule
\multirow{3}{*}{$J_1(x)$}
  & $[sd][dc]\bar{c}$ & $-$ & $\phantom{-}0.04$ & $-0.02$ & $-1.23$ & $-1.21$ \\
  & $[sd][uc]\bar{c}$ & $-0.04$ & $\phantom{-}0.02$ & $-0.02$ & $-1.23$ & $-1.27$ \\
  & $[su][uc]\bar{c}$ & $-0.07$ & $-$ & $-0.02$ & $-1.23$ & $-1.32$ \\[4pt]
\multirow{3}{*}{$J_2(x)$}
  & $[sd][dc]\bar{c}$ & $-$ & $\phantom{-}2.40$ & $\phantom{-}0.01$ & $\phantom{-}1.09$ & $\phantom{-}3.50$ \\
  & $[sd][uc]\bar{c}$ & $-2.40$ & $\phantom{-}1.20$ & $\phantom{-}0.01$ & $\phantom{-}1.09$ & $-0.10$ \\
  & $[su][uc]\bar{c}$ & $-4.80$ & $-$ & $\phantom{-}0.01$ & $\phantom{-}1.09$ & $-3.70$ \\[4pt]
\multirow{3}{*}{$J_3(x)$}
  & $[sd][dc]\bar{c}$ & $-$ & $\phantom{-}3.72$ & $\phantom{-}0.04$ & $\phantom{-}0.00$ & $\phantom{-}3.76$ \\
  & $[sd][uc]\bar{c}$ & $-3.72$ & $\phantom{-}1.86$ & $\phantom{-}0.04$ & $\phantom{-}0.00$ & $-1.82$ \\
  & $[su][uc]\bar{c}$ & $-7.44$ & $-$ & $\phantom{-}0.04$ & $\phantom{-}0.00$ & $-7.40$ \\[4pt]
\multirow{3}{*}{$J_4(x)$}
  & $[sd][dc]\bar{c}$ & $-$ & $\phantom{-}0.05$ & $-0.03$ & $-1.79$ & $-1.77$ \\
  & $[sd][uc]\bar{c}$ & $-0.05$ & $\phantom{-}0.03$ & $-0.03$ & $-1.79$ & $-1.85$ \\
  & $[su][uc]\bar{c}$ & $-0.10$ & $-$ & $-0.03$ & $-1.79$ & $-1.92$ \\[4pt]
\multirow{3}{*}{$J_5(x)$}
  & $[sd][dc]\bar{c}$ & $-$ & $\phantom{-}0.05$ & $-0.02$ & $-1.43$ & $-1.40$ \\
  & $[sd][uc]\bar{c}$ & $-0.05$ & $\phantom{-}0.03$ & $-0.02$ & $-1.43$ & $-1.47$ \\
  & $[su][uc]\bar{c}$ & $-0.09$ & $-$ & $-0.02$ & $-1.43$ & $-1.54$ \\[4pt]
\multirow{3}{*}{$J_6(x)$}
  & $[sd][dc]\bar{c}$ & $-$ & $\phantom{-}3.40$ & $\phantom{-}0.87$ & $\phantom{-}0.00$ & $\phantom{-}4.27$ \\
  & $[sd][uc]\bar{c}$ & $-3.41$ & $\phantom{-}1.70$ & $\phantom{-}0.87$ & $\phantom{-}0.00$ & $-0.84$ \\
  & $[su][uc]\bar{c}$ & $-6.81$ & $-$ & $\phantom{-}0.87$ & $\phantom{-}0.00$ & $-5.94$ \\
\bottomrule
\end{tabular}
\end{table}

\paragraph{Comparison with the molecular picture:}
The molecular constituent quark model calculations of \cite{Li:2024wxr} provide predictions for the magnetic moments of $P^{\Sigma}_{\psi s}$ states in the $8_{1f}$ and  $8_{2f}$ flavor representations:
\[
\renewcommand{\arraystretch}{1.2}
\begin{array}{lcc}
\toprule
\text{State} & 8_{1f}~(\mu_N) & 8_{2f}~(\mu_N) \\
\midrule
P^{\Sigma^+}_{\psi s} & ~~1.81 & 0.38 \\
P^{\Sigma^0}_{\psi s} & ~~0.26 & 0.38 \\
P^{\Sigma^-}_{\psi s} & -1.29 & 0.38 \\
\bottomrule
\end{array}
\]
A comparison between our results and the molecular predictions reveals both structural similarities and fundamental differences that trace directly to the underlying physical assumptions of each approach.

The $8_{2f}$ representation yields charge-independent moments ($\mu = 0.38\,\mu_N$) because in pure $\Xi_c\bar{D}$ configurations the light antiquark of the $\bar{D}$ meson carries no net magnetic moment in the $S$-wave approximation, and the entire moment is determined by the charm sector of the $\Xi_c$ baryon. Our scalar-diquark currents exhibit an analogous structure --- charm-sector dominance and flavor independence --- but with a negative sign and larger magnitude ($-1.92$ to $-1.21\,\mu_N$). The sign difference between the two approaches has a clear methodological origin. In the constituent quark model, the charm contribution is computed from the valence charm quark within the $\Xi_c$ baryon with its specific spatial and spin wave function; in the LCSR approach, the charm contribution $\mu_c$ is obtained from the charm propagator in the external field and represents the net response of the charm sector to the electromagnetic field as encoded by the operator $\Gamma_3 = C$. The two methods couple the charm sector to the external field through fundamentally different mechanisms, and agreement in sign is not guaranteed even when both predict charm dominance. The magnitude difference ($0.38$ vs.\ $\sim 1.5\,\mu_N$) reflects an additional distinction: the compact diquark correlation in our currents, operative at the $\sim \Lambda_{\rm QCD}^{-1} < 1 $~fm scale, produces a larger effective coupling of the charm sector to the external field than the loosely bound $\Xi_c\bar{D}$ pair at the molecular scale ($r_{\rm mol} \sim 1/\sqrt{2\mu B_E} >  1$~fm). This scale separation between the compact diquark and molecular configurations is the most fundamental physical difference between the two approaches and directly accounts for the magnitude discrepancy.

The $8_{1f}$ representation shows strong charge dependence ($-1.29$ to $+1.81\,\mu_N$) arising from the baryon isospin multiplet structure of the mixed $\Sigma_c\bar{D}_s$/$\Xi_c'\bar{D}$ configurations. Our axial-vector-diquark currents also exhibit strong charge dependence, but through a qualitatively different mechanism: the $e_u/e_d = -2$ charge ratio produces a sign reversal between $[sd][dc]\bar{c}$ and $[su][uc]\bar{c}$ rather than the linear isospin progression of the $\Sigma_c$ multiplet. The two charge-dependence patterns are therefore mechanistically distinct. In the molecular picture, the charge dependence follows from the isospin structure of the constituent baryon, which
changes continuously from $\Sigma^-$ to $\Sigma^0$ to $\Sigma^+$; in our axial-vector-diquark currents, the change from $d$ to $u$ produces an abrupt sign reversal driven by the factor of $-2$ in the charge ratio. These two patterns could in principle be distinguished experimentally if the magnetic moments of multiple charge states of the multiplet can be
measured, since the molecular prediction of a sign change between $\Sigma^-$ and $\Sigma^+$  passing through $\Sigma^0$ is qualitatively different from our prediction of a sign change between $[sd][dc]\bar{c}$ 
($+3.50\,\mu_N$ for $J_2(x)$) and $[su][uc]\bar{c}$ 
($-3.70\,\mu_N$ for $J_2(x)$), with the intermediate 
$[sd][uc]\bar{c}$ state yielding $-0.10\,\mu_N$, a value 
that is near-zero but negative — in contrast to the 
molecular $8_{1f}$ prediction of $+0.26\,\mu_N$ for 
$\Sigma^0$.

\subsection{Spin-$3/2$ sector}\label{subsec:spin32}

For the $J^P = \frac{3}{2}^-$ channels, the magnetic dipole ($\mu$),
electric quadrupole ($\mathcal{Q}$), and magnetic octupole ($\mathcal{O}$)
moments are presented in Tables~\ref{tab:multipole_results_spin32}, \ref{tab:quark_magnetic_spin32},  \ref{tab:quadrupole_spin32}
and~\ref{tab:octupole_spin32}.

\begin{table}[htbp]
\centering
\caption{Magnetic dipole moments $\mu$ (in $\mu_N$), electric 
quadrupole moments $\mathcal{Q}$ (in $10^{-2}$~fm$^2$), and 
magnetic octupole moments $\mathcal{O}$ (in $10^{-3}$~fm$^3$) 
of the $P^{\Sigma^\ast}_{\psi s}$ pentaquarks in the 
spin-$\tfrac{3}{2}$ sector, computed from the LCSR for each 
interpolating current and flavor configuration.}
\label{tab:multipole_results_spin32}
\renewcommand{\arraystretch}{1.25}
\setlength{\tabcolsep}{9pt}
\begin{tabular}{llccc}
\toprule
& & Magnetic Dipole & Electric Quadrupole & Magnetic Octupole \\
Current & Configuration & $\mu_{P^{\Sigma^\ast}_{\psi s}}$ 
  & $\mathcal{Q}_{P^{\Sigma^\ast}_{\psi s}}$
  & $\mathcal{O}_{P^{\Sigma^\ast}_{\psi s}}$ \\
\midrule
\multirow{3}{*}{$J_\mu^1(x)$}
  & $[sd][dc]\bar{c}$ & $-0.41 \pm 0.11$ & $-1.40 \pm 0.29$ & $-0.27 \pm 0.05$ \\
  & $[sd][uc]\bar{c}$ & $-0.47 \pm 0.12$ & $-1.42 \pm 0.29$ & $-0.27 \pm 0.05$ \\
  & $[su][uc]\bar{c}$ & $-0.53 \pm 0.13$ & $-1.43 \pm 0.29$ & $-0.27 \pm 0.05$ \\[4pt]
\multirow{3}{*}{$J_\mu^2(x)$}
  & $[sd][dc]\bar{c}$ & $-0.30 \pm 0.07$ & $-1.17 \pm 0.24$ & $-0.23 \pm 0.05$ \\
  & $[sd][uc]\bar{c}$ & $-0.37 \pm 0.09$ & $-1.19 \pm 0.25$ & $-0.23 \pm 0.05$ \\
  & $[su][uc]\bar{c}$ & $-0.44 \pm 0.11$ & $-1.20 \pm 0.25$ & $-0.23 \pm 0.05$ \\[4pt]
\multirow{3}{*}{$J_\mu^3(x)$}
  & $[sd][dc]\bar{c}$ & $-1.19 \pm 0.30$ & $\phantom{-}3.14 \pm 0.63$ & $-0.94 \pm 0.19$ \\
  & $[sd][uc]\bar{c}$ & $-1.16 \pm 0.30$ & $\phantom{-}3.14 \pm 0.63$ & $-0.94 \pm 0.19$ \\
  & $[su][uc]\bar{c}$ & $-1.13 \pm 0.29$ & $\phantom{-}3.14 \pm 0.63$ & $-0.94 \pm 0.19$ \\[4pt]
\multirow{3}{*}{$J_\mu^4(x)$}
  & $[sd][dc]\bar{c}$ & $-3.49 \pm 0.91$ & $\phantom{-}4.83 \pm 0.97$ & $-1.42 \pm 0.29$ \\
  & $[sd][uc]\bar{c}$ & $-2.95 \pm 0.77$ & $\phantom{-}0.26 \pm 0.05$ & $-0.43 \pm 0.09$ \\
  & $[su][uc]\bar{c}$ & $-2.41 \pm 0.91$ & $-4.32 \pm 0.86$ & $\phantom{-}0.56 \pm 0.11$ \\[4pt]
\multirow{3}{*}{$J_\mu^5(x)$}
  & $[sd][dc]\bar{c}$ & $-0.26 \pm 0.07$ & $-1.11 \pm 0.29$ & $-0.22 \pm 0.04$ \\
  & $[sd][uc]\bar{c}$ & $-0.32 \pm 0.08$ & $-1.13 \pm 0.30$ & $-0.22 \pm 0.04$ \\
  & $[su][uc]\bar{c}$ & $-0.38 \pm 0.10$ & $-1.14 \pm 0.30$ & $-0.22 \pm 0.04$ \\[4pt]
\multirow{3}{*}{$J_\mu^6(x)$}
  & $[sd][dc]\bar{c}$ & $-1.10 \pm 0.29$ & $\phantom{-}2.91 \pm 0.58$ & $-0.88 \pm 0.18$ \\
  & $[sd][uc]\bar{c}$ & $-1.39 \pm 0.36$ & $\phantom{-}2.16 \pm 0.43$ & $-0.73 \pm 0.15$ \\
  & $[su][uc]\bar{c}$ & $-1.68 \pm 0.44$ & $\phantom{-}1.41 \pm 0.29$ & $-0.58 \pm 0.12$ \\[4pt]
\multirow{3}{*}{$J_\mu^7(x)$}
  & $[sd][dc]\bar{c}$ & $-3.37 \pm 0.87$ & $\phantom{-}4.51 \pm 0.90$ & $-1.39 \pm 0.28$ \\
  & $[sd][uc]\bar{c}$ & $-2.83 \pm 0.74$ & $\phantom{-}0.96 \pm 0.19$ & $-0.58 \pm 0.12$ \\
  & $[su][uc]\bar{c}$ & $-2.29 \pm 0.60$ & $-2.60 \pm 0.52$ & $\phantom{-}0.23 \pm 0.04$ \\
\bottomrule
\end{tabular}
\end{table}
\subsubsection{Magnetic dipole moments}\label{subsubsec:mu32}

The spin-$3/2$ magnetic moments are uniformly negative and reproduce the
same diquark-spin dependence established in the spin-$1/2$ sector.

Currents $J^1_\mu(x)$, $J^2_\mu(x)$, $J^3_\mu(x)$, and $J^5_\mu(x)$ couple the diquark system to the charm sector via $\Gamma_3 = C$, a scalar coupling. Their moments are small ($|\mu| \lesssim 1.2\,\mu_N$), charm-dominated, and flavor-insensitive, in direct analogy with $J_1(x)$, $J_4(x)$, $J_5(x)$ in the spin-$1/2$ sector. The $\Gamma_3 = C$ coupling plays the same role in both spin sectors: it projects onto a channel in which the diquark spin degrees of freedom are not efficiently coupled to the external field, leaving the charm sector as the primary magnetic source.

The moderately larger magnitude of $J^3_\mu(x)$ ($-1.19\,\mu_N$) compared
to $J^1_\mu(x)$, $J^2_\mu(x)$, $J^5_\mu(x)$ ($-0.53$ to $-0.26\,\mu_N$) is
explained by its operator structure: $J^3_\mu(x)$ contains a light axial-vector diquark $[q^T\Gamma_2 q]$ providing the external Lorentz index and a
scalar heavy-light diquark $[s^T\Gamma_1 c]$. The light axial-vector diquark
contributes $\mu_s = 0.50\,\mu_N$. This makes $J^3_\mu(x)$ a partial exception: although the $\Gamma_3$ coupling still dominates the magnetic response through the charm sector ($\mu_c = -1.67\,\mu_N$), the 
light axial-vector diquark partially circumvents the scalar-coupling suppression by activating a non-negligible strange-quark contribution ($\mu_s = +0.50\,\mu_N$), absent in the other $\Gamma_3$-coupled currents.

Currents $J^4_\mu(x)$, $J^6_\mu(x)$, and $J^7_\mu(x)$ contain two axial-vector diquarks, $\Gamma_2$ (external index $\mu$) and $\Gamma_5$ (internal index $\alpha$), and couple to the charm sector via $\Gamma_6 = \gamma_5\gamma^\alpha C$. Their moments are substantially larger ($-3.49$ to $-1.10\,\mu_N$). The structural difference between $J^4_\mu(x)$ 
($-3.49$, $-2.95$, $-2.41\,\mu_N$) and $J^6_\mu(x)$ ($-1.10$, $-1.39$, $-1.68\,\mu_N$) is particularly instructive and would be missed by any analysis that does not examine the operator structure
in detail. In $J^4_\mu(x)$, the external Lorentz index $\mu$ is carried by
the heavy-light diquark $[s^T\Gamma_2 c]$, while the internal index
$\alpha$ is carried by the light diquark $[q^T\Gamma_5 q]$. In $J^6_\mu(x)$,
the assignment is reversed: the external index is carried by the light
diquark $[q^T\Gamma_2 q]$ and the internal index by the heavy-light
diquark $[s^T\Gamma_5 c]$. This exchange of which diquark provides the
external index changes the way the spin degrees of freedom couple to the
external electromagnetic field: when the heavy-light diquark carries the
external index, the charm sector is more directly coupled to the field
and the charm contribution ($\mu_c = -3.09\,\mu_N$ for $J^4_\mu(x)$)
dominates; when the light diquark carries the external index, the coupling
is mediated through the light-quark sector first, producing a different
effective polarization ($\mu_c = -1.59\,\mu_N$ for $J^6_\mu(x)$). The
factor-of-three difference between the charm contributions of $J^4_\mu(x)$
and $J^6_\mu(x)$ ($-3.09$ vs.\ $-1.59\,\mu_N$) is entirely a consequence
of this Lorentz-index assignment and demonstrates that the index structure
of the operator is not merely a bookkeeping device but has direct physical
consequences for the electromagnetic response. 
The near-equality of $J^4_\mu(x)$ ($-3.49$, $-2.95$, $-2.41\,\mu_N$) 
and $J^7_\mu(x)$ ($-3.37$, $-2.83$, $-2.29\,\mu_N$) is consistent with $J^7_\mu(x)$ being the isospin-antisymmetric counterpart of $J^4_\mu(x)$: the $3\%$ difference shows that the isospin projection modifies the result only marginally, just as $J_5(x)$ modifies $J_4(x)$ by only $8\%$ in the spin-$1/2$ sector. This near-independence of the isospin projection is a consistent feature across both spin sectors and both operator types, suggesting that it is a structural property of LCSR calculations for this system rather than a current-specific coincidence.

\paragraph{Comparison with the molecular picture:} 
The molecular decuplet representation~\cite{Li:2025ddx} predicts
$\mu = +1.94\,\mu_N$ ($\Sigma^+$), $-0.23\,\mu_N$ ($\Sigma^0$),
$-2.40\,\mu_N$ ($\Sigma^-$), a monotonic progression reflecting the
linear isospin dependence of the $\Sigma_c^{(*)}$ baryon multiplet.

Our results differ qualitatively. Currents $J^1_\mu(x)$, $J^2_\mu(x)$,
$J^5_\mu(x)$ are essentially flavor-insensitive ($-0.53$ to $-0.26\,\mu_N$
for both configurations), which is incompatible with the strong charge
dependence of the molecular prediction. If the physical $P^{\Sigma^\ast}_{\psi s}$ has a dominant scalar-coupled diquark structure, its magnetic moment would be nearly the same for the $\Sigma^+$, $\Sigma^0$, and $\Sigma^-$ charge states — a pattern that is fundamentally at odds with the monotonic molecular progression. The two-axial-vector-diquark currents ($J^4_\mu(x)$, $J^6_\mu(x)$, $J^7_\mu(x)$) are not flavor-insensitive, but their charge dependence follows a different pattern. Taking $J^4_\mu(x)$: the moment is $-3.49\,\mu_N$ for 
$[sd][dc]\bar{c}$, $-2.95\,\mu_N$ for $[sd][uc]\bar{c}$, 
and $-2.41\,\mu_N$ for $[su][uc]\bar{c}$, all negative with the $d$-quark configuration more negative than the $u$-quark configuration. The molecular pattern is the opposite: the $\Sigma^-$ ($d$-quark) state is most negative ($-2.40$) while the $\Sigma^+$ ($u$-quark) state is strongly positive ($+1.94$).

The physical origin of this reversal lies in the mechanism of charge
dependence in the two approaches. In the molecular picture, the progression from $\Sigma^-$ to $\Sigma^+$ involves a change in the baryon isospin wave function that directly flips the spin polarization of the $u$-$d$ subsystem within the $\Sigma_c$ baryon; this produces a large, nearly linear change in the moment including a sign reversal. In our two-axial-vector-diquark currents, the $d \to u$ substitution changes only the charge weight of the light-quark contribution: $\mu_u = -2\mu_d$, so the light-quark contribution partially cancels the charm contribution for the $u$-quark configuration rather than reinforcing it. No sign reversal occurs because the charm contribution always dominates. The key experimental discriminant is therefore the sign of the moment for the $[su][uc]\bar{c}$ ($\Sigma^+$-like) state: a positive value strongly favors the molecular picture; a negative value, comparable in magnitude to the $\Sigma^-$ state, strongly favors the compact diquark structure.

\begin{table}[htbp]
\centering
\caption{Decomposition of the magnetic dipole moments of the
$P^{\Sigma^\ast}_{\psi s}$ pentaquarks into quark-flavor 
contributions (in $\mu_N$) for the spin-$\tfrac{3}{2}$ sector, 
evaluated for the $[sd][dc]\bar{c}$, $[sd][uc]\bar{c}$, and 
$[su][uc]\bar{c}$ configurations across all interpolating 
currents.}
\label{tab:quark_magnetic_spin32}
\renewcommand{\arraystretch}{1.25}
\setlength{\tabcolsep}{9pt}
\begin{tabular}{llccccc}
\toprule
& & \multicolumn{4}{c}{Quark Contributions} & Total \\
\cmidrule(lr){3-6}
Current & Configuration
  & $\mu_u$ & $\mu_d$ & $\mu_s$ & $\mu_c$ & $\mu_{\mathrm{tot}}$ \\
\midrule
\multirow{3}{*}{$J_\mu^1(x)$}
  & $[sd][dc]\bar{c}$ & $-$ & $\phantom{-}0.04$ & $\phantom{-}0.01$ & $-0.46$ & $-0.41$ \\
  & $[sd][uc]\bar{c}$ & $-0.04$ & $\phantom{-}0.02$ & $\phantom{-}0.01$ & $-0.46$ & $-0.47$ \\
  & $[su][uc]\bar{c}$ & $-0.08$ & $-$ & $\phantom{-}0.01$ & $-0.46$ & $-0.53$ \\[4pt]
\multirow{3}{*}{$J_\mu^2(x)$}
  & $[sd][dc]\bar{c}$ & $-$ & $\phantom{-}0.05$ & $\sim 0$ & $-0.35$ & $-0.30$ \\
  & $[sd][uc]\bar{c}$ & $-0.05$ & $\phantom{-}0.03$ & $\sim 0$ & $-0.35$ & $-0.37$ \\
  & $[su][uc]\bar{c}$ & $-0.09$ & $-$ & $\sim 0$ & $-0.35$ & $-0.44$ \\[4pt]
\multirow{3}{*}{$J_\mu^3(x)$}
  & $[sd][dc]\bar{c}$ & $-$ & $-0.02$ & $\phantom{-}0.50$ & $-1.67$ & $-1.19$ \\
  & $[sd][uc]\bar{c}$ & $\phantom{-}0.02$ & $-0.01$ & $\phantom{-}0.50$ & $-1.67$ & $-1.16$ \\
  & $[su][uc]\bar{c}$ & $\phantom{-}0.04$ & $-$ & $\phantom{-}0.50$ & $-1.67$ & $-1.13$ \\[4pt]
\multirow{3}{*}{$J_\mu^4(x)$}
  & $[sd][dc]\bar{c}$ & $-$ & $-0.36$ & $-0.04$ & $-3.09$ & $-3.49$ \\
  & $[sd][uc]\bar{c}$ & $\phantom{-}0.36$ & $-0.18$ & $-0.04$ & $-3.09$ & $-2.95$ \\
  & $[su][uc]\bar{c}$ & $\phantom{-}0.72$ & $-$ & $-0.04$ & $-3.09$ & $-2.41$ \\[4pt]
\multirow{3}{*}{$J_\mu^5(x)$}
  & $[sd][dc]\bar{c}$ & $-$ & $\phantom{-}0.04$ & $\sim 0$ & $-0.30$ & $-0.26$ \\
  & $[sd][uc]\bar{c}$ & $-0.04$ & $\phantom{-}0.02$ & $\sim 0$ & $-0.30$ & $-0.32$ \\
  & $[su][uc]\bar{c}$ & $-0.08$ & $-$ & $\sim 0$ & $-0.30$ & $-0.38$ \\[4pt]
\multirow{3}{*}{$J_\mu^6(x)$}
  & $[sd][dc]\bar{c}$ & $-$ & $\phantom{-}0.19$ & $\phantom{-}0.29$ & $-1.59$ & $-1.10$ \\
  & $[sd][uc]\bar{c}$ & $-0.19$ & $\phantom{-}0.10$ & $\phantom{-}0.29$ & $-1.59$ & $-1.39$ \\
  & $[su][uc]\bar{c}$ & $-0.38$ & $-$ & $\phantom{-}0.29$ & $-1.59$ & $-1.68$ \\[4pt]
\multirow{3}{*}{$J_\mu^7(x)$}
  & $[sd][dc]\bar{c}$ & $-$ & $-0.36$ & $-0.14$ & $-2.87$ & $-3.37$ \\
  & $[sd][uc]\bar{c}$ & $\phantom{-}0.36$ & $-0.18$ & $-0.14$ & $-2.87$ & $-2.83$ \\
  & $[su][uc]\bar{c}$ & $\phantom{-}0.72$ & $-$ & $-0.14$ & $-2.87$ & $-2.29$ \\
\bottomrule
\end{tabular}
\end{table}

\subsubsection{Electric quadrupole moments and intrinsic deformation}
\label{subsubsec:Q32}

The electric quadrupole moments are the most structurally informative
observable in this work and, to our knowledge, constitute the first
systematic predictions of this quantity for  $P^{\Sigma^\ast}_{\psi s}$ pentaquarks. No counterpart predictions exist in the molecular or constituent quark model approaches: in the simplest $S$-wave molecular approximation, the two constituent hadrons carry no orbital angular momentum relative to each other, the individual hadrons are treated as spherical, and the electric quadrupole moment of the composite system vanishes identically. A non-zero measurement of $\mathcal{Q}$ would therefore constitute model-independent evidence for an internal structure that goes beyond a simple $S$-wave two-body bound state.

To extract the physical shape information encoded in $\mathcal{Q}$, we
compute the intrinsic quadrupole moment \cite{Buchmann:2001gj}
\begin{equation}
Q_0 = \frac{(J+1)(2J+3)}{J(2J-1)}\,\mathcal{Q}
\;=\; \frac{5}{3}\,\mathcal{Q} \quad (J = 3/2),
\label{eq:Q0}
\end{equation}
which describes the deformation of the charge distribution in the body-fixed frame. A negative $Q_0$ corresponds to an oblate (disk-like) distribution; a positive $Q_0$ corresponds to a prolate (elongated) distribution.
\begin{table}[htbp]
\centering
\caption{Decomposition of the electric quadrupole moments 
$\mathcal{Q}$ (in $10^{-2}$~fm$^2$) of the 
$P^{\Sigma^\ast}_{\psi s}$ pentaquarks into quark-flavor 
contributions for the spin-$\tfrac{3}{2}$ sector, evaluated 
for the $[sd][dc]\bar{c}$, $[sd][uc]\bar{c}$, and 
$[su][uc]\bar{c}$ configurations.}
\label{tab:quadrupole_spin32}
\renewcommand{\arraystretch}{1.25}
\setlength{\tabcolsep}{9pt}
\begin{tabular}{llccccc}
\toprule
& & \multicolumn{4}{c}{Quark Contributions} & Total \\
\cmidrule(lr){3-6}
Current & Configuration
  & $\mathcal{Q}_u$ & $\mathcal{Q}_d$
  & $\mathcal{Q}_s$ & $\mathcal{Q}_c$
  & $\mathcal{Q}_{\mathrm{tot}}$ \\
\midrule
\multirow{3}{*}{$J_\mu^1(x)$}
  & $[sd][dc]\bar{c}$ & $-$ & $\phantom{-}0.01$ & $\sim 0$ & $-1.41$ & $-1.40$ \\
  & $[sd][uc]\bar{c}$ & $-0.01$ & $\phantom{-}0.01$ & $\sim 0$ & $-1.41$ & $-1.42$ \\
  & $[su][uc]\bar{c}$ & $-0.02$ & $-$ & $\sim 0$ & $-1.41$ & $-1.43$ \\[4pt]
\multirow{3}{*}{$J_\mu^2(x)$}
  & $[sd][dc]\bar{c}$ & $-$ & $\phantom{-}0.01$ & $\sim 0$ & $-1.18$ & $-1.17$ \\
  & $[sd][uc]\bar{c}$ & $-0.01$ & $\phantom{-}0.01$ & $\sim 0$ & $-1.18$ & $-1.19$ \\
  & $[su][uc]\bar{c}$ & $-0.02$ & $-$ & $\sim 0$ & $-1.18$ & $-1.20$ \\[4pt]
\multirow{3}{*}{$J_\mu^3(x)$}
  & $[sd][dc]\bar{c}$ & $-$ & ${\sim}0$ & $\phantom{-}1.63$ & $\phantom{-}1.51$ & $\phantom{-}3.14$ \\
  & $[sd][uc]\bar{c}$ & ${\sim}0$ & ${\sim}0$ & $\phantom{-}1.63$ & $\phantom{-}1.51$ & $\phantom{-}3.14$ \\
  & $[su][uc]\bar{c}$ & ${\sim}0$ & $-$ & $\phantom{-}1.63$ & $\phantom{-}1.51$ & $\phantom{-}3.14$ \\[4pt]
\multirow{3}{*}{$J_\mu^4(x)$}
  & $[sd][dc]\bar{c}$ & $-$ & $\phantom{-}3.05$ & $\sim 0$ & $\phantom{-}1.78$ & $\phantom{-}4.83$ \\
  & $[sd][uc]\bar{c}$ & $-3.05$ & $\phantom{-}1.53$ & $\sim 0$ & $\phantom{-}1.78$ & $\phantom{-}0.26$ \\
  & $[su][uc]\bar{c}$ & $-6.10$ & $-$ & $\sim 0$ & $\phantom{-}1.78$ & $-4.32$ \\[4pt]
\multirow{3}{*}{$J_\mu^5(x)$}
  & $[sd][dc]\bar{c}$ & $-$ & $\phantom{-}0.01$ & ${\sim}0$ & $-1.12$ & $-1.11$ \\
  & $[sd][uc]\bar{c}$ & $-0.01$ & $\phantom{-}0.01$ & ${\sim}0$ & $-1.12$ & $-1.13$ \\
  & $[su][uc]\bar{c}$ & $-0.02$ & $-$ & ${\sim}0$ & $-1.12$ & $-1.14$ \\[4pt]
\multirow{3}{*}{$J_\mu^6(x)$}
  & $[sd][dc]\bar{c}$ & $-$ & $\phantom{-}0.50$ & $\phantom{-}1.00$ & $\phantom{-}1.41$ & $\phantom{-}2.91$ \\
  & $[sd][uc]\bar{c}$ & $-0.50$ & $\phantom{-}0.25$ & $\phantom{-}1.00$ & $\phantom{-}1.41$ & $\phantom{-}2.16$ \\
  & $[su][uc]\bar{c}$ & $-1.00$ & $-$ & $\phantom{-}1.00$ & $\phantom{-}1.41$ & $\phantom{-}1.41$ \\[4pt]
\multirow{3}{*}{$J_\mu^7(x)$}
  & $[sd][dc]\bar{c}$ & $-$ & $\phantom{-}2.37$ & $\phantom{-}0.46$ & $\phantom{-}1.68$ & $\phantom{-}4.51$ \\
  & $[sd][uc]\bar{c}$ & $-2.37$ & $\phantom{-}1.19$ & $\phantom{-}0.46$ & $\phantom{-}1.68$ & $\phantom{-}0.96$ \\
  & $[su][uc]\bar{c}$ & $-4.74$ & $-$ & $\phantom{-}0.46$ & $\phantom{-}1.68$ & $-2.60$ \\
\bottomrule
\end{tabular}
\end{table}

\begin{figure}[htbp]
\centering
\includegraphics[width=0.75\textwidth]{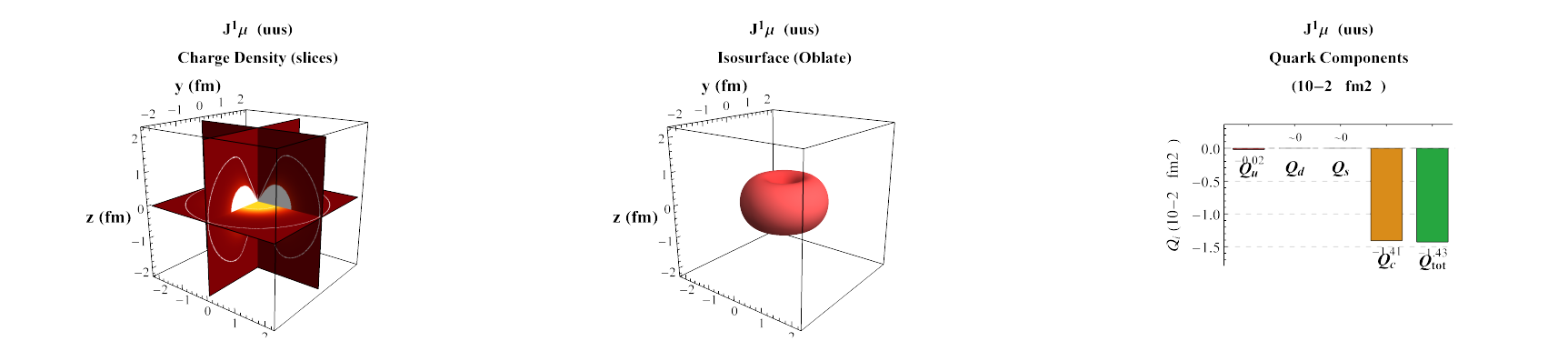} \\
\includegraphics[width=0.75\textwidth]{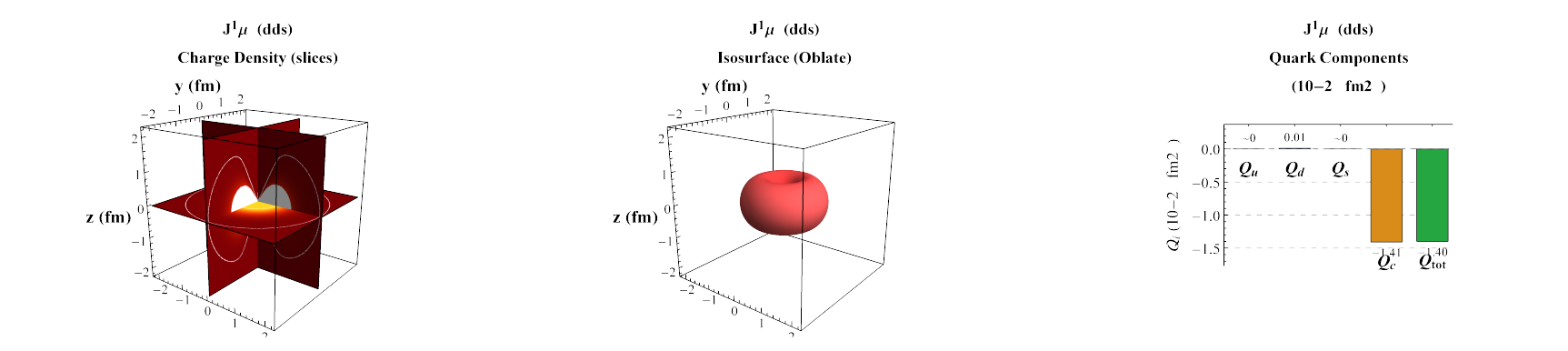}\\
\includegraphics[width=0.75\textwidth]{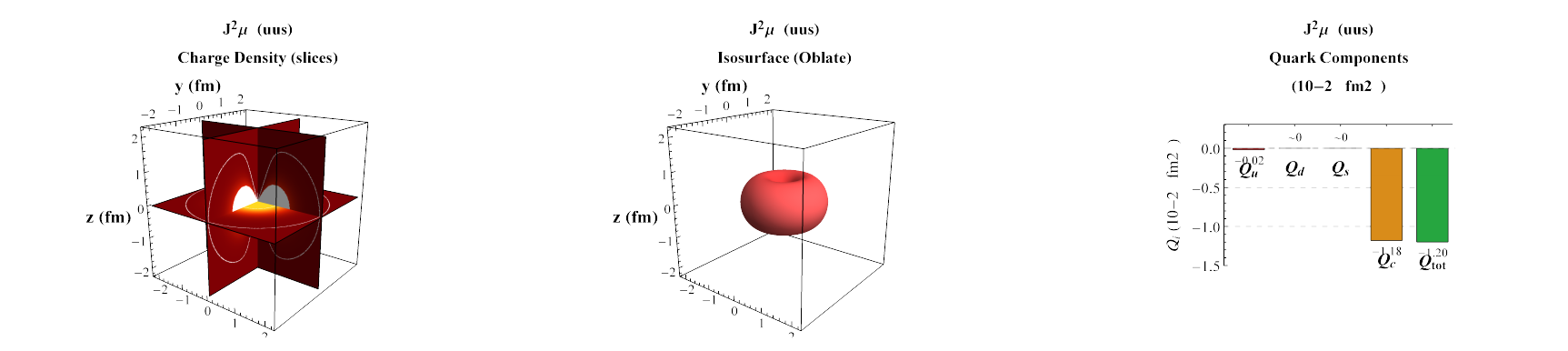} \\
\includegraphics[width=0.75\textwidth]{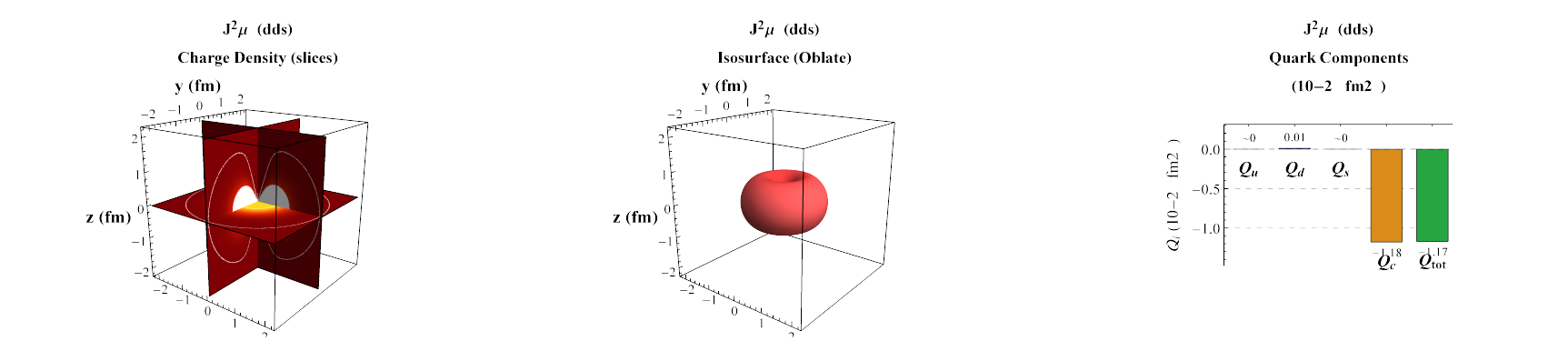}\\
\includegraphics[width=0.75\textwidth]{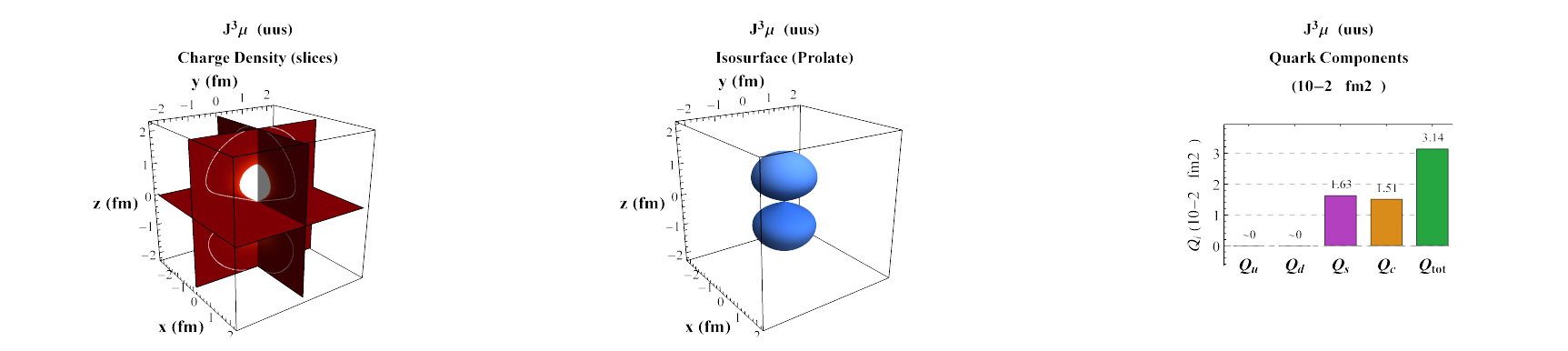} \\
\includegraphics[width=0.75\textwidth]{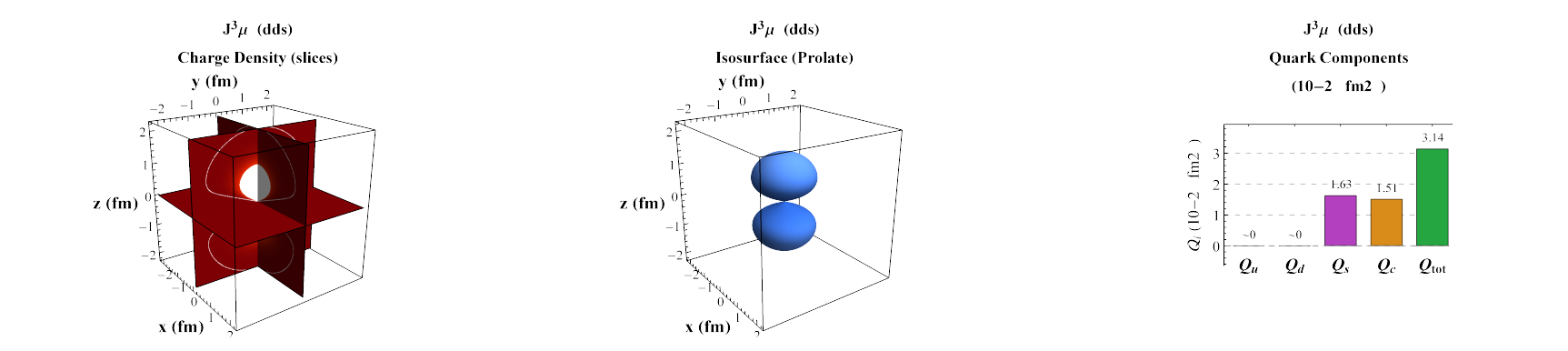}\\
\caption{Electric quadrupole moment analysis of the $P^{\Sigma^\ast}_{\psi s}$ pentaquarks for $J^{1}_{\mu}(x)$, $J^{2}_{\mu}(x)$ and $J^{3}_{\mu}(x)$ interpolating currents and both diquark configurations, $[su][uc]\bar{c}$ (uus) and $[sd][dc]\bar{c}$ (dds). For each state, three panels are shown: (\textit{left}) the three-dimensional electric charge density $\rho(\mathbf{r})$ evaluated on the orthogonal midplanes $x=0$, $y=0$, and $z=0$, displayed with a SolarColors heat map in which brighter regions indicate higher charge concentration; (\textit{center}) the charge-density isosurface at a fixed threshold, coloured blue for prolate and red for oblate  deformation, whose shape directly reflects the sign and magnitude of $\mathcal{Q}_{\mathrm{tot}}$; and (\textit{right}) the quark-flavour decomposition bar chart showing the individual contributions $\mathcal{Q}_{u}$, $\mathcal{Q}_{d}$, $\mathcal{Q}_{s}$, $\mathcal{Q}_{c}$, and the total $\mathcal{Q}_{\mathrm{tot}}$, with numeric values annotated on each bar. A positive (negative) total quadrupole moment corresponds to a prolate, cigar-shaped (oblate, disk-shaped) charge distribution elongated along (compressed along) the symmetry axis $z$. All quadrupole moments are given in units of $10^{-2}$~fm$^{2}$ and all spatial coordinates in fm.}
\label{fig:quadrupole_all}
\end{figure}

\begin{figure}[htbp]
\centering
\includegraphics[width=0.70\textwidth]{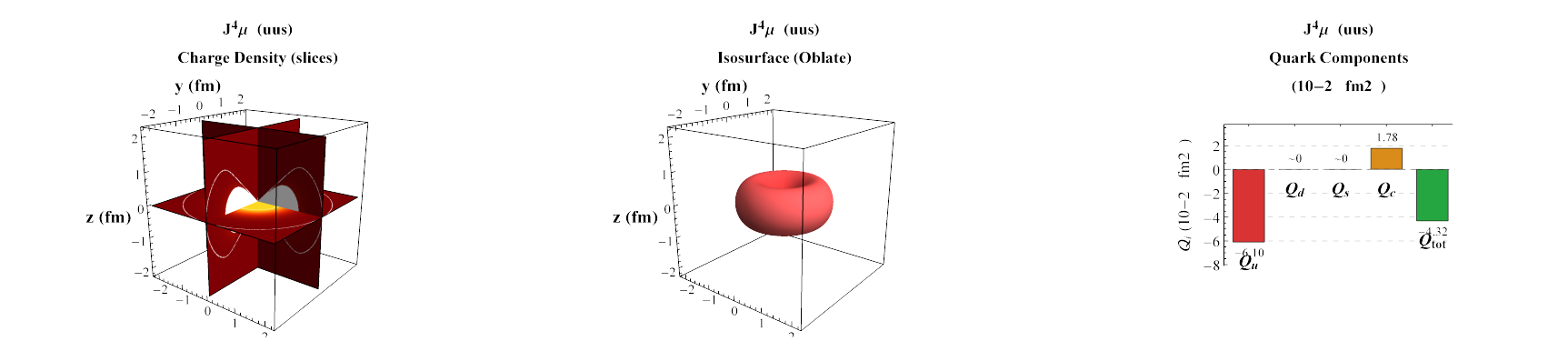} \\
\includegraphics[width=0.70\textwidth]{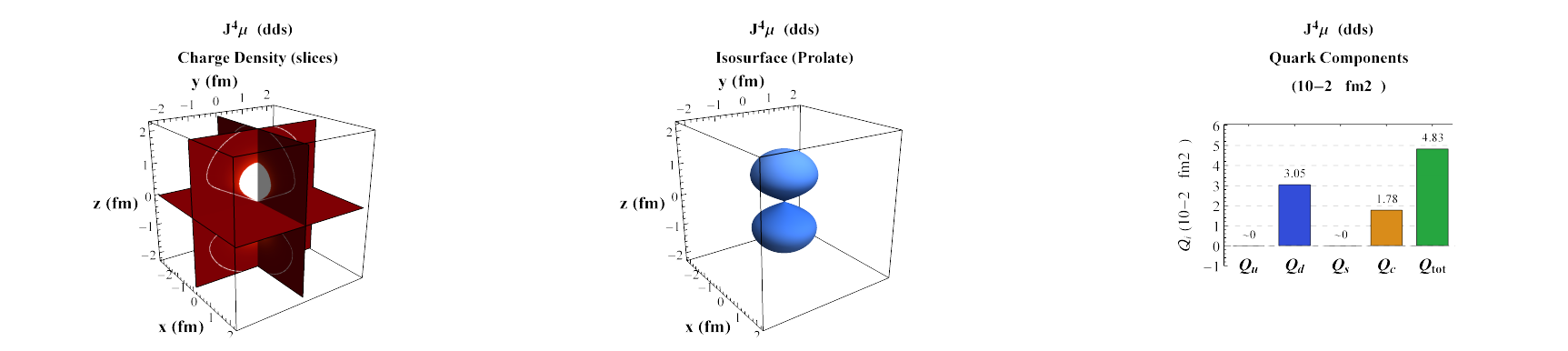}\\
\includegraphics[width=0.70\textwidth]{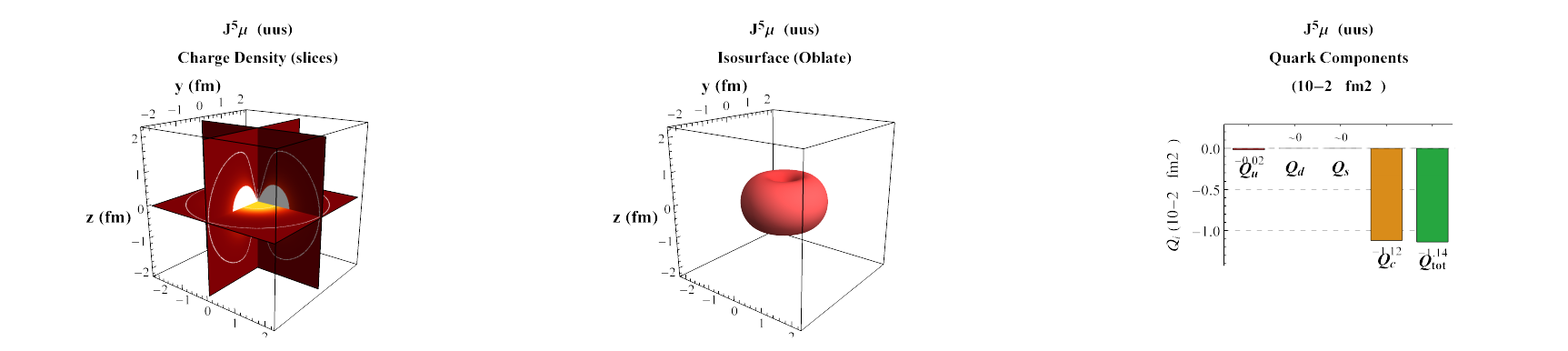} \\
\includegraphics[width=0.70\textwidth]{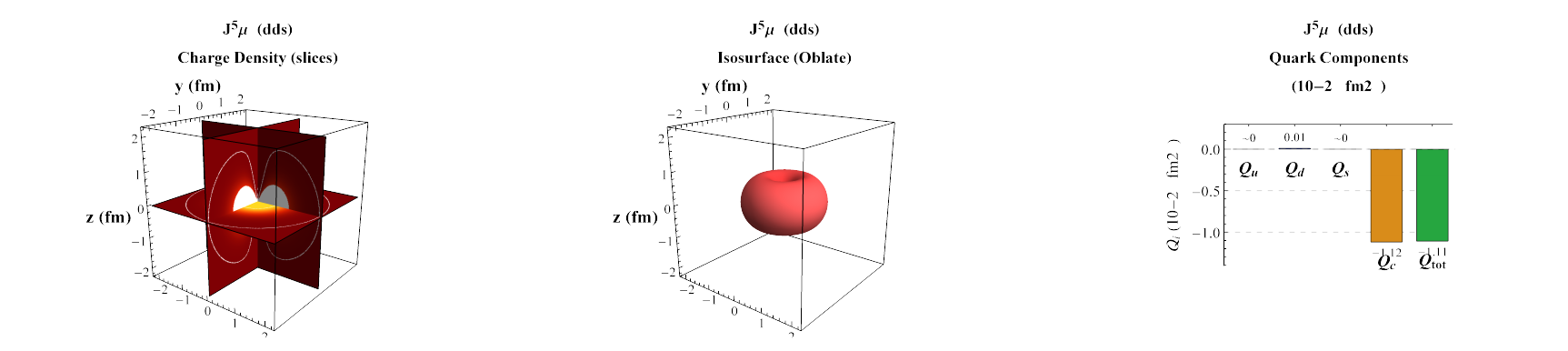}\\
\includegraphics[width=0.70\textwidth]{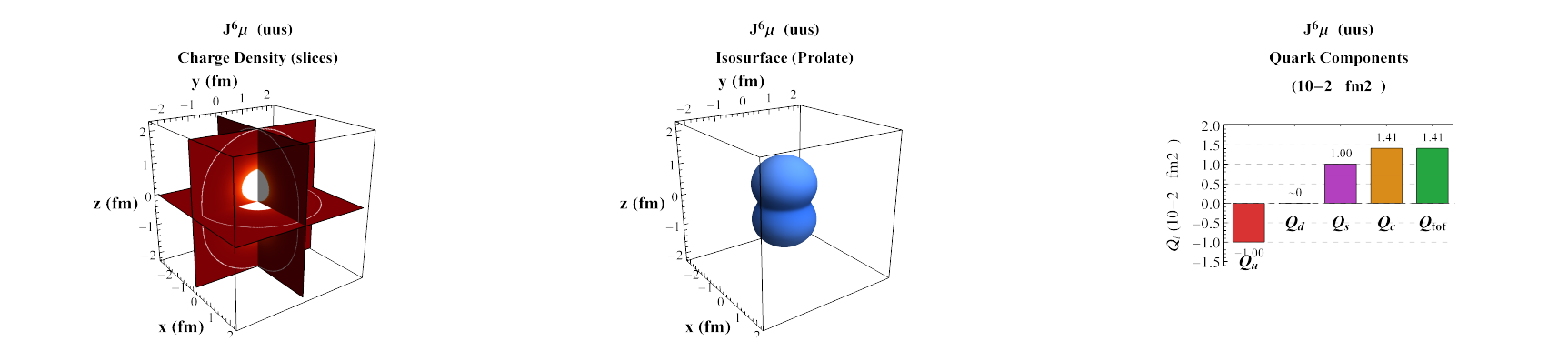} \\
\includegraphics[width=0.70\textwidth]{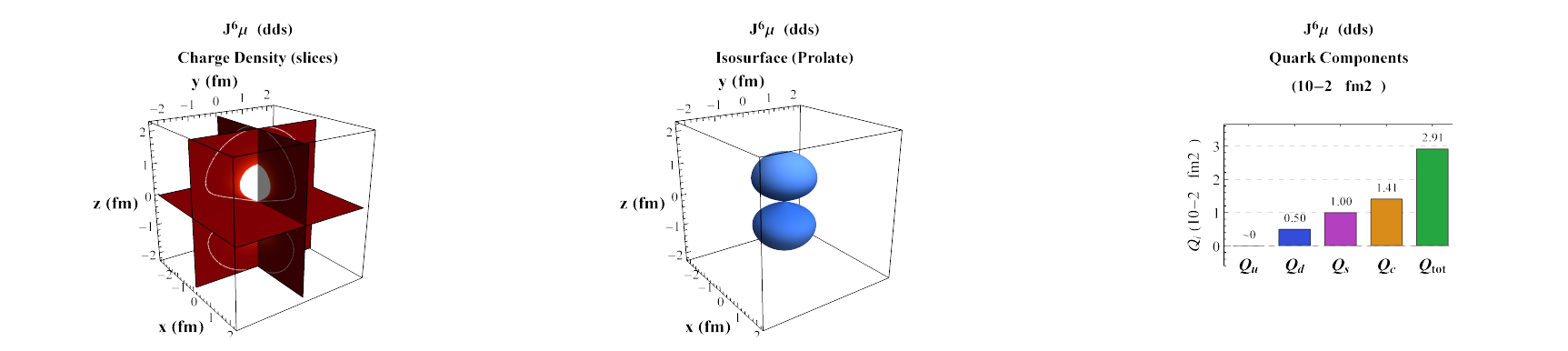}\\
\includegraphics[width=0.70\textwidth]{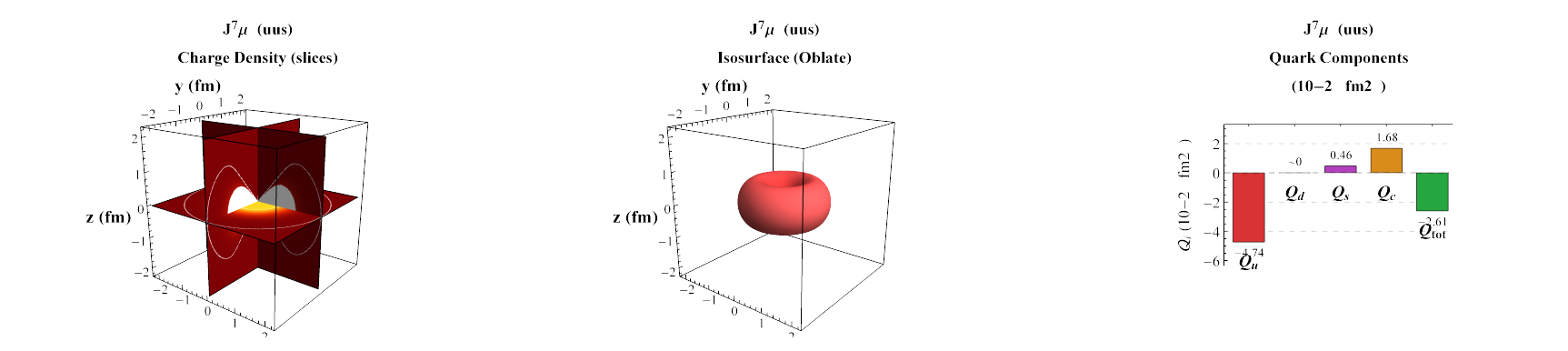} \\
\includegraphics[width=0.70\textwidth]{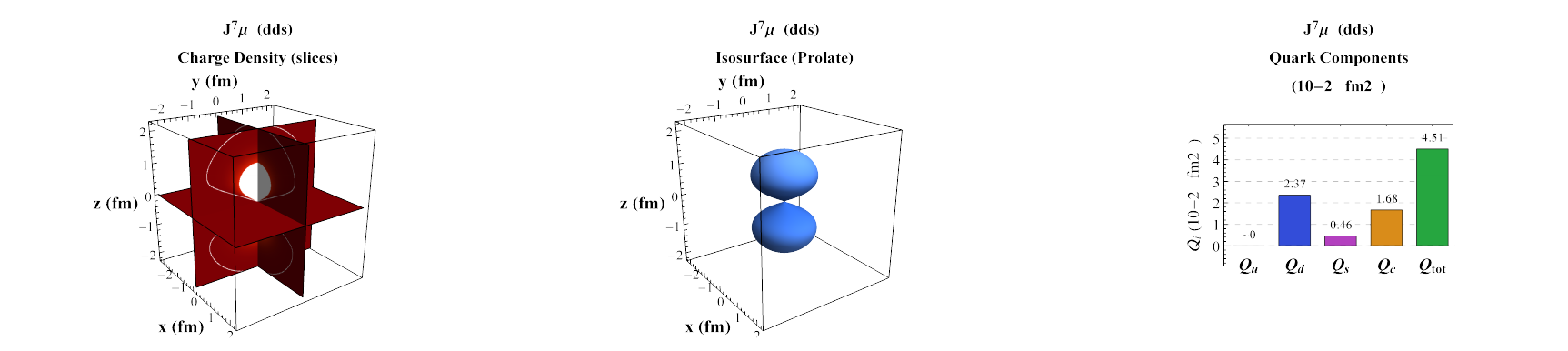}\\

\caption{Same as  Fig.~\ref{fig:quadrupole_all}, but for $J^{4}_{\mu}(x)$, $J^{5}_{\mu}(x)$, $J^{6}_{\mu}(x)$ and $J^{7}_{\mu}(x)$ interpolating currents.}
\label{fig:quadrupole_all2}
\end{figure}

The numerical results of Table~\ref{tab:quadrupole_spin32} are
complemented by the three-dimensional visualizations presented in
Figs.~\ref{fig:quadrupole_all} and~\ref{fig:quadrupole_all2}, which
display for each state the electric charge density on the three
orthogonal midplanes, the deformed charge-density isosurface, and the
quark-flavour decomposition bar chart. The isosurfaces make the sign of $\mathcal{Q}_{\mathrm{tot}}$ immediately recognizable: the blue prolate (cigar-shaped) surfaces of $J^{3}_{\mu}(x)$, $J^{4}_{\mu}(x)$, $J^{6}_{\mu}(x)$, and $J^{7}_{\mu}(x)$ (dds configuration) contract to the
red oblate (disk-shaped) surfaces of the charm-dominated currents
$J^{1}_{\mu}(x)$, $J^{2}_{\mu}(x)$, and $J^{5}_{\mu}(x)$, and to the
strongly oblate surface of $J^{4}_{\mu}(x)$ (uus). The charge-density midplane plots further reveal that the asymmetry between the $z$-axis and the equatorial plane is largest precisely for the states with the largest $|Q_0|$, while the nearly spherical midplane pattern of, e.g.,
$J^{1}_{\mu}(x)$ visible in Figs.~\ref{fig:quadrupole_all}
and~\ref{fig:quadrupole_all2} is consistent with the small
intrinsic deformation $Q_0 \approx -2.3\times 10^{-2}$~fm$^2$
of that state.

\paragraph{Charm-dominated currents: oblate deformation:}
Currents $J^1_\mu(x)$, $J^2_\mu(x)$, and $J^5_\mu(x)$ yield
$\mathcal{Q} \in [-1.43,\,-1.11]\times 10^{-2}~\mathrm{fm}^2$ and thus
$Q_0 \in [-2.4,\,-1.9]\times 10^{-2}~\mathrm{fm}^2$. The quark
decomposition (Table~\ref{tab:quadrupole_spin32}) shows that the charm
contribution accounts for $\sim 99\%$ of the total
($\mathcal{Q}_c \approx (-1.1\text{ to }-1.4)\times 10^{-2}~\mathrm{fm}^2$), with light-quark contributions at the
$\lesssim 0.02\times 10^{-2}~\mathrm{fm}^2$ level. The physical
interpretation is consistent with the operator structure: when the light diquarks are in spin-singlet states, their contribution to the quadrupole operator --- which measures the spatial anisotropy of the charge distribution --- is suppressed just as their spin contribution is suppressed. The quadrupole moment then reflects only the spatial coupling of the charm sector to the external field as encoded by $\Gamma_3 = C$, which produces a negative (oblate) result. The near-universality of $\mathcal{Q}_c$ across three structurally different currents ($-1.41$, $-1.18$, $-1.12$ $\times 10^{-2}~\mathrm{fm}^2$ for $J^1_\mu(x)$, $J^2_\mu(x)$, $J^5_\mu(x)$ respectively) further supports the interpretation that the $\Gamma_3 = C$ coupling, shared by all three currents, is the dominant factor determining the charm contribution to the quadrupole.

\paragraph{Two-axial-vector-diquark currents: prolate deformation and flavor reversal:}
Currents $J^3_\mu(x)$, $J^4_\mu(x)$, $J^6_\mu(x)$, and $J^7_\mu(x)$ yield large positive quadrupole moments for the $[sd][dc]\bar{c}$ configuration: $\mathcal{Q} = +3.14,\,+4.83,\,+2.91,\,+4.51\times 10^{-2}~\mathrm{fm}^2$, corresponding to $Q_0 = +5.2,\,+8.0,\,+4.9,\,+7.5\times 10^{-2}~\mathrm{fm}^2$. 
The intrinsic quadrupole moments obtained for the two-axial-vector-diquark currents reach values up to $Q_0 \approx +8.0 \times 10^{-2}$~fm$^2$ (Table~\ref{tab:multipole_results_spin32}), approximately four times larger in magnitude than the oblate deformations predicted by the scalar-diquark currents 
($Q_0 \approx -2\times10^{-2}$~fm$^2$). The latter are 
governed entirely by the charm sector, indicating a charge 
distribution compressed along the spin axis rather than 
elongated. This stark contrast between the two diquark 
configurations directly reflects the organizing principle 
established throughout this work: the spin content of the 
diquarks governs not only the magnitude but also the 
geometry of the electromagnetic response. The prolate and 
oblate isosurfaces visible in 
Figs.~\ref{fig:quadrupole_all} and~\ref{fig:quadrupole_all2} 
provide a direct geometric representation of this conclusion.

The mechanism producing the prolate deformation is qualitatively different
from the oblate case. In $J^3_\mu(x)$ with $[sd][dc]\bar{c}$, the quark
decomposition gives $\mathcal{Q}_s = +1.63$ and
$\mathcal{Q}_c = +1.51\times 10^{-2}~\mathrm{fm}^2$, with both contributions
positive and comparable in magnitude. This constructive addition requires
that the spatial distributions of both the strange quark and the charm
sector are elongated along the same axis, which in turn requires a
coherent spin-spatial correlation between the two sectors. In the
two-axial-vector-diquark operator structure, both $\Gamma_2$ and $\Gamma_5$
carry Lorentz indices that couple the spin and spatial degrees of freedom,
and the $\Gamma_6 = \gamma_5\gamma^\alpha C$ coupling contracts the
internal index while leaving the external index $\mu$ free. This specific
index structure correlates the spin orientations of the two diquarks in
a way that produces coherent contributions to the quadrupole operator,
which is the underlying reason for the large positive values.

In the $[su][uc]\bar{c}$ configuration, $J^4_\mu(x)$ and 
$J^7_\mu(x)$ exhibit a sign reversal: $\mathcal{Q} = -4.32$ 
and $-2.60\times 10^{-2}~\mathrm{fm}^2$, while $J^3_\mu(x)$ 
and $J^6_\mu(x)$ remain positive. The intermediate 
$[sd][uc]\bar{c}$ configuration yields $\mathcal{Q} = +0.26$ 
and $+0.96\times 10^{-2}~\mathrm{fm}^2$ for $J^4_\mu(x)$ 
and $J^7_\mu(x)$ respectively, confirming the continuous 
transition from prolate to oblate geometry as $d\to u$. The reversal in $J^4_\mu(x)$ and $J^7_\mu(x)$ is driven by $\mathcal{Q}_u = -6.10\times 10^{-2}~\mathrm{fm}^2 
= -2\,\mathcal{Q}_d$, which overcomes the positive charm contribution
($+1.78\times 10^{-2}~\mathrm{fm}^2$). The fact that the reversal occurs
in $J^4_\mu(x)$ and $J^7_\mu(x)$ but not in $J^3_\mu(x)$ and $J^6_\mu(x)$ reflects again the Lorentz-index assignment: in $J^4_\mu(x)$, the heavy-light diquark $[s^T\Gamma_2 c]$ carries the external index, making the charm contribution more sensitive to the light-quark flavor change. In $J^3_\mu(x)$, the light diquark $[q^T\Gamma_2 q]$ carries the external index, and the $s$-quark contribution ($\mathcal{Q}_s = +1.63\times 10^{-2}~\mathrm{fm}^2$) is sufficiently large to prevent reversal even when $u$ replaces $d$. This flavor-driven sign reversal is clearly captured in
Figs.~\ref{fig:quadrupole_all} and~\ref{fig:quadrupole_all2}, where the
transition from a prolate isosurface in the dds configuration to an oblate
one in the uus configuration of $J^4_\mu(x)$ and $J^7_\mu(x)$ is directly visible. The deformation geometry of the pentaquark is therefore sensitive not only to the diquark spin content but also to the specific index structure of the operator, providing a layer of discrimination between different diquark configurations that is unique to higher multipole observables.

\subsubsection{Magnetic octupole moments}
\label{subsubsec:O32}

The magnetic octupole moments $\mathcal{O}$ are an order of magnitude
smaller than the quadrupole moments, as expected from the multipole
hierarchy, and constitute new predictions with no molecular or quark
model counterpart. They range from
$-1.42\times 10^{-3}~\mathrm{fm}^3$ ($J^4_\mu(x)$, $[sd][dc]\bar{c}$)
to $+0.56\times 10^{-3}~\mathrm{fm}^3$ ($J^4_\mu(x)$, $[su][uc]\bar{c}$).
\begin{table}[htbp]
\centering
\caption{Decomposition of the magnetic octupole moments 
$\mathcal{O}$ (in $10^{-3}$~fm$^3$) of the 
$P^{\Sigma^\ast}_{\psi s}$ pentaquarks into quark-flavor 
contributions for the spin-$\tfrac{3}{2}$ sector, evaluated 
for the $[sd][dc]\bar{c}$, $[sd][uc]\bar{c}$, and 
$[su][uc]\bar{c}$ configurations.}
\label{tab:octupole_spin32}
\renewcommand{\arraystretch}{1.25}
\setlength{\tabcolsep}{9pt}
\begin{tabular}{llccccc}
\toprule
& & \multicolumn{4}{c}{Quark Contributions} & Total \\
\cmidrule(lr){3-6}
Current & Configuration
  & $\mathcal{O}_u$ & $\mathcal{O}_d$ &
  $\mathcal{O}_s$ & $\mathcal{O}_c$ &
  $\mathcal{O}_{\mathrm{tot}}$ \\
\midrule
\multirow{3}{*}{$J_\mu^1(x)$}
  & $[sd][dc]\bar{c}$ & $-$ & ${\sim}0$ & ${\sim}0$ & $-0.27$ & $-0.27$ \\
  & $[sd][uc]\bar{c}$ & ${\sim}0$ & ${\sim}0$ & ${\sim}0$ & $-0.27$ & $-0.27$ \\
  & $[su][uc]\bar{c}$ & ${\sim}0$ & $-$ & ${\sim}0$ & $-0.27$ & $-0.27$ \\[4pt]
\multirow{3}{*}{$J_\mu^2(x)$}
  & $[sd][dc]\bar{c}$ & $-$ & $\sim 0$ & ${\sim}0$ & $-0.23$ & $-0.23$ \\
  & $[sd][uc]\bar{c}$ & $\sim 0$ & $\sim 0$ & ${\sim}0$ & $-0.23$ & $-0.23$ \\
  & $[su][uc]\bar{c}$ & $\sim 0$ & $-$ & ${\sim}0$ & $-0.23$ & $-0.23$ \\[4pt]
\multirow{3}{*}{$J_\mu^3(x)$}
  & $[sd][dc]\bar{c}$ & $-$ & ${\sim}0$ & $-0.32$ & $-0.62$ & $-0.94$ \\
  & $[sd][uc]\bar{c}$ & ${\sim}0$ & ${\sim}0$ & $-0.32$ & $-0.62$ & $-0.94$ \\
  & $[su][uc]\bar{c}$ & ${\sim}0$ & $-$ & $-0.32$ & $-0.62$ & $-0.94$ \\[4pt]
\multirow{3}{*}{$J_\mu^4(x)$}
  & $[sd][dc]\bar{c}$ & $-$ & $-0.66$ & ${\sim}0$ & $-0.76$ & $-1.42$ \\
  & $[sd][uc]\bar{c}$ & $\phantom{-}0.66$ & $-0.33$ & ${\sim}0$ & $-0.76$ & $-0.43$ \\
  & $[su][uc]\bar{c}$ & $\phantom{-}1.32$ & $-$ & ${\sim}0$ & $-0.76$ & $\phantom{-}0.56$ \\[4pt]
\multirow{3}{*}{$J_\mu^5(x)$}
  & $[sd][dc]\bar{c}$ & $-$ & ${\sim}0$ & ${\sim}0$ & $-0.22$ & $-0.22$ \\
  & $[sd][uc]\bar{c}$ & ${\sim}0$ & ${\sim}0$ & ${\sim}0$ & $-0.22$ & $-0.22$ \\
  & $[su][uc]\bar{c}$ & ${\sim}0$ & $-$ & ${\sim}0$ & $-0.22$ & $-0.22$ \\[4pt]
\multirow{3}{*}{$J_\mu^6(x)$}
  & $[sd][dc]\bar{c}$ & $-$ & $-0.10$ & $-0.20$ & $-0.58$ & $-0.88$ \\
  & $[sd][uc]\bar{c}$ & $\phantom{-}0.10$ & $-0.05$ & $-0.20$ & $-0.58$ & $-0.73$ \\
  & $[su][uc]\bar{c}$ & $\phantom{-}0.20$ & $-$ & $-0.20$ & $-0.58$ & $-0.58$ \\[4pt]
\multirow{3}{*}{$J_\mu^7(x)$}
  & $[sd][dc]\bar{c}$ & $-$ & $-0.54$ & $-0.11$ & $-0.74$ & $-1.39$ \\
  & $[sd][uc]\bar{c}$ & $\phantom{-}0.54$ & $-0.27$ & $-0.11$ & $-0.74$ & $-0.58$ \\
  & $[su][uc]\bar{c}$ & $\phantom{-}1.08$ & $-$ & $-0.11$ & $-0.74$ & $\phantom{-}0.23$ \\
\bottomrule
\end{tabular}
\end{table}

\begin{figure}[htbp]
\centering
\includegraphics[width=0.7\textwidth]{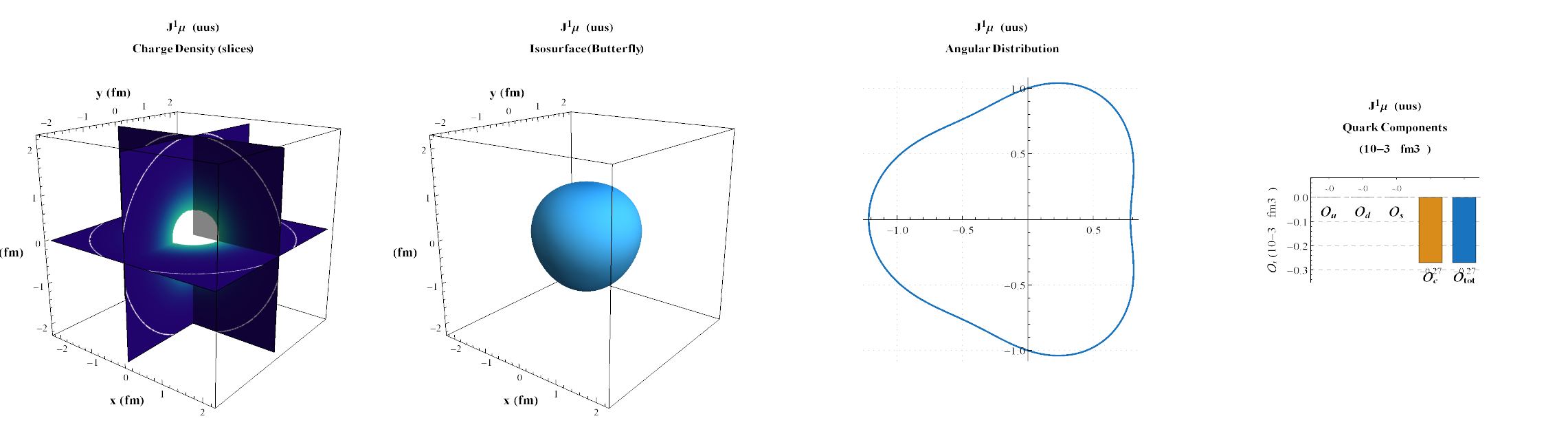} \\
\includegraphics[width=0.7\textwidth]{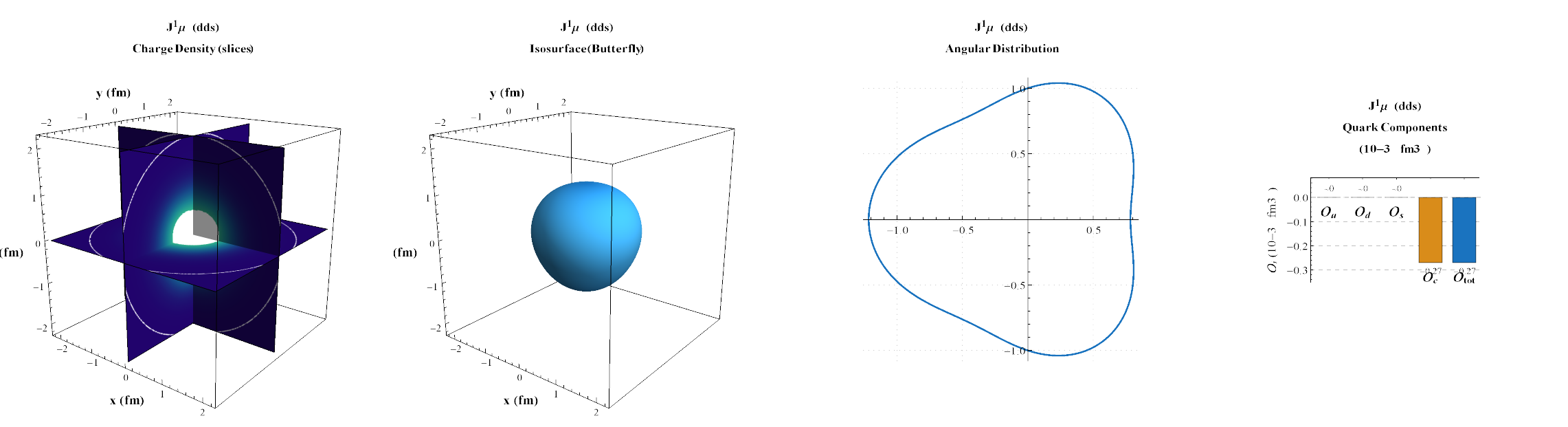}\\
\includegraphics[width=0.7\textwidth]{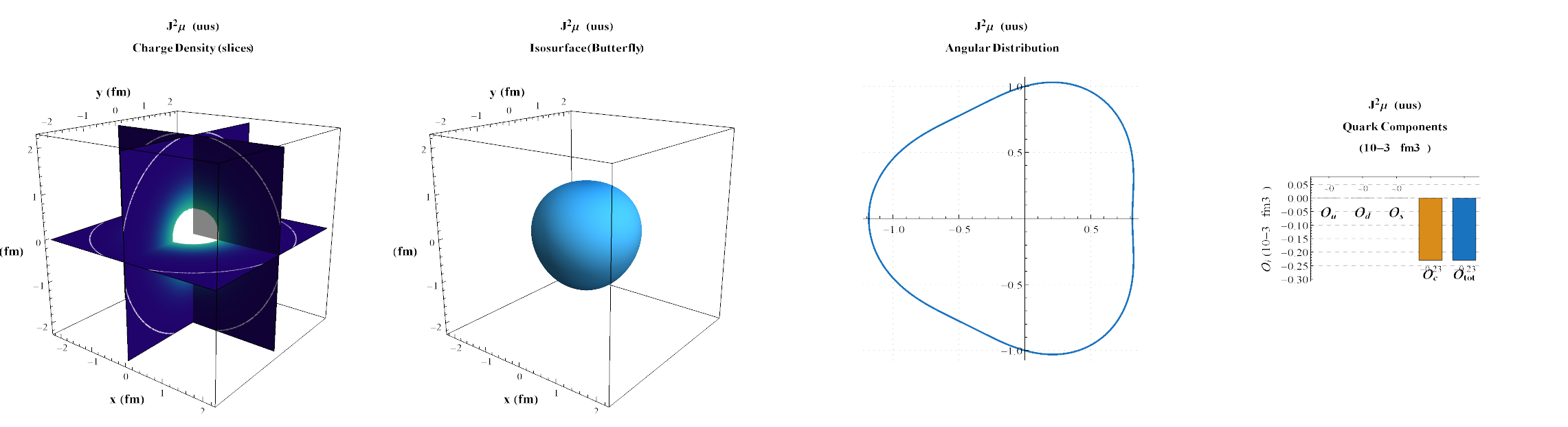} \\
\includegraphics[width=0.7\textwidth]{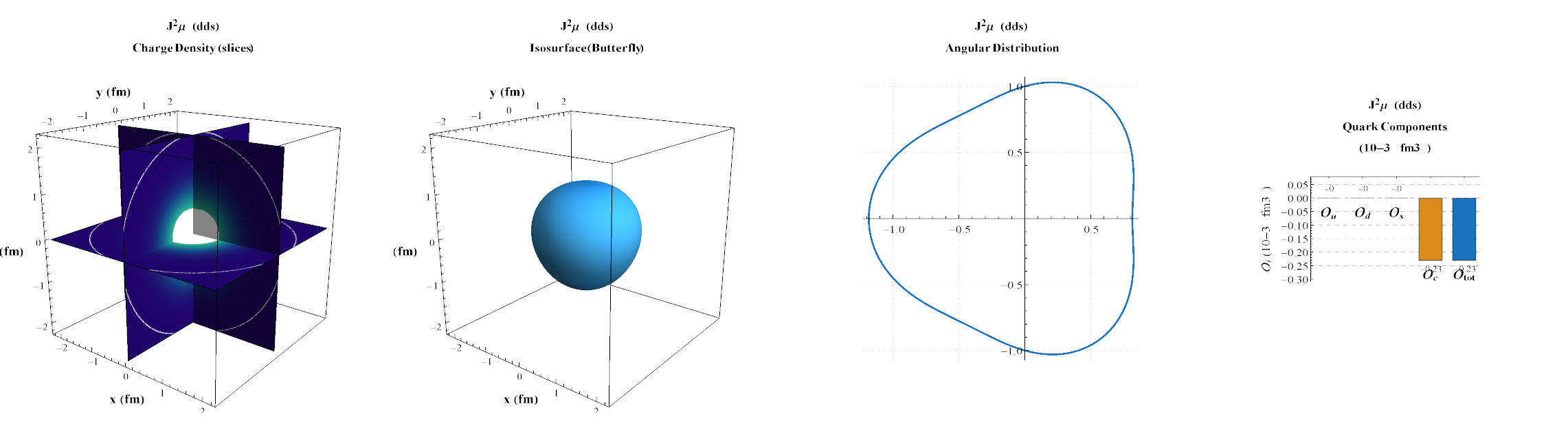}\\
\includegraphics[width=0.7\textwidth]{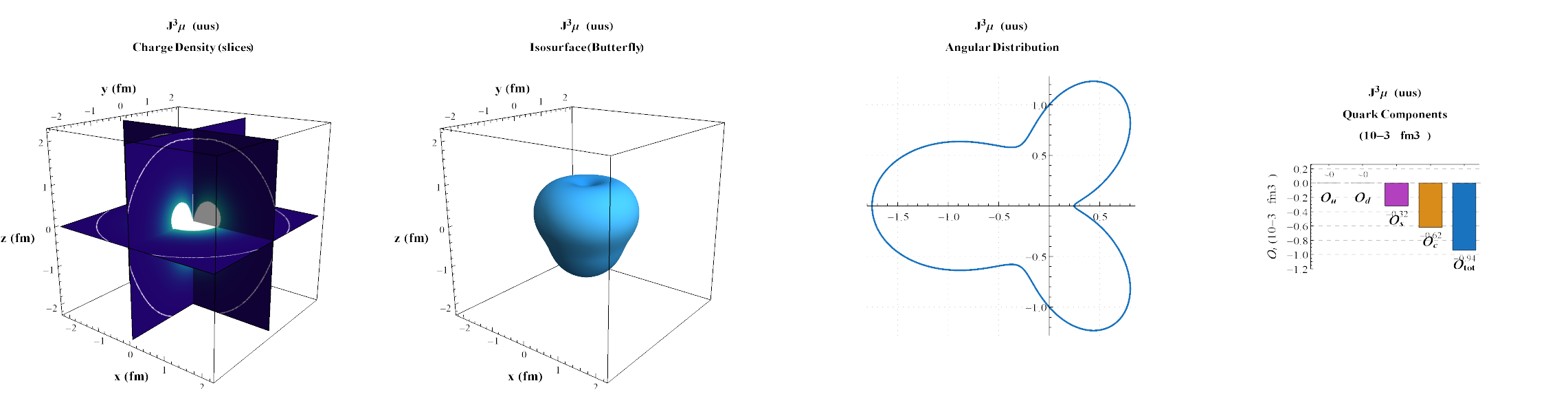} \\
\includegraphics[width=0.7\textwidth]{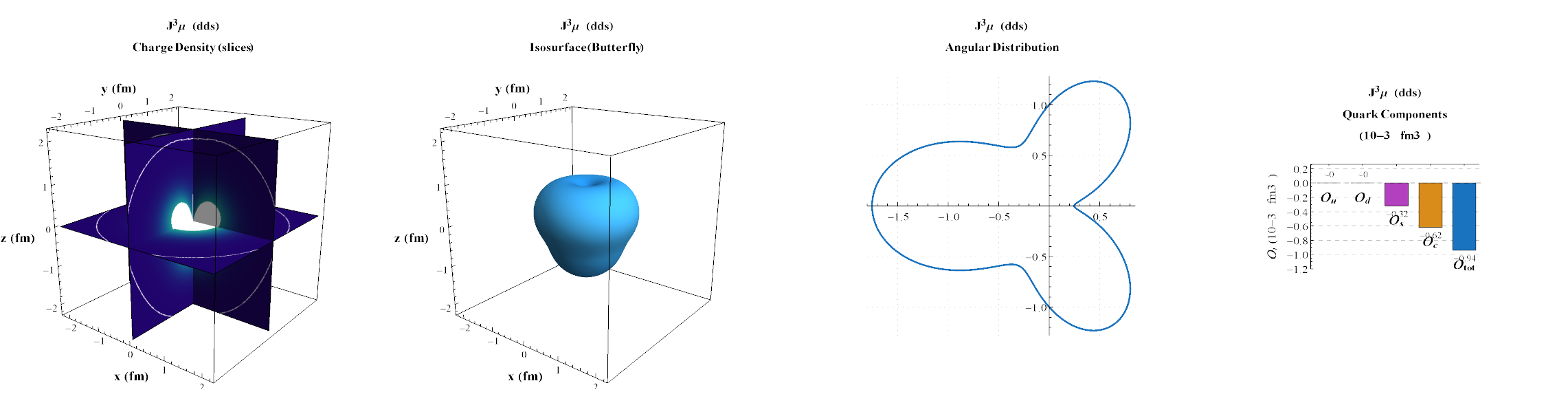}\\
\caption{
Magnetic octupole moment analysis of the $P^{\Sigma^\ast}_{\psi s}$ pentaquarks for $J^{1}_{\mu}(x)$, $J^{2}_{\mu}(x)$ and $J^{3}_{\mu}(x)$ interpolating currents and both diquark configurations, $[su][uc]\bar{c}$ (uus) and $[sd][dc]\bar{c}$ (dds). For each state, four panels are shown: (\textit{first}) the three-dimensional charge density $\rho(\mathbf{r})$ on the orthogonal midplanes $x=0$, $y=0$, and $z=0$, coloured with a blue-green heat map under a global normalisation so that the colour scale is consistent across all states;
(\textit{second}) the charge-density isosurface at a fixed threshold, coloured purple for a positive (pear-shaped) and teal for a negative (butterfly-shaped) octupole deformation, reflecting the dominant $Y_{30}$ angular
structure; (\textit{third}) the polar angular distribution
$r(\theta)\propto 1+\alpha\,\mathcal{O}_{\mathrm{tot}}
(5\cos^{3}\theta - 3\cos\theta)$, which illustrates the characteristic pear, butterfly, or spherical shape associated with a positive, negative, or
vanishing total octupole moment, respectively; and (\textit{fourth}) the quark-flavour decomposition bar chart showing the contributions
$\mathcal{O}_{u}$, $\mathcal{O}_{d}$, $\mathcal{O}_{s}$,
$\mathcal{O}_{c}$, and $\mathcal{O}_{\mathrm{tot}}$,
with numeric values annotated on each bar. All octupole moments are given in units of $10^{-3}$~fm$^{3}$ and all spatial coordinates in fm.}
\label{fig:octupole_all}
\end{figure}

\begin{figure}[htbp]
\centering
\includegraphics[width=0.62\textwidth]{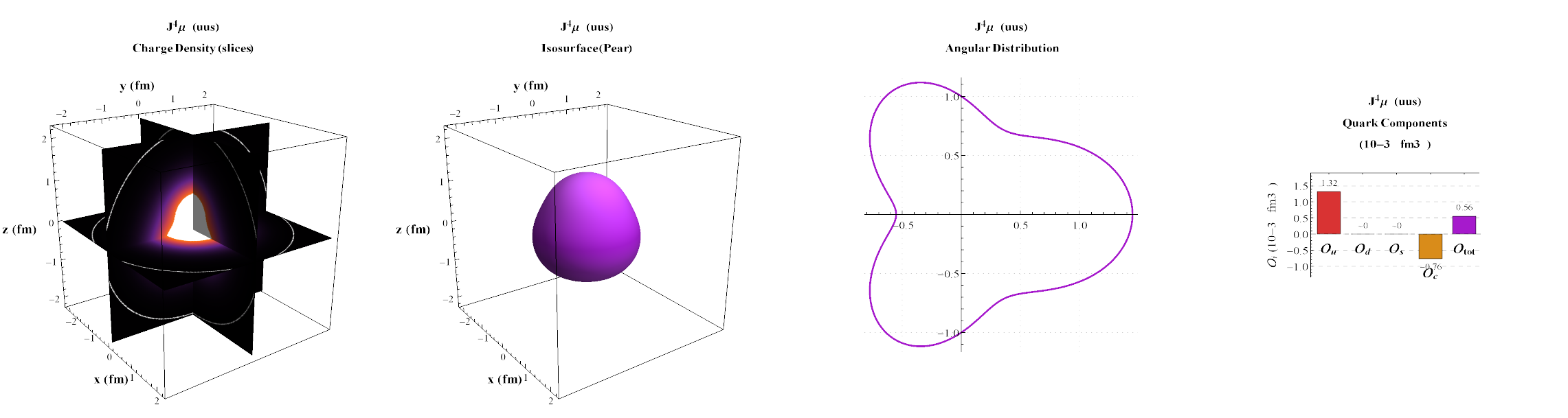} \\
\includegraphics[width=0.62\textwidth]{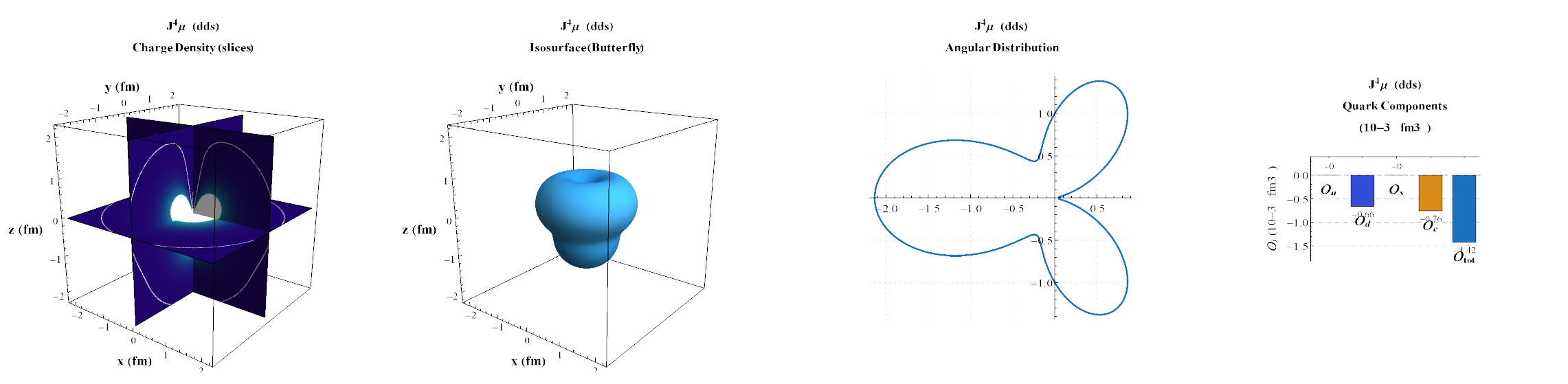}\\
\includegraphics[width=0.62\textwidth]{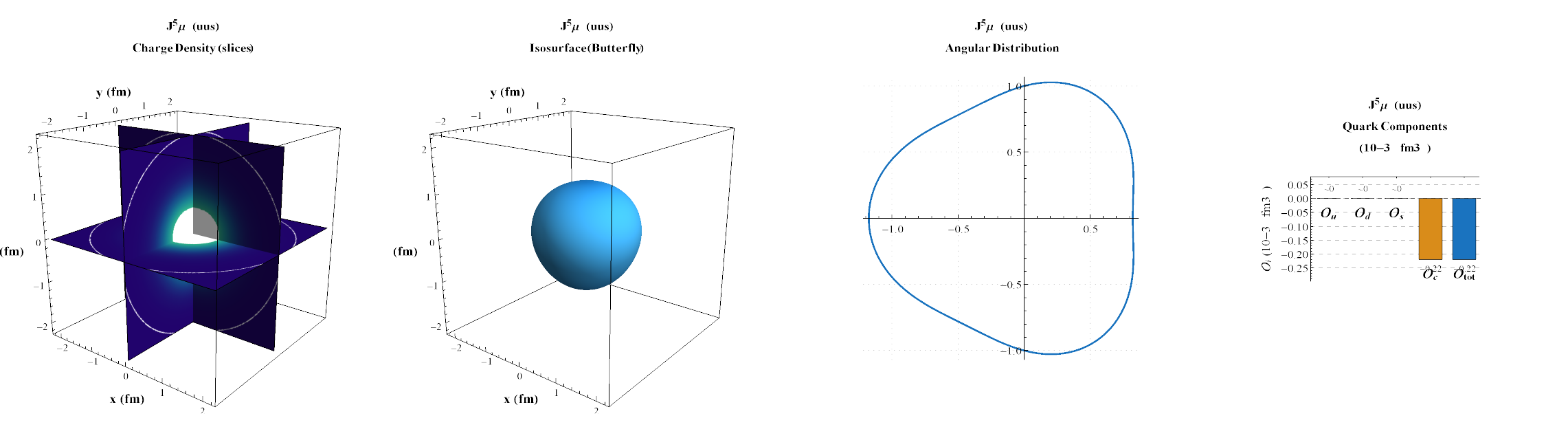} \\
\includegraphics[width=0.62\textwidth]{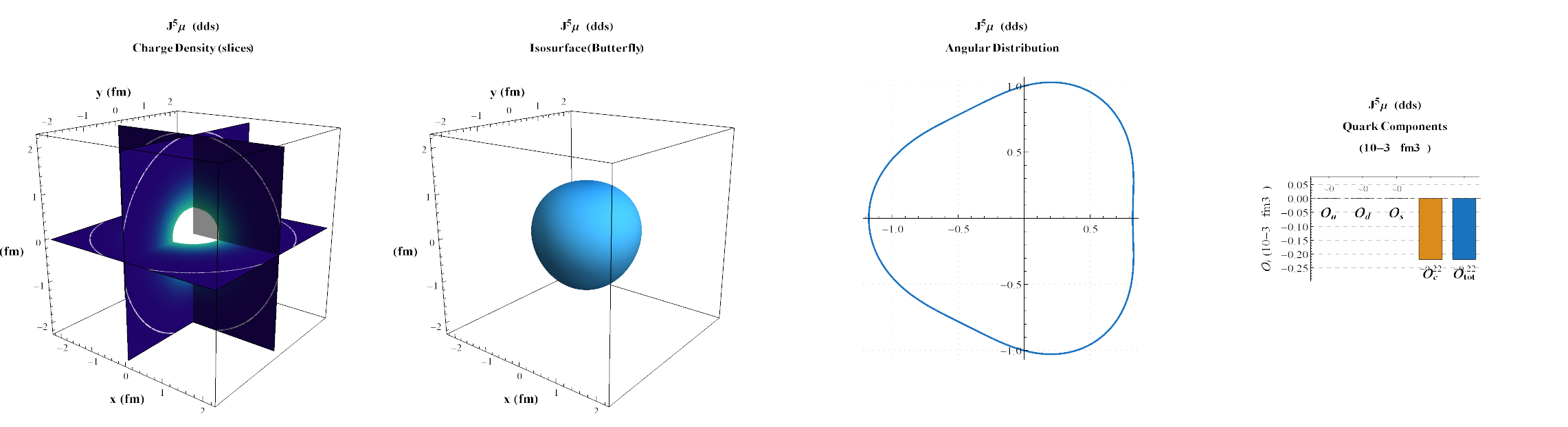}\\
\includegraphics[width=0.62\textwidth]{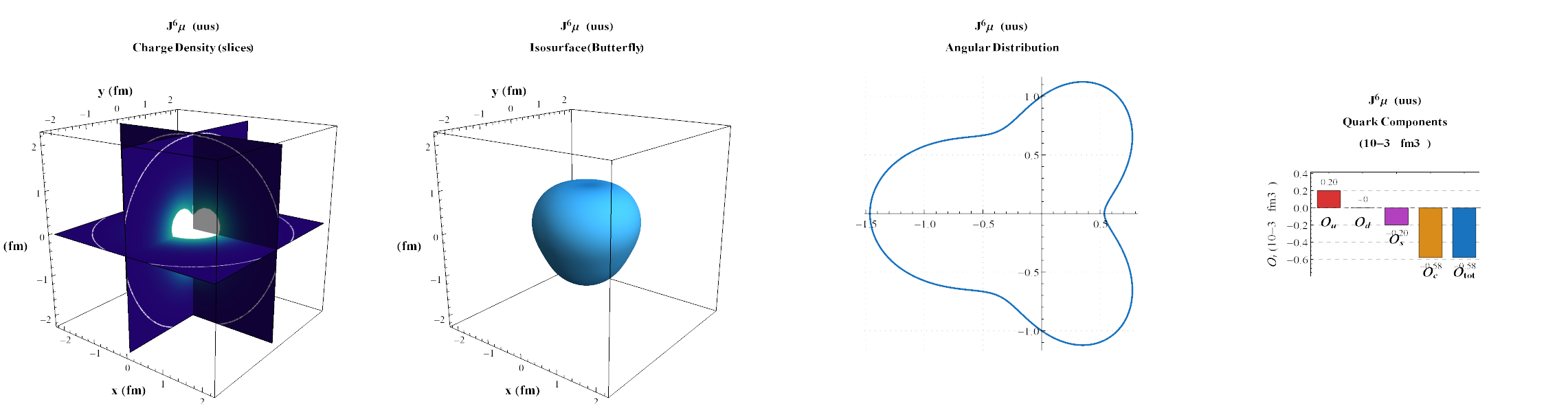} \\
\includegraphics[width=0.62\textwidth]{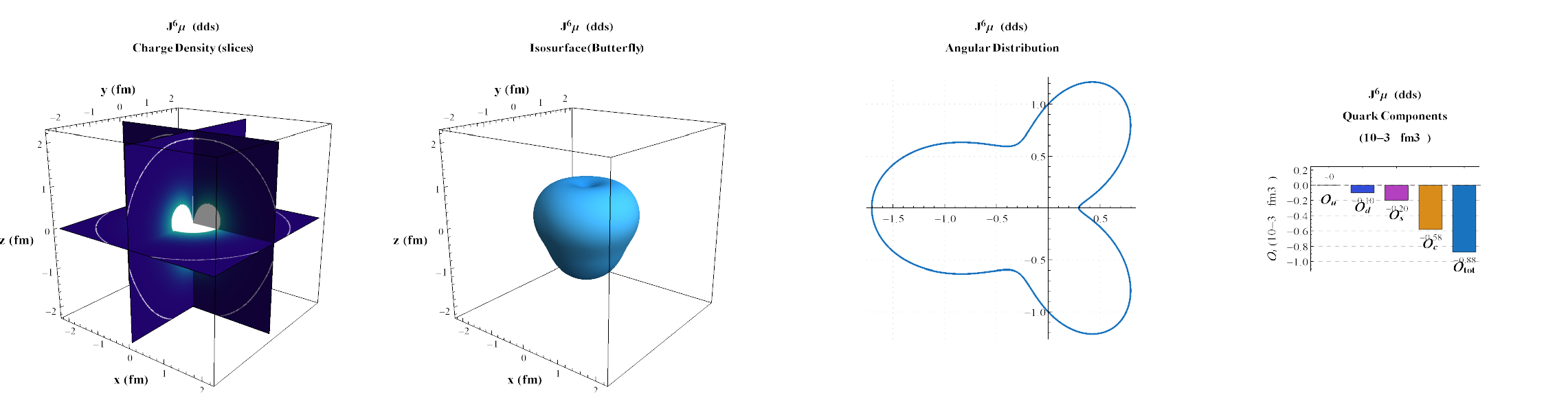}\\
\includegraphics[width=0.62\textwidth]{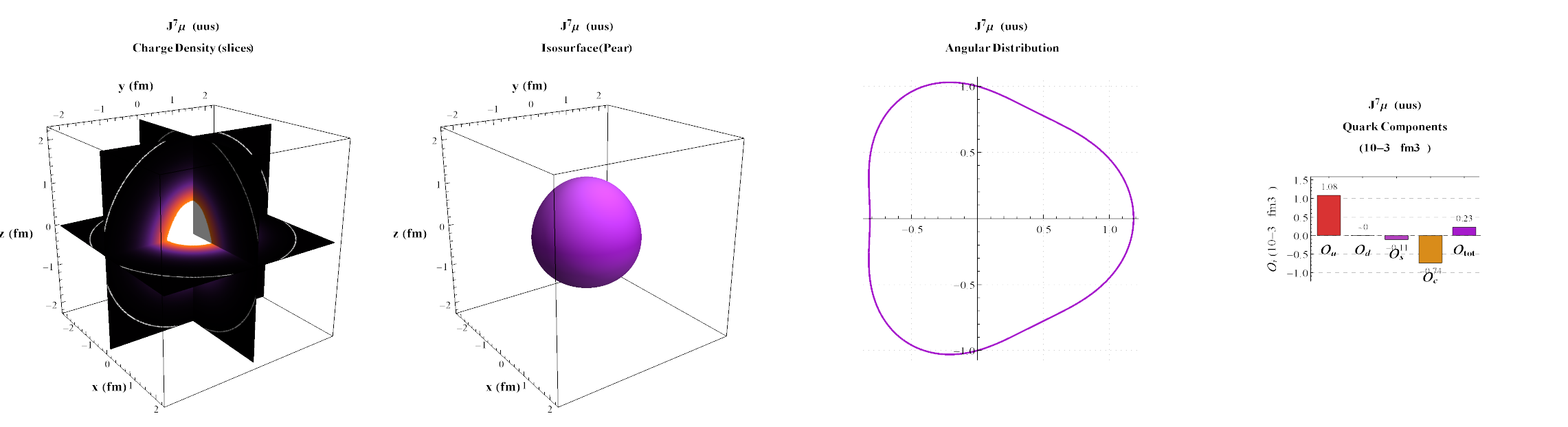} \\
\includegraphics[width=0.62\textwidth]{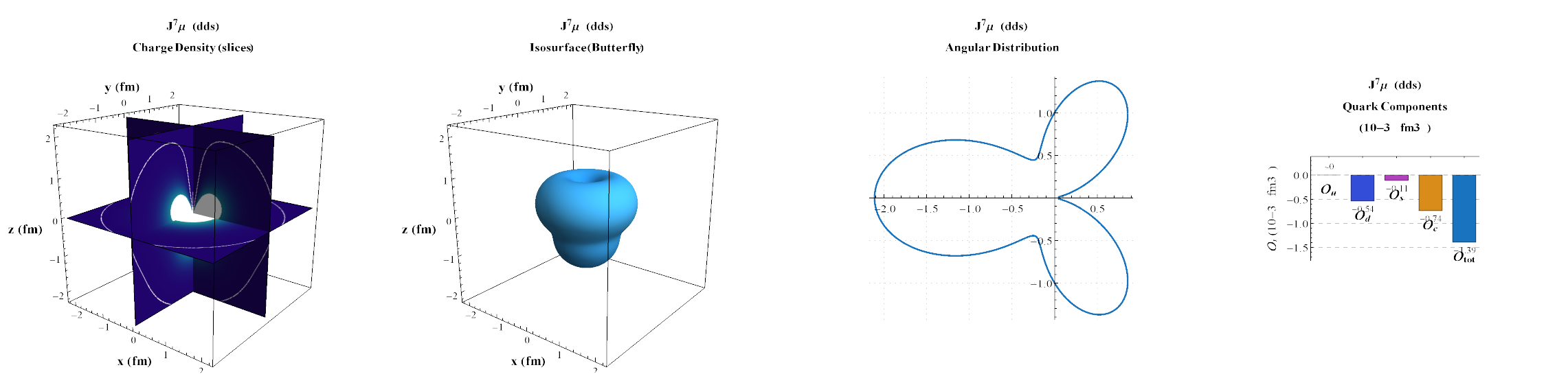}\\
\caption{Same as Fig.~\ref{fig:octupole_all}, but for $J^{4}_{\mu}(x)$, $J^{5}_{\mu}(x)$, $J^{6}_{\mu}(x)$ and $J^{7}_{\mu}(x)$ interpolating currents.}
\label{fig:octupole_all2}
\end{figure}

The numerical results of Table~\ref{tab:octupole_spin32} are complemented by the four-panel visualizations presented in Figs.~\ref{fig:octupole_all} and~\ref{fig:octupole_all2}, which display for each state the charge density on the three orthogonal midplanes, the deformed charge-density isosurface, the polar angular distribution $r(\theta)\propto 1+\alpha\,\mathcal{O}_{\mathrm{tot}} (5\cos^3\theta-3\cos\theta)$, and the quark-flavour decomposition bar chart. The isosurfaces and polar plots together make the sign and magnitude of $\mathcal{O}_{\mathrm{tot}}$ immediately accessible: states dominated by the charm sector ($J^1_\mu(x)$, $J^2_\mu(x)$, $J^5_\mu(x)$) display compact, nearly spherical surfaces with a mild teal tint indicating a small negative octupole, whereas $J^4_\mu(x)$ (dds) and $J^7_\mu(x)$ (dds) exhibit strongly asymmetric butterfly-shaped surfaces consistent with their large negative values $-1.42$ and $-1.39\times10^{-3}$~fm$^3$, respectively. The sign reversal between the dds and uus configurations of $J^4_\mu(x)$ and $J^7_\mu(x)$ is directly visible as a colour transition from teal (butterfly, negative) to purple (pear, positive) in the isosurface panels of Figs.~\ref{fig:octupole_all} and~\ref{fig:octupole_all2}.

\paragraph{Near-universal charm dominance and its physical significance:}
For currents $J^1_\mu(x)$, $J^2_\mu(x)$, and $J^5_\mu(x)$, the octupole moment is dominated by the charm contribution and is nearly identical across
all three currents and both flavor configurations:
\begin{equation}
\mathcal{O}_{J^1_\mu(x)} \;\approx\; \mathcal{O}_{J^2_\mu(x)} \;\approx\;
\mathcal{O}_{J^5_\mu(x)} \;\approx\;
-(0.22\text{--}0.27)\times 10^{-3}~\mathrm{fm}^3.
\label{eq:O_universal}
\end{equation}
This near-universality is more than a numerical coincidence: it has a structural explanation. All three currents share the $\Gamma_3 = C$ coupling, which suppresses light-quark contributions at every multipole order. The octupole operator, like the quadrupole operator, probes the spatial anisotropy of the magnetization distribution, and when the light diquarks are in spin-singlet configurations, this anisotropy is governed entirely by the charm-sector coupling through $\Gamma_3$. The mild variation of $\mathcal{O}_c$ across the three currents ($-0.27$, $-0.23$, $-0.22\times 10^{-3}~\mathrm{fm}^3$) reflects the different light-diquark structures in $J^1_\mu(x)$, $J^2_\mu(x)$, $J^5_\mu(x)$ having a small but non-zero effect on the charm propagator through subleading OPE terms. The value $\mathcal{O} \approx -0.25\times 10^{-3}~\mathrm{fm}^3$ is therefore a prediction that depends primarily on the $\Gamma_3 = C$ coupling and is insensitive to the detailed diquark topology. 
It constitutes a theoretical target for future lattice QCD 
calculations that can be tested without resolving the 
interpolating-operator ambiguity. We note that extracting 
the octupole moment on the lattice will require substantially 
higher statistical precision than the dipole or quadrupole, 
and that extending existing tetraquark 
calculations~\cite{Vujmilovic:2025czt} to pentaquark systems 
represents a significant additional computational challenge. 
Nevertheless, the operator-independence of this prediction 
makes it a well-defined benchmark once such calculations 
become feasible.  
The near-spherical isosurfaces and nearly circular polar plots of these three currents in Figs.~\ref{fig:octupole_all} and~\ref{fig:octupole_all2} provide a direct geometric confirmation of this universality.

\paragraph{Two-axial-vector-diquark currents and the breakdown of universality:}
For $J^4_\mu(x)$, $J^6_\mu(x)$, and $J^7_\mu(x)$, light-quark contributions to the octupole moment are significant: $|\mathcal{O}_{u,d}|$ reaches $1.32\times 10^{-3}$~fm$^3$ for $J^4_\mu(x)$ in the $[su][uc]\bar{c}$ configuration, a value 
five times larger than the typical light-quark contribution in the $\Gamma_3$-coupled currents. Among these three currents, $J^4_\mu(x)$ and $J^7_\mu(x)$ exhibit sign reversals between $[sd][dc]\bar{c}$ and $[su][uc]\bar{c}$ via the same $Q_u/Q_d = -2$ mechanism as in the quadrupole, confirming that the charge ratio governs the light-quark electromagnetic response at every multipole order once the spin-singlet suppression is lifted. In contrast, $J^6_\mu(x)$ does not 
exhibit a sign reversal: both configurations remain negative ($-0.88$ and $-0.58\times 10^{-3}$~fm$^3$), because the strange-quark contribution $\mathcal{O}_s = -0.20\times 10^{-3}$~fm$^3$ reinforces the charm term and prevents the $u$-quark contribution from reversing the total sign. Notably, while $J^6_\mu(x)$ also belongs to the two-axial-vector-diquark class, its $(\mathcal{Q},\mathcal{O})$ sign correlation remains $(+,-)$ for both flavor 
configurations, unlike $J^4_\mu(x)$ and $J^7_\mu(x)$ where the correlation switches from $(+,-)$ to $(-,+)$ upon $d\to u$ substitution. This distinction traces directly to the strange-quark contribution stabilizing the sign of the 
total octupole moment in $J^6_\mu(x)$, and mirrors the analogous pattern observed in the quadrupole sector. This breakdown of universality is strikingly visible in 
Table~\ref{tab:octupole_spin32} and Figs.~\ref{fig:octupole_all} and~\ref{fig:octupole_all2}: while the dds configurations of $J^4_\mu(x)$ and $J^7_\mu(x)$ show strongly elongated butterfly isosurfaces in Fig.~\ref{fig:octupole_all2}, their uus counterparts display a pear-shaped or nearly spherical distribution, reflecting the large positive $u$-quark contributions $\mathcal{O}_u = +1.32$ and $+1.08\times 10^{-3}$~fm$^3$ 
that partially cancel or reverse the negative charm and $d/s$-quark terms. $J^6_\mu(x)$, by contrast, retains a negative octupole moment in both flavor configurations (Table~\ref{tab:octupole_spin32}), and displays a butterfly 
isosurface in both cases as seen in Fig.~\ref{fig:octupole_all2}, consistent with the stable $(\mathcal{Q},\mathcal{O}) = (+,-)$ sign correlation across 
the entire isospin multiplet.

\paragraph{Sign correlation between the electric quadrupole moment
$\mathcal{Q}$ and the magnetic octupole moment $\mathcal{O}$:}
The relative sign between $\mathcal{Q}$ and $\mathcal{O}$ reveals a
systematic structural pattern that provides an additional experimental
discriminant independent of the magnitudes, summarized in
Table~\ref{tab:sign_correlation}.

\begin{table}[htbp]
\centering
\caption{Sign correlation between the electric quadrupole 
moment $\mathcal{Q}$ and the magnetic octupole moment 
$\mathcal{O}$ for the spin-$3/2$ currents. The entries give 
$(\mathrm{sgn}\,\mathcal{Q},\,\mathrm{sgn}\,\mathcal{O})$.}
\label{tab:sign_correlation}
\renewcommand{\arraystretch}{1.2}
\setlength{\tabcolsep}{8pt}
\begin{tabular}{lcccc}
\toprule
Current & $[sd][dc]\bar{c}$ & $[sd][uc]\bar{c}$ 
  & $[su][uc]\bar{c}$ & Pattern \\
\midrule
$J^1_\mu(x)$ & $(-,-)$ & $(-,-)$ & $(-,-)$ & Same sign \\
$J^2_\mu(x)$ & $(-,-)$ & $(-,-)$ & $(-,-)$ & Same sign \\
$J^3_\mu(x)$ & $(+,-)$ & $(+,-)$ & $(+,-)$ & Opposite sign \\
$J^4_\mu(x)$ & $(+,-)$ & $(+,-)$ & $(-,+)$ & Opposite sign \\
$J^5_\mu(x)$ & $(-,-)$ & $(-,-)$ & $(-,-)$ & Same sign \\
$J^6_\mu(x)$ & $(+,-)$ & $(+,-)$ & $(+,-)$ & Opposite sign \\
$J^7_\mu(x)$ & $(+,-)$ & $(+,-)$ & $(-,+)$ & Opposite sign \\
\bottomrule
\end{tabular}
\end{table}

The same-sign pattern for $J^1_\mu(x)$, $J^2_\mu(x)$, $J^5_\mu(x)$ (both  $\mathcal{Q}$ and $\mathcal{O}$ negative) reflects the fact that when the charm sector dominates both multipoles, the charge and magnetization distributions carry the same spatial orientation. The opposite-sign pattern for the two-axial-vector-diquark currents ($\mathcal{Q} > 0$, $\mathcal{O} < 0$ for $[sd][dc]\bar{c}$) signals a misalignment between the charge and magnetization distributions. For $J^4_\mu(x)$ and $J^7_\mu(x)$, this misalignment further  reverses upon $d\to u$ substitution: the $[su][uc]\bar{c}$ configuration yields $(\mathcal{Q} < 0,\,\mathcal{O} > 0)$, driven by the large negative $u$-quark quadrupole contribution $\mathcal{Q}_u = -6.10\times 10^{-2}$~fm$^2$ overwhelming 
the positive charm term, while the positive $u$-quark octupole $\mathcal{O}_u = +1.32\times 10^{-3}$~fm$^3$ reverses the sign of the total octupole. In contrast, $J^3_\mu(x)$ and $J^6_\mu(x)$ retain the same $(\mathcal{Q} > 0,\,\mathcal{O} < 0)$ pattern in both configurations, stabilized by the strange-quark contributions. The physical origin is the different quark-property weightings: the electric quadrupole weights $Q_q r_q^2$, while the magnetic octupole weights $\mu_q r_q^2 \propto (Q_q/m_q) r_q^2$. The additional $1/m_q$ factor means that the spatial distribution is weighted differently for the charge and magnetic cases: the contribution of a given quark to $\mathcal{O}$ relative to $\mathcal{Q}$ is
suppressed by $m_q^{-1}$, so heavier quarks contribute less to the octupole than to the quadrupole. When multiple quarks contribute comparably to the quadrupole (as in the two-axial-vector-diquark currents), the reweighting by $1/m_q$ can change the dominant contribution and hence the sign. In $J^3_\mu(x)$ with $[sd][dc]\bar{c}$, both $\mathcal{Q}_s = +1.63$ and $\mathcal{Q}_c = +1.51\times 10^{-2}~\mathrm{fm}^2$ are positive, giving a positive total quadrupole; but for the octupole, $\mathcal{O}_s = -0.32$ and $\mathcal{O}_c = -0.62\times 10^{-3}~\mathrm{fm}^3$ are both negative, giving a negative total. The $1/m_s$ reweighting effectively changes the sign of the strange contribution relative to
the quadrupole, producing the opposite-sign alignment. This sign-correlation pattern is therefore a direct probe of the $1/m_q$ quark-mass dependence of the magnetization distribution and provides an observable that is qualitatively different from what either the dipole or the quadrupole alone can reveal. The two-panel comparison of the quadrupole and octupole isosurfaces in Figs.~\ref{fig:quadrupole_all}, \ref{fig:quadrupole_all2},
\ref{fig:octupole_all}, and~\ref{fig:octupole_all2} makes this misalignment between charge and magnetization distributions visually transparent: states that appear prolate in Figs.~\ref{fig:quadrupole_all} and~\ref{fig:quadrupole_all2}
consistently display a butterfly octupole pattern in Figs.~\ref{fig:octupole_all} and~\ref{fig:octupole_all2}, confirming that the $(\mathcal{Q},\mathcal{O})$ sign correlation is a robust geometric feature of the two-axial-vector-diquark current
structure. 

\subsection{Theoretical implications} \label{subsec:theory}

The systematic results presented above permit several conclusions that go beyond the individual current-by-current observations.
\begin{itemize}
\item The most fundamental finding is that the entire pattern of electromagnetic multipole moments --- across all currents, all multipole orders, and both flavor configurations --- is organized by a single physical parameter: the spin content of the diquarks. When the light-quark diquarks are spin-singlet ($J = 0$), the 
electromagnetic response of the system at every multipole order is dominated by the charm sector, is flavor-insensitive, and is suppressed in magnitude. When the light-quark diquarks are spin-triplet ($J = 1$), the light quarks carry substantial electromagnetic response, the moments are large, flavor-sensitive, 
and exhibit sign reversals governed by the $e_u/e_d = -2$ charge ratio. This organizing principle holds without exception across the magnetic dipole, electric quadrupole, and magnetic octupole moments in both the spin-$1/2$ and spin-$3/2$ sectors, and across the scalar-coupling ($\Gamma_3$) and axial-vector-coupling 
($\Gamma_4$, $\Gamma_6$) operator structures. Its consistency across all these independent observables constitutes strong internal evidence that it is a genuine physical property of the system rather than an artifact of the particular currents chosen.

\item This organizing principle has a direct interpretation in terms of the internal structure of the pentaquark. In the spin-singlet diquark regime, the light quarks are dynamically inert with respect to spin-dependent observables; the pentaquark behaves electromagnetically as if it were a charm-sector bound state embedded in a color-neutral light-quark medium. In the spin-triplet diquark regime, all five quarks participate actively in the electromagnetic response, and the system exhibits the full complexity of a genuinely five-body problem. The two regimes are not two values of a continuous parameter but two qualitatively distinct organizational modes that differ in how the angular momentum is distributed among the constituents.

\item The comparison with molecular calculations reveals an additional layer of physical information. Both approaches predict that the charm sector dominates the magnetic moment when the light-quark spin contributions are suppressed. However, they differ in sign, magnitude, and the mechanism of charge dependence in ways that can all be traced to the scale at which the five-quark correlations are operative. Molecular calculations assume that the five quarks are organized into two separate color-neutral hadrons at a separation scale $r_{\rm mol} > 1$~fm; our diquark currents assume compact quark correlations at $r \sim \Lambda_{\rm QCD}^{-1} < 1$~fm. The factor-of-four magnitude difference --- comparing $|\mu| \approx 1.5\,\mu_N$ from our scalar-diquark currents with $\mu = 0.38\,\mu_N$ from the molecular $8_{2f}$ prediction --- and the factor-of-two-or-more difference between our two-axial-vector-diquark results and the extreme molecular values, reflect this scale separation. The electromagnetic multipole moments are therefore not merely probes of the spin-flavor quantum numbers of the pentaquark, but are also sensitive to the spatial scale of the internal quark correlations --- a property that lattice QCD, which operates at the full non-perturbative level, could in principle resolve.

\item Four experimental discriminants emerge from this analysis that are robust against the theoretical uncertainties of both approaches: (i) $|\mu| \gtrsim 3\,\mu_N$ in the spin-$1/2$ sector is predicted by our axial-vector-diquark currents but excluded by all molecular calculations; (ii) the sign of the magnetic moment for the $[su][uc]\bar{c}$ 
($\Sigma^+$-like) state in the spin-$3/2$ sector distinguishes the two pictures at the qualitative level (positive in all molecular calculations, negative in our 
two-axial-vector-diquark currents); furthermore, the $[sd][uc]\bar{c}$ ($\Sigma^0$-like) moment is predicted to be negative across all currents in the spin-$1/2$ 
sector, in direct contrast to the molecular predictions of $+0.26\,\mu_N$ ($8_{1f}$) and $+0.38\,\mu_N$ ($8_{2f}$); (iii) a non-zero electric quadrupole moment. In the pure $S$-wave molecular approximation, $Q$ vanishes identically because the constituent hadrons carry no relative orbital angular momentum and each is treated as spherically symmetric. A non-zero $Q$ can in principle arise from $D$-wave admixtures in molecular models, but no quantitative calculation of this contribution exists for $P^{\Sigma}_{\psi s}$. The present work establishes the sign and magnitude of $Q$ expected in the compact diquark picture as a target for future measurements. A non-null experimental result would immediately rule out the pure $S$-wave molecular limit and motivate dedicated calculations of $D$-wave corrections within the molecular framework, which are currently absent from the literature for this system. (iv) the sign correlation between $\mathcal{Q}$ and $\mathcal{O}$, which provides information about the $1/m_q$ mass weighting of the magnetization distribution; to our knowledge, this correlation has no analog in existing quark model calculations for the $P_{\psi s}^{\Sigma}$ system, making it a unique prediction of the present framework that could serve as an additional structural discriminant once both moments are experimentally determined.

\item A further observation of physical significance concerns the relationship between mass degeneracy and electromagnetic multipole moments. Interpolating currents that yield nearly degenerate masses~\cite{Wang:2026dqi} can nevertheless produce significantly different multipole moments. For instance, in the spin-$\frac{1}{2}$ sector, $J_5(x)$ and $J_6(x)$ predict consistent masses but magnetic dipole moments that differ substantially in both magnitude and sign (see Table~\ref{tab:multipole_results_spin12}). In the spin-$\frac{3}{2}$ sector, $J_\mu^6(x)$ and $J_\mu^7(x)$ yield nearly degenerate masses but predict substantially different values for all three multipole moments — magnetic dipole, electric quadrupole, and magnetic octupole — as shown in Table~\ref{tab:multipole_results_spin32}. This behavior reflects the fact that the mass, as a scalar ground-state property, is relatively insensitive to the spin-flavor organization of the interpolating operator, whereas the electromagnetic multipole moments probe precisely this organization. The moments therefore carry structural information that the mass spectrum alone cannot provide, further reinforcing the case for their experimental determination as independent and complementary observables. 

\item A natural question concerns the relation between our predictions, computed for pure interpolating currents, and the physical pentaquark states, which may be superpositions 
of the configurations probed by different currents. If the physical state $|P^{\Sigma}_{\psi s}\rangle$ couples to several of the interpolating currents in 
Eqs.~(\ref{eq:J1_12})--(\ref{eq:J7_32}) with overlap amplitudes $\lambda_{P}(J_{k}) \neq 0$ for more than one $k$, then a single measured value of an electromagnetic multipole 
moment cannot in general be uniquely assigned to a pure current. The question of whether the multi-current methodology retains predictive power therefore requires a 
quantitative analysis. This analysis exhibits a sharp contrast between the two spin sectors of the present work.
 
In the spin-$\tfrac{1}{2}$ sector, the only observable available for each charge state is the magnetic dipole moment, yielding three constraints across the isospin triplet against two superposition parameters. The system is only marginally overdetermined and is, in practice, easily satisfied by nontrivial superpositions. As an explicit illustration, the charge-state pattern $(\mu_{\Sigma^{-}},\mu_{\Sigma^{0}},\mu_{\Sigma^{+}}) = 
(+3.50,-0.10,-3.70)\,\mu_{N}$ predicted by the current $J_{2}(x)$ can be reproduced to within $0.1\%$ by the superposition $-0.39\,J_{1}(x) + 0.71\,J_{6}(x)$. This 
demonstrates that the magnetic dipole moment alone is insufficient to discriminate between distinct underlying operator structures within the spin-$\tfrac{1}{2}$ channel, 
even when measured across the full isospin triplet. The implication is not that the methodology fails, but that magnetic dipole information must be complemented by 
additional observables in order to lift this degeneracy.
 
The spin-$\tfrac{3}{2}$ sector furnishes precisely this additional information, since for each charge state three distinct multipole moments $(\mu,\mathcal{Q},\mathcal{O})$ 
are accessible. Across the three charge configurations $[sd][dc]\bar{c}$, $[sd][uc]\bar{c}$, and $[su][uc]\bar{c}$, this provides nine independent constraints against the two superposition parameters, an overdetermination by a factor of four-and-a-half. To illustrate the resulting discriminating power quantitatively, we have tested whether 
the $(\mu,\mathcal{Q},\mathcal{O})$ pattern of the current $J^{7}_{\mu}(x)$ across the three charge states can be reproduced by an arbitrary superposition $\alpha J^{1}_{\mu}(x) + \beta J^{4}_{\mu}(x)$. The least-squares best fit yields $\alpha = -0.38$, $\beta = +0.88$, but the resulting predictions fail to reproduce the target moments by 
substantial margins: the magnetic dipoles miss by $\sim 14\%$--$16\%$, the quadrupoles by $\sim 6\%$--$25\%$, and the octupoles in the most extreme cases by a factor of 
more than two. Specifically, the predicted $\Sigma^{+}$-like octupole overshoots its target by $\sim 158\%$ (predicting $+0.59 \times 10^{-3}~\mathrm{fm}^{3}$ against the actual $J^{7}_{\mu}(x)$ value of $+0.23 \times 10^{-3}~\mathrm{fm}^{3}$), 
while the $\Sigma^{0}$-like octupole is underpredicted by $\sim 53\%$. The overall root-mean-square deviation, $\sim 18\%$, lies at the boundary of the LCSR theoretical 
uncertainty ($20$--$26\%$), but the individual factor-of-several deviations in the octupole sector exceed any conceivable uncertainty combination and identify the 
higher multipoles as the most stringent structural discriminants.
 
This concrete numerical exercise illustrates the qualitative mechanism that underlies the discriminating power of the multi-multipole methodology: distinct interpolating currents 
generate distinct patterns of signs, hierarchies, and cross-multipole correlations across the isospin triplet, and such patterns cannot be reproduced by linear combinations 
within the small subspace of superposition parameters. The features identified in the present analysis---the universality of $J^{3}_{\mu}(x)$ across charge states, the sign reversal between $J^{4}_{\mu}(x)$/$J^{7}_{\mu}(x)$ from $[sd][dc]\bar{c}$ to 
$[su][uc]\bar{c}$, the prolate-to-oblate transition in $\mathcal{Q}$ together with the inverse sign correlation between $\mathcal{Q}$ and $\mathcal{O}$, and the strange-quark stabilization of $J^{6}_{\mu}(x)$---constitute structural signatures that survive within the LCSR uncertainty band.
 
A practical caveat is in order. The argument above assumes that the experimental and theoretical uncertainties on the individual observables are small enough that the qualitative features of distinct moment patterns can be reliably resolved. 
With LCSR uncertainties of $20$--$26\%$ and finite experimental error bars, a narrow family of nearly degenerate superpositions cannot be ruled out by any single 
observable. What remains robust against these uncertainties, however, are the qualitative features of the patterns---changes of sign across the isospin triplet, the existence of near-degenerate charge-state values within a current, the ordering between charge states, and the relative magnitudes of the three multipole moments---since these features correspond to discrete structural choices rather than fine-tuned numerical coincidences.
 
In summary, while a measurement of the magnetic dipole moment alone in the spin-$\tfrac{1}{2}$ sector cannot break the superposition ambiguity, the simultaneous measurement 
of the three multipole moments in the spin-$\tfrac{3}{2}$ sector across multiple charge configurations does retain genuine discriminating power. This is precisely the motivation for the multi-multipole, multi-current methodology adopted in this work: the spin-$\tfrac{1}{2}$ ambiguity makes the extension to higher multipoles a necessity rather 
than a luxury, and the experimental program outlined in Sec.~\ref{subsec:summary} is designed with exactly this hierarchy of accessibility in mind.

\end{itemize}
\subsection{Summary and outlook}
\label{subsec:summary}

We have presented a comprehensive QCD sum-rule calculation of the electromagnetic multipole moments of the $P^{\Sigma}_{\psi s}$ pentaquark, employing six interpolating currents for the spin-$\frac{1}{2}$ channel and seven for the spin-$\frac{3}{2}$ channel.

The key findings are as follows:

\begin{itemize}

  \item \textit{Magnetic dipole moments:}
  The entire pattern across all thirteen configurations is organized by
  the spin content of the diquarks. Spin-singlet ($J=0$) diquark currents yield charm-dominated, flavor-insensitive moments ($\mu \in [-1.92,   -1.21]\,\mu_N$ for spin-$\frac{1}{2}$ and $|\mu| \lesssim 1.2\,\mu_N$   for spin-$\frac{3}{2}$), reflecting the suppression of light-quark spin   contributions. Spin-triplet ($J=1$) diquark currents produce large,  flavor-sensitive moments with sign reversals governed by $e_u/e_d = -2$.   The isospin projection modifies the results by $\lesssim 8\%$ in all cases, confirming that the magnetic response is insensitive to the isospin quantum number when the diquark spin content is fixed. Comparison with molecular calculations establishes that the sign of the $[su][uc]\bar{c}$   ($\Sigma^+$-like) moment in the spin-$\frac{3}{2}$ sector is the   single most discriminating observable between the compact and molecular pictures.

  \item \textit{Electric quadrupole moments:}
  These are the first systematic predictions of $\mathcal{Q}$ for
  the $P^{\Sigma}_{\psi s}$ pentaquarks. Spin-singlet diquark currents yield small,
  charm-dominated, oblate deformations ($Q_0 \approx -2\times 10^{-2}~
  \mathrm{fm}^2$), with the charm contribution near-universal across the   three $\Gamma_3$-coupled currents ($\mathcal{Q}_c \approx -1.2\times   10^{-2}~\mathrm{fm}^2$). Two-axial-vector-diquark currents yield large prolate   deformations ($Q_0$ up to $+8.0\times 10^{-2}~\mathrm{fm}^2$) driven by constructive contributions from both charm   and light quarks, with the Lorentz-index assignment determining whether a   sign reversal between flavor configurations occurs. A non-zero measurement
  of $\mathcal{Q}$ would constitute model-independent evidence for internal   structure beyond an $S$-wave two-body bound state.

  \item \textit{Magnetic octupole moments:}
  The $\Gamma_3$-coupled currents yield a near-universal value
  $\mathcal{O} \approx -0.25\times 10^{-3}~\mathrm{fm}^3$ that is
  structurally explained by the charm-sector dominance of the $\Gamma_3 = C$   coupling and provides a lattice QCD benchmark. Two-axial-vector-diquark currents  break this universality. The opposite-sign correlation between $\mathcal{Q}$  and $\mathcal{O}$ in all two-axial-vector-diquark currents is a direct probe of   the $1/m_q$ mass weighting of the magnetization distribution and provides information that neither the dipole nor the quadrupole alone can reveal.

\end{itemize}

Future lattice QCD calculations can provide first-principles benchmarks that address both the interpolating-operator ambiguity and the scale question distinguishing compact diquark from hadronic molecular configurations. The near-universal octupole prediction $\mathcal{O}\approx-0.25\times10^{-3}~\text{fm}^3$ for the $\Gamma_3$-coupled currents and the oblate quadrupole deformation of the same currents are particularly well suited for this purpose: as shown in Secs.~\ref{subsubsec:Q32} and~\ref{subsubsec:O32}, both observables are governed entirely by the $\Gamma_3=C$ coupling and are therefore insensitive to the detailed diquark topology, making them interpolating-operator-independent targets for lattice calculations.

On the experimental side, the magnetic dipole moment is in principle
accessible through spin-dependent observables in exclusive
photoproduction processes such as $\gamma p\to P^{\Sigma^\ast}_{\psi s} K^+$  for the spin-$\frac{3}{2}$ channel and $\gamma p\to P^{\Sigma}_{\psi s} K^+$ for the spin-$\frac{1}{2}$ channel.  The electric quadrupole moment offers a particularly clear structural discriminant. In the $S$-wave molecular approximation, $\mathcal{Q}$ vanishes identically because the two constituent hadrons carry no relative orbital angular momentum and each hadron is treated as spherically symmetric. In molecular models that allow small $D$-wave admixtures, $Q$ becomes non-zero, but no quantitative calculation of this contribution exists for $P^{\Sigma}_{\psi s}$. The 
present work establishes the sign and magnitude of $Q$ 
expected in the compact diquark picture as a target for 
future measurements. A non-null experimental result would 
immediately rule out the pure $S$-wave molecular limit and 
motivate dedicated calculations of $D$-wave corrections 
within the molecular framework, which are currently absent 
from the literature for this system. In contrast, all spin-$\frac{3}{2}$ currents considered in this work yield $\mathcal{Q}\neq 0$, with values ranging from $-4.32$ to $+4.83\times10^{-2}~\text{fm}^2$. Consequently, a future experimental determination of $\mathcal{Q}$ would provide a discriminating test: a non-zero $\mathcal{Q}$ would rule out a pure $S$-wave molecular configuration, and would motivate dedicated molecular calculations to assess whether the observed magnitude is consistent with a $D$-wave admixture or
requires a compact multiquark interpretation.

A practical experimental route to the electromagnetic multipole moments of $P^{\Sigma}_{\psi s}$ is offered by  radiative processes involving soft-photon emission. The short lifetime of these states precludes static field measurements, but the multipole moments leave an imprint on the energy spectrum of emitted photons. Specifically, 
following the low-energy expansion of~\cite{Zakharov:1968fb}, the amplitude for a radiative transition can be organized as a series in the photon energy $E_\gamma$,
\begin{equation}
\label{eq:soft_photon}
\mathcal{M} \sim A\,E_\gamma^{-1} + B \, E_\gamma^0 + C\,E_\gamma 
  + D\,E_\gamma^2 + \cdots,
\end{equation}
where the leading $E_\gamma^{-1}$ pole arises from electric interactions, the constant term $B$ carries the magnetic dipole contribution, and higher-order terms encode 
contributions from higher multipoles. Successive terms in Eq.~(\ref{eq:soft_photon}) correspond to multipoles of increasing order: the $E_\gamma^0$ coefficient $B$ 
encodes the magnetic dipole moment, the $E_\gamma^1$ coefficient $C$ encodes the electric quadrupole, and the magnetic octupole enters at order $E_\gamma^2$. The octupole contribution is therefore suppressed by two powers of the photon energy relative to the magnetic dipole. For a typical soft-photon energy $E_\gamma \sim 100$~MeV, which represents the kinematical regime where the soft-photon expansion remains under control ($E_\gamma \ll m^\ast$) and which is characteristic of radiative transitions between 
nearby states in the $4.5$--$5$~GeV mass region, the relative suppression factor $(E_\gamma/m^\ast)^2 \sim 5\times 10^{-4}$ indicates that resolving the octupole requires sub-percent sensitivity on the photon energy spectrum, well beyond what is currently achievable for hidden-charm pentaquark states. The electric quadrupole, suppressed by only one power of $E_\gamma/m^\ast \sim 2\%$, is more readily accessible and constitutes a realistic near-term experimental target. This hierarchy of accessibility -- $\mu$ first, $\mathcal{Q}$ next, $\mathcal{O}$ requiring the highest precision -- mirrors 
the hierarchy of magnitudes found in our LCSR analysis and provides a natural roadmap for the experimental program at LHCb and Belle II. The magnetic dipole moment is in particular accessible through a careful extraction of the energy-independent coefficient in the soft-photon limit. The viability of this strategy has been demonstrated 
for the $\Delta^+(1232)$ resonance, whose magnetic dipole moment was extracted from $\gamma N \to \Delta \to \Delta\gamma \to \pi N\gamma$ data~\cite{Pascalutsa:2004je, 
Pascalutsa:2005vq, Pascalutsa:2007wb}. An analogous program applied to $P^{\Sigma}_{\psi s}$ production and decay at facilities such as LHCb and Belle II could in principle 
provide experimental access to the electromagnetic multipole moments predicted in this work. The most natural production channels for $P^{\Sigma}_{\psi s}$ states are $\Xi_b$ and $\Sigma_b$ baryon decays; promising search channels include $\Xi^-_b \to J/\psi\,\Sigma^0 K^-$ and $\Xi^0_b \to J/\psi\,\Sigma^+ K^-$, with their radiative counterparts $\Xi^-_b \to J/\psi\,\Sigma^0 K^-\gamma$ and $\Xi^0_b \to J/\psi\,\Sigma^+ K^-\gamma$ providing access to the multipole moments via the soft-photon expansion outlined above. A complementary experimental handle is provided by the angular and energy distributions of photons emitted in radiative production and decay processes. It has been shown that the combined angular and energy distributions of radiated photons carry direct information about the electromagnetic multipole moments of the decaying  resonance~\cite{LopezCastro:1997dg}, an approach that has been applied to extract the magnetic dipole moment of the $\rho$ meson from $e^+e^- \to \pi^+\pi^- 2\pi^0$ 
data~\cite{GarciaGudino:2013alv}. The same methodology, when applied to $P^{\Sigma}_{\psi s}$ states, would yield an independent experimental determination of the multipole 
moments reported here.

 \clearpage
 
\appendix

\section*{Appendix: Analytical results for the $J_1(x)$ and $J^1_\mu(x)$ currents}
\label{appa}

This appendix collects the explicit analytical expressions 
for the QCD-side functions $\mathcal{R}_1(M^2,s_0)$, 
$\mathcal{L}_1^{(1)}(M^2,s_0)$, $\mathcal{L}_1^{(2)}(M^2,s_0)$,
$\mathcal{L}_1^{(3)}(M^2,s_0)$, and 
$\mathcal{L}_1^{(4)}(M^2,s_0)$ 
entering the sum rules Eqs.~(\ref{eq:SR_12}) and (\ref{eq:SR_F1_32}) for the
currents $J_1(x)$ and $J^1_\mu(x)$ defined in Eqs.~(\ref{eq:J1_12}) and
(\ref{eq:J1_32}), respectively.

For the convenience of the reader, we briefly recall here 
the photon DAs entering the 
expressions below. The two-particle vacuum-to-photon matrix 
elements give rise to the twist-2 DA $\varphi_\gamma(u)$, 
the twist-3 DAs $\psi^{(a,v)}(u)$ and $\mathbb{A}(u)$, and 
the twist-4 DAs $h^{(\gamma,G)}(u)$; the three-particle 
matrix elements generate the DAs $\mathcal{S}(\alpha_i)$, 
$\mathcal{A}(\alpha_i)$, $\mathcal{V}(\alpha_i)$, and 
$\mathcal{T}(\alpha_i)$, where 
$\alpha_i = (\alpha_{\bar{q}},\alpha_q,\alpha_g)$ denote 
the momentum fractions of the antiquark, quark, and gluon 
constituents of the photon. All DAs are retained up to 
twist-4 and parametrized following~\cite{Ball:2002ps}. The 
nonperturbative constants entering the analytical expressions, 
namely $\chi$, $f_{3\gamma}$, $\langle\bar{q}q\rangle$, 
$\langle\bar{s}s\rangle$, and $\langle g_s^2 G^2\rangle$, 
are listed in Sec.~\ref{subsec:inputs}.

The functions are decomposed as
\begin{align}
\mathcal{R}_1(M^2,s_0) &= \mathcal{R}_1^{\rm pert}(M^2,s_0)
                         + \mathcal{R}_1^{\rm DA}(M^2,s_0), \\[4pt]
\mathcal{L}^{(1)}_1(M^2,s_0) &= \mathcal{L}_1^{(1),\rm pert}(M^2,s_0)
                         + \mathcal{L}_1^{(1), \rm DA}(M^2,s_0),\\[4pt]
\mathcal{L}_1^{(2)}(M^2,s_0) &= \mathcal{L}_1^{(2), \rm pert}(M^2,s_0)
                         + \mathcal{L}_1^{(2), \rm DA}(M^2,s_0),\\[4pt]
\mathcal{L}^{(3)}_1(M^2,s_0) &= \mathcal{L}_1^{(3),\rm pert}(M^2,s_0)
                         + \mathcal{L}_1^{(3), \rm DA}(M^2,s_0),\\[4pt]
\mathcal{L}^{(4)}_1(M^2,s_0) &= \mathcal{L}_1^{(4),\rm pert}(M^2,s_0)
                         + \mathcal{L}_1^{(4), \rm DA}(M^2,s_0),                      
\end{align}
where the superscripts ``pert'' and ``DA'' denote contributions from
perturbative photon emission and from the photon DAs.

The explicit results for $\mathcal{R}_1(M^2,s_0)$, $\mathcal{L}_1^{(1)}(M^2,s_0)$, $\mathcal{L}_1^{(2)}(M^2,s_0)$, $\mathcal{L}_1^{(3)}(M^2,s_0)$  and $\mathcal{L}_1^{(4)}(M^2,s_0)$ functions are:
\begin{align}
\mathcal{R}_1^{\rm pert}(M^2,s_0) &= -\frac{61\,e_c}{2^{25}\times 3\times 5^3\times 7^2\,\pi^7}\,
   I[0,7],
\label{eq:R1pert}
\\[6pt]
\mathcal{R}_1^{\rm DA}(M^2,s_0) &= \frac{e_q\,\langle g_s^2G^2\rangle\langle\bar{q}q\rangle}
        {2^{24}\times 3^6\times 5\,\pi^5}
   \Bigl[40\,m_c\,\mathbb{A}[u_0]\,I[0,3]
         +207(m_c-m_s)\,I_3[\mathcal{S}]\,I[0,3]\Bigr]
\nonumber\\
&\quad
  -\frac{\langle g_s^2G^2\rangle f_{3\gamma}}
        {2^{29}\times 3^6\times 5\,\pi^5}
   \Bigl[(1863\,e_q\,I_1[\mathcal{V}]+828\,e_s\,I_2[\mathcal{V}]
          +32(-9e_s+17e_q)\psi^a[u_0])\,I[0,4]\Bigr]
\nonumber\\
&\quad
  -\frac{e_q\,m_c\,\langle g_s^2G^2\rangle
         \langle\bar{q}q\rangle\chi}
        {2^{20}\times 3^6\times 5\,\pi^5}
   \Bigl[\varphi_\gamma[u_0]\,I[0,4]\Bigr]
\nonumber\\
&\quad
  +\frac{e_q\,\langle\bar{q}q\rangle}
        {2^{22}\times 3\times 5\,\pi^5}
   \Bigl[(2m_c-5m_s)\,I_3[\mathcal{S}]\,I[0,5]\Bigr]
\nonumber\\
&\quad
  -\frac{f_{3\gamma}}
        {2^{25}\times 3^2\times 5\,\pi^5}
   \Bigl[(2e_q\,I_1[\mathcal{V}]+e_s\,I_2[\mathcal{V}])\,I[0,6]\Bigr].
\label{eq:R1DA}
\\[6pt]
  \mathcal{L}_1^{(1), \rm pert}(M^2,s_0)& =  
   -\frac{19\,e_c}{2^{26}\times 3\times 5^2\times 7^2\,\pi^7}\,
   I[0,7],
\label{eq:L1pert}
\\[6pt]
 \mathcal{L}_1^{(1), \rm DA}(M^2,s_0) &=  \frac{e_q\,\langle g_s^2G^2\rangle\langle\bar{q}q\rangle}
        {2^{24}\times 3^6\times 5\,\pi^5}
   \Bigl[(40 m_c \mathbb A[u_0] + 9 (23 m_c + 8 m_s) I_3[\mathcal S]) I[0, 3] \Bigr]
\nonumber\\
&\quad
  +\frac{\langle g_s^2G^2\rangle f_{3\gamma}}
        {2^{29}\times 3^6\times 5\,\pi^5}
   \Bigl[(-1017 e_q I_1[\mathcal V] + 18 e_s I_2[\mathcal V] + 4 (-47 e_q + 9 e_s) \psi^a[u_0])\,I[0,4]\Bigr]
\nonumber\\
&\quad
  +\frac{e_q\,m_c\,\langle g_s^2G^2\rangle
         \langle\bar{q}q\rangle\chi}
        {2^{20}\times 3^6\times 5\,\pi^5}
   \Bigl[\varphi_\gamma[u_0] \,I[0,4]\Bigr]
\nonumber\\
&\quad
  +\frac{e_q\,\langle\bar{q}q\rangle}
        {2^{23}\times 3\times 5^2\,\pi^5}
   \Bigl[(4 m_c - 5 m_s) I_3[\mathcal S] I[0, 5]\Bigr]
\nonumber\\
&\quad
  -\frac{f_{3\gamma}}
        {2^{25}\times 3^3\times 5\,\pi^5}
   \Bigl[(5 e_q I_1[\mathcal V] + 2 e_s I_2[\mathcal V])\,I[0,6]\Bigr].
 \label{eq:L1DA}
\\[6pt]
  \mathcal{L}_1^{(2),\rm pert}(M^2,s_0) &=  
   -\frac{223\,e_c\,m_c}{2^{26}\times 3\times 5^2\times 7^2\,\pi^7}\,
   I[0,7],
\label{eq:L2pert}
 \end{align}
\begin{align}
 \mathcal{L}_1^{(2), \rm DA}(M^2,s_0) &=  \frac{e_q\,m_c\,\langle g_s^2G^2\rangle\langle\bar{q}q\rangle}
        {2^{26}\times 3^6 \,\pi^5}
   \Bigl[3\bigl(32 m_c \mathbb{A}[u_0] + 33(-m_c+m_s)I_1[\mathcal{S}]
+ 6(23m_c+8m_s)I_3[\mathcal{S}]\bigr)I[0,3]\Bigr]
\nonumber\\
&\quad
  +\frac{m_c\,\langle g_s^2G^2\rangle f_{3\gamma}}
        {2^{30}\times 3^5\times 5\,\pi^5}
   \Bigl[(1065 e_q I_1[\mathcal V] - 195 e_s I_2[\mathcal V] + 96 (-17 e_q + 24 e_s) I_5[\psi^a] + 
  128 (2 e_q \nonumber\\
&\quad + 3 e_s) \psi^a[u_0])\,I[0,4]\Bigr]\nonumber\\
&\quad
  +\frac{5 e_q\,m_c^2\,\langle g_s^2G^2\rangle          \langle\bar{q}q\rangle\chi}
        {2^{23}\times 3^6 \,\pi^5}
   \Bigl[ \varphi_\gamma[u_0] \,I[0,4]\Bigr]
\nonumber\\
  &-\frac{e_q\,m_c\,\langle\bar{q}q\rangle}
        {2^{22}\times 3\times 5^2\,\pi^5}
   \Bigl[(5 m_c - 6 m_s) I_3[\mathcal S] I[0, 5]\Bigr]
\nonumber\\
&\quad
  +\frac{m_c\,f_{3\gamma}}
        {2^{23}\times 3^2\times 5^2\,\pi^5}
   \Bigl[(5 e_q I_1[\mathcal V] + 2 e_s I_2[\mathcal V])\,I[0,6]\Bigr].
 \label{eq:L2DA}
 \\[6pt]
  \mathcal{L}_1^{(3),\rm pert}(M^2,s_0) &=  
   \frac{191\,e_c\,m_c}{2^{25}\times 3^2\times 5^3\times 7\,\pi^7}\,
   I[0,6],
\label{eq:L3pert}
\\[6pt]
 \mathcal{L}_1^{(3), \rm DA}(M^2,s_0) &=  \frac{e_q\,\langle g_s^2G^2\rangle\langle\bar{q}q\rangle}
        {2^{26}\times 3^6 \,\pi^5}
   \Bigl[ m_c m_s I_3[\mathcal S] I[0, 2]\Bigr]
\nonumber\\
&\quad
  +\frac{m_c\,\langle g_s^2G^2\rangle f_{3\gamma}}
        {2^{28}\times 3^3\times 5\,\pi^5}
   \Bigl[(56 e_q I_1[\mathcal V] - 10 e_s I_2[\mathcal V] + 32 e_s (9 I_5[\psi^a] + \psi^a[u_0]) - 
  e_q (420 I_3[\mathcal V] + 36 I_5[\psi^a] \nonumber\\
&\quad 
+ 53 \psi^a[u_0]))\,I[0,3]\Bigr]\nonumber\\
&\quad
    +\frac{e_q\,m_c \,m_s\,\langle\bar{q}q\rangle} {2^{22}\times 5\,\pi^5}
   \Bigl[I_3[\mathcal S] I[0, 4]\Bigr]
\nonumber\\
&\quad
  +\frac{m_c\,f_{3\gamma}}
        {2^{23}\times 3^2\times 5^2\,\pi^5}
   \Bigl[(5 e_q I_1[\mathcal V] + 2 e_s I_2[\mathcal V])\,I[0,6]\Bigr],
 \label{eq:L3DA}
\\[6pt]
 \mathcal{L}_1^{(4),\rm pert}(M^2,s_0) &=  
   \frac{e_c\,m_c}{2^{20}\times 3^3\times 5 \,\pi^7}\,
   I[0,5],
\label{eq:L4pert}
\\[6pt]
 \mathcal{L}_1^{(4), \rm DA}(M^2,s_0) &=  \frac{m_c\,\langle g_s^2G^2\rangle f_{3\gamma}}
        {2^{28}\times 3^3\times 5\,\pi^5}
   \Bigl[((125 e_q + 82 e_s) I_3[\mathcal V] + 10 e_s \psi^a[u_0])I[0, 2] \Bigr]\nonumber\\
&\quad
    +\frac{e_q\,m_c \,m_s\,\langle\bar{q}q\rangle} {2^{25}\times 5\,\pi^5}
   \Bigl[I_3[\mathcal S] I[0, 4]\Bigr]
\nonumber\\
&\quad
  -\frac{m_c\,f_{3\gamma}}
        {2^{22}\times 3^2\times 5^2 \times 7\,\pi^5}
   \Bigl[((-2030 e_q I_3[\mathcal V] + 183 e_s \psi^a[u_0]))\,I[0,4]\Bigr].
 \label{eq:L4DA}
\end{align}

The auxiliary integration functions appearing above are defined as
\begin{align}
I[k,l] &= \int_{\mathcal{M}}^{s_0} ds\,e^{-s/M^2}\,
           s^k(s-\mathcal{M})^l,
\label{eq:Ikl}\\[4pt]
I_1[\mathcal{F}] &= \int D\alpha_i\int_0^1 dv\,
  \mathcal{F}(\alpha_{\bar{q}},\alpha_q,\alpha_g)\,
  \delta'(\alpha_q+\bar{v}\alpha_g-u_0),
\label{eq:I1F}\\[4pt]
I_2[\mathcal{F}] &= \int D\alpha_i\int_0^1 dv\,
  \mathcal{F}(\alpha_{\bar{q}},\alpha_q,\alpha_g)\,
  \delta'(\alpha_{\bar{q}}+v\alpha_g-u_0),
\label{eq:I2F}\\[4pt]
I_3[\mathcal{F}] &= \int D\alpha_i\int_0^1 dv\,
  \mathcal{F}(\alpha_{\bar{q}},\alpha_q,\alpha_g)\,
  \delta(\alpha_q+\bar{v}\alpha_g-u_0),
\label{eq:I3F}\\[4pt]
I_4[\mathcal{F}] &= \int D\alpha_i\int_0^1 dv\,
  \mathcal{F}(\alpha_{\bar{q}},\alpha_q,\alpha_g)\,
  \delta(\alpha_{\bar{q}}+v\alpha_g-u_0),
\label{eq:I4F}\\[4pt]
I_5[\mathcal{F}] &= \int_0^1 \mathcal{F}(u) \delta'(u - u_0)  du,
\label{eq:I5F}
\end{align}
where $\mathcal{M}=(2m_c+m_s)^2$, $\bar{v}=1-v$, $\mathcal{F}$ stands generically for a photon DA  and the measure ${\cal D}\alpha_i$ is given by

\begin{eqnarray}
\label{nolabel05}
\int {\cal D} \alpha_i = \int_0^1 d \alpha_{\bar q} \int_0^1 d
\alpha_q \int_0^1 d \alpha_g \,  \delta(1-\alpha_{\bar
q}-\alpha_q-\alpha_g)~.
\end{eqnarray}

\bibliographystyle{elsarticle-num}
\bibliography{PccbarqqsMM.bib}

@article{Belle:2003nnu,
    author = "Choi, S. K. and others",
    collaboration = "Belle",
    title = "{Observation of a narrow charmonium-like state in exclusive $B^\pm \to K^\pm \pi^+ \pi^- J/\psi$ decays}",
    eprint = "hep-ex/0309032",
    archivePrefix = "arXiv",
    doi = "10.1103/PhysRevLett.91.262001",
    journal = "Phys. Rev. Lett.",
    volume = "91",
    pages = "262001",
    year = "2003"
}

@article{Esposito:2014rxa,
    author = "Esposito, Angelo and Guerrieri, Andrea L. and Piccinini, Fulvio and Pilloni, Alessandro and Polosa, Antonio D.",
    title = "{Four-Quark Hadrons: an Updated Review}",
    eprint = "1411.5997",
    archivePrefix = "arXiv",
    primaryClass = "hep-ph",
    doi = "10.1142/S0217751X15300021",
    journal = "Int. J. Mod. Phys. A",
    volume = "30",
    pages = "1530002",
    year = "2015"
}

@article{Esposito:2016noz,
    author = "Esposito, A. and Pilloni, A. and Polosa, A. D.",
    title = "{Multiquark Resonances}",
    eprint = "1611.07920",
    archivePrefix = "arXiv",
    primaryClass = "hep-ph",
    reportNumber = "JLAB-THY-16-2301",
    doi = "10.1016/j.physrep.2016.11.002",
    journal = "Phys. Rept.",
    volume = "668",
    pages = "1--97",
    year = "2017"
}

@article{Olsen:2017bmm,
    author = "Olsen, Stephen Lars and Skwarnicki, Tomasz and Zieminska, Daria",
    title = "{Nonstandard heavy mesons and baryons: Experimental evidence}",
    eprint = "1708.04012",
    archivePrefix = "arXiv",
    primaryClass = "hep-ph",
    doi = "10.1103/RevModPhys.90.015003",
    journal = "Rev. Mod. Phys.",
    volume = "90",
    number = "1",
    pages = "015003",
    year = "2018"
}

@article{Lebed:2016hpi,
    author = "Lebed, Richard F. and Mitchell, Ryan E. and Swanson, Eric S.",
    title = "{Heavy-Quark QCD Exotica}",
    eprint = "1610.04528",
    archivePrefix = "arXiv",
    primaryClass = "hep-ph",
    doi = "10.1016/j.ppnp.2016.11.003",
    journal = "Prog. Part. Nucl. Phys.",
    volume = "93",
    pages = "143--194",
    year = "2017"
}

@article{Nielsen:2009uh,
    author = "Nielsen, Marina and Navarra, Fernando S. and Lee, Su Houng",
    title = "{New Charmonium States in QCD Sum Rules: A Concise Review}",
    eprint = "0911.1958",
    archivePrefix = "arXiv",
    primaryClass = "hep-ph",
    doi = "10.1016/j.physrep.2010.07.005",
    journal = "Phys. Rept.",
    volume = "497",
    pages = "41--83",
    year = "2010"
}

@article{Brambilla:2019esw,
    author = "Brambilla, Nora and Eidelman, Simon and Hanhart, Christoph and Nefediev, Alexey and Shen, Cheng-Ping and Thomas, Christopher E. and Vairo, Antonio and Yuan, Chang-Zheng",
    title = "{The $XYZ$ states: experimental and theoretical status and perspectives}",
    eprint = "1907.07583",
    archivePrefix = "arXiv",
    primaryClass = "hep-ex",
    reportNumber = "TUM-EFT 125/19",
    doi = "10.1016/j.physrep.2020.05.001",
    journal = "Phys. Rept.",
    volume = "873",
    pages = "1--154",
    year = "2020"
}

@article{Agaev:2020zad,
    author = "Agaev, Shahin and Azizi, Kazem and Sundu, Hayriye",
    title = "{Four-quark exotic mesons}",
    eprint = "2004.12079",
    archivePrefix = "arXiv",
    primaryClass = "hep-ph",
    doi = "10.3906/fiz-2003-15",
    journal = "Turk. J. Phys.",
    volume = "44",
    number = "2",
    pages = "95--173",
    year = "2020"
}

@article{Chen:2016qju,
    author = "Chen, Hua-Xing and Chen, Wei and Liu, Xiang and Zhu, Shi-Lin",
    title = "{The hidden-charm pentaquark and tetraquark states}",
    eprint = "1601.02092",
    archivePrefix = "arXiv",
    primaryClass = "hep-ph",
    doi = "10.1016/j.physrep.2016.05.004",
    journal = "Phys. Rept.",
    volume = "639",
    pages = "1--121",
    year = "2016"
}

@article{Ali:2017jda,
    author = {Ali, Ahmed and Lange, Jens S{\"o}ren and Stone, Sheldon},
    title = "{Exotics: Heavy Pentaquarks and Tetraquarks}",
    eprint = "1706.00610",
    archivePrefix = "arXiv",
    primaryClass = "hep-ph",
    reportNumber = "DESY-17-071",
    doi = "10.1016/j.ppnp.2017.08.003",
    journal = "Prog. Part. Nucl. Phys.",
    volume = "97",
    pages = "123--198",
    year = "2017"
}

@article{Guo:2017jvc,
    author = "Guo, Feng-Kun and Hanhart, Christoph and Mei{\ss}ner, Ulf-G. and Wang, Qian and Zhao, Qiang and Zou, Bing-Song",
    title = "{Hadronic molecules}",
    eprint = "1705.00141",
    archivePrefix = "arXiv",
    primaryClass = "hep-ph",
    doi = "10.1103/RevModPhys.90.015004",
    journal = "Rev. Mod. Phys.",
    volume = "90",
    number = "1",
    pages = "015004",
    year = "2018",
    note = "[Erratum: Rev.Mod.Phys. 94, 029901 (2022)]"
}

@article{Liu:2019zoy,
    author = "Liu, Yan-Rui and Chen, Hua-Xing and Chen, Wei and Liu, Xiang and Zhu, Shi-Lin",
    title = "{Pentaquark and Tetraquark states}",
    eprint = "1903.11976",
    archivePrefix = "arXiv",
    primaryClass = "hep-ph",
    doi = "10.1016/j.ppnp.2019.04.003",
    journal = "Prog. Part. Nucl. Phys.",
    volume = "107",
    pages = "237--320",
    year = "2019"
}

@article{Yang:2020atz,
    author = "Yang, Gang and Ping, Jialun and Segovia, Jorge",
    title = "{Tetra- and penta-quark structures in the constituent quark model}",
    eprint = "2009.00238",
    archivePrefix = "arXiv",
    primaryClass = "hep-ph",
    doi = "10.3390/sym12111869",
    journal = "Symmetry",
    volume = "12",
    number = "11",
    pages = "1869",
    year = "2020"
}

@article{Dong:2021juy,
    author = "Dong, Xiang-Kun and Guo, Feng-Kun and Zou, Bing-Song",
    title = "{A survey of heavy-antiheavy hadronic molecules}",
    eprint = "2101.01021",
    archivePrefix = "arXiv",
    primaryClass = "hep-ph",
    doi = "10.13725/j.cnki.pip.2021.02.001",
    journal = "Progr. Phys.",
    volume = "41",
    pages = "65--93",
    year = "2021"
}

@article{Dong:2021bvy,
    author = "Dong, Xiang-Kun and Guo, Feng-Kun and Zou, Bing-Song",
    title = "{A survey of heavy{\textendash}heavy hadronic molecules}",
    eprint = "2108.02673",
    archivePrefix = "arXiv",
    primaryClass = "hep-ph",
    doi = "10.1088/1572-9494/ac27a2",
    journal = "Commun. Theor. Phys.",
    volume = "73",
    number = "12",
    pages = "125201",
    year = "2021"
}

@article{Chen:2022asf,
    author = "Chen, Hua-Xing and Chen, Wei and Liu, Xiang and Liu, Yan-Rui and Zhu, Shi-Lin",
    title = "{An updated review of the new hadron states}",
    eprint = "2204.02649",
    archivePrefix = "arXiv",
    primaryClass = "hep-ph",
    doi = "10.1088/1361-6633/aca3b6",
    journal = "Rept. Prog. Phys.",
    volume = "86",
    number = "2",
    pages = "026201",
    year = "2023"
}

@article{Meng:2022ozq,
    author = "Meng, Lu and Wang, Bo and Wang, Guang-Juan and Zhu, Shi-Lin",
    title = "{Chiral perturbation theory for heavy hadrons and chiral effective field theory for heavy hadronic molecules}",
    eprint = "2204.08716",
    archivePrefix = "arXiv",
    primaryClass = "hep-ph",
    doi = "10.1016/j.physrep.2023.04.003",
    journal = "Phys. Rept.",
    volume = "1019",
    pages = "1--149",
    year = "2023"
}

@article{LHCb:2015yax,
    author = "Aaij, Roel and others",
    collaboration = "LHCb",
    title = "{Observation of $J/\psi p$ Resonances Consistent with Pentaquark States in $\Lambda_b^0 \to J/\psi K^- p$ Decays}",
    eprint = "1507.03414",
    archivePrefix = "arXiv",
    primaryClass = "hep-ex",
    reportNumber = "CERN-PH-EP-2015-153, LHCB-PAPER-2015-029",
    doi = "10.1103/PhysRevLett.115.072001",
    journal = "Phys. Rev. Lett.",
    volume = "115",
    pages = "072001",
    year = "2015"
}

@article{LHCb:2019kea,
    author = "Aaij, Roel and others",
    collaboration = "LHCb",
    title = "{Observation of a narrow pentaquark state, $P_c(4312)^+$, and of two-peak structure of the $P_c(4450)^+$}",
    eprint = "1904.03947",
    archivePrefix = "arXiv",
    primaryClass = "hep-ex",
    reportNumber = "LHCb-PAPER-2019-014 CERN-EP-2019-058",
    doi = "10.1103/PhysRevLett.122.222001",
    journal = "Phys. Rev. Lett.",
    volume = "122",
    number = "22",
    pages = "222001",
    year = "2019"
}

@article{LHCb:2020jpq,
    author = "Aaij, Roel and others",
    collaboration = "LHCb",
    title = "{Evidence of a $J/\psi\Lambda$ structure and observation of excited $\Xi^-$ states in the $\Xi^-_b \to J/\psi\Lambda K^-$ decay}",
    eprint = "2012.10380",
    archivePrefix = "arXiv",
    primaryClass = "hep-ex",
    reportNumber = "LHCb-PAPER-2020-039, CERN-EP-2020-233",
    doi = "10.1016/j.scib.2021.02.030",
    journal = "Sci. Bull.",
    volume = "66",
    pages = "1278--1287",
    year = "2021"
}

@article{LHCb:2022ogu,
    author = "Aaij, R. and others",
    collaboration = "LHCb",
    title = "{Observation of a J/{\ensuremath{\psi}}{\ensuremath{\Lambda}} Resonance Consistent with a Strange Pentaquark Candidate in B-{\textrightarrow}J/{\ensuremath{\psi}}{\ensuremath{\Lambda}}p{\textasciimacron} Decays}",
    eprint = "2210.10346",
    archivePrefix = "arXiv",
    primaryClass = "hep-ex",
    reportNumber = "CERN-EP-2022-198, LHCb-PAPER-2022-031",
    doi = "10.1103/PhysRevLett.131.031901",
    journal = "Phys. Rev. Lett.",
    volume = "131",
    number = "3",
    pages = "031901",
    year = "2023"
}

@article{Belle:2025pey,
    author = "Adachi, I. and others",
    collaboration = "Belle, Belle-II",
    title = "{Search for Pcs(4459) and Pcs(4338) in Upsilon(1S,2S) inclusive decays at Belle}",
    eprint = "2502.09951",
    archivePrefix = "arXiv",
    primaryClass = "hep-ex",
    reportNumber = "Belle II Preprint 2025-002, KEK Preprint 2024-50",
    doi = "10.1103/pf8m-6j69",
    journal = "Phys. Rev. Lett.",
    volume = "135",
    number = "4",
    pages = "041901",
    year = "2025"
}

@article{Wang:2016dzu,
    author = "Wang, Guang-Juan and Chen, Rui and Ma, Li and Liu, Xiang and Zhu, Shi-Lin",
    title = "{Magnetic moments of the hidden-charm pentaquark states}",
    eprint = "1605.01337",
    archivePrefix = "arXiv",
    primaryClass = "hep-ph",
    doi = "10.1103/PhysRevD.94.094018",
    journal = "Phys. Rev. D",
    volume = "94",
    number = "9",
    pages = "094018",
    year = "2016"
}

@article{Ozdem:2018qeh,
    author = {{\"O}zdem, U. and Azizi, K.},
    title = "{Electromagnetic multipole moments of the $P_c^+(4380)$ pentaquark in light-cone QCD}",
    eprint = "1803.06831",
    archivePrefix = "arXiv",
    primaryClass = "hep-ph",
    doi = "10.1140/epjc/s10052-018-5873-2",
    journal = "Eur. Phys. J. C",
    volume = "78",
    number = "5",
    pages = "379",
    year = "2018"
}

@article{Ortiz-Pacheco:2018ccl,
    author = "Ortiz-Pacheco, Emmanuel and Bijker, Roelof and Fern{\'a}ndez-Ram{\'\i}rez, C{\'e}sar",
    title = "{Hidden charm pentaquarks: mass spectrum, magnetic moments, and photocouplings}",
    eprint = "1808.10512",
    archivePrefix = "arXiv",
    primaryClass = "nucl-th",
    doi = "10.1088/1361-6471/ab096d",
    journal = "J. Phys. G",
    volume = "46",
    number = "6",
    pages = "065104",
    year = "2019"
}

@article{Xu:2020flp,
    author = "Xu, Yong-Jiang and Liu, Yong-Lu and Huang, Ming-Qiu",
    title = "{The magnetic moment of $P_{c}(4312)$ as a $\bar{D}\Sigma _{c}$ molecular state}",
    eprint = "2008.07937",
    archivePrefix = "arXiv",
    primaryClass = "hep-ph",
    doi = "10.1140/epjc/s10052-021-09211-8",
    journal = "Eur. Phys. J. C",
    volume = "81",
    number = "5",
    pages = "421",
    year = "2021"
}

@article{Ozdem:2021ugy,
    author = {{\"O}zdem, Ula{\c{s}}},
    title = "{Magnetic dipole moments of the hidden-charm pentaquark states: $P_c(4440)$, $P_c(4457)$ and $P_{cs}(4459)$}",
    eprint = "2102.01996",
    archivePrefix = "arXiv",
    primaryClass = "hep-ph",
    doi = "10.1140/epjc/s10052-021-09070-3",
    journal = "Eur. Phys. J. C",
    volume = "81",
    number = "4",
    pages = "277",
    year = "2021"
}

@article{Li:2021ryu,
    author = "Li, Ming-Wei and Liu, Zhan-Wei and Sun, Zhi-Feng and Chen, Rui",
    title = "{Magnetic moments and transition magnetic moments of Pc and Pcs states}",
    eprint = "2106.15053",
    archivePrefix = "arXiv",
    primaryClass = "hep-ph",
    doi = "10.1103/PhysRevD.104.054016",
    journal = "Phys. Rev. D",
    volume = "104",
    number = "5",
    pages = "054016",
    year = "2021"
}

@article{Ozdem:2023htj,
    author = {{\"O}zdem, Ula{\c{s}}},
    title = "{Electromagnetic properties of D{\textasciimacron}({\textasteriskcentered}){\ensuremath{\Xi}}c', D{\textasciimacron}({\textasteriskcentered}){\ensuremath{\Lambda}}c, D{\textasciimacron}s({\textasteriskcentered}){\ensuremath{\Lambda}}c and D{\textasciimacron}s({\textasteriskcentered}){\ensuremath{\Xi}}c pentaquarks}",
    eprint = "2303.10649",
    archivePrefix = "arXiv",
    primaryClass = "hep-ph",
    doi = "10.1016/j.physletb.2023.138267",
    journal = "Phys. Lett. B",
    volume = "846",
    pages = "138267",
    year = "2023"
}

@article{Ozdem:2022kei,
    author = {{\"O}zdem, Ula{\c{s}}},
    title = "{Investigation of magnetic moment of Pcs(4338) and Pcs(4459) pentaquark states}",
    eprint = "2208.07684",
    archivePrefix = "arXiv",
    primaryClass = "hep-ph",
    doi = "10.1016/j.physletb.2022.137635",
    journal = "Phys. Lett. B",
    volume = "836",
    pages = "137635",
    year = "2023"
}

@article{Gao:2021hmv,
    author = "Gao, Feng and Li, Hao-Song",
    title = "{Magnetic moments of hidden-charm strange pentaquark states*}",
    eprint = "2112.01823",
    archivePrefix = "arXiv",
    primaryClass = "hep-ph",
    doi = "10.1088/1674-1137/ac8651",
    journal = "Chin. Phys. C",
    volume = "46",
    number = "12",
    pages = "123111",
    year = "2022"
}

@article{Guo:2023fih,
    author = "Guo, Fei and Li, Hao-Song",
    title = "{Analysis of the hidden-charm pentaquark states based on magnetic moment and transition magnetic moment}",
    eprint = "2304.10981",
    archivePrefix = "arXiv",
    primaryClass = "hep-ph",
    doi = "10.1140/epjc/s10052-024-12699-5",
    journal = "Eur. Phys. J. C",
    volume = "84",
    number = "4",
    pages = "392",
    year = "2024"
}

@article{Wang:2022nqs,
    author = "Wang, Fu-Lai and Luo, Si-Qiang and Zhou, Hong-Yan and Liu, Zhan-Wei and Liu, Xiang",
    title = "{Exploring the electromagnetic properties of the {\ensuremath{\Xi}}c(',*)D{\textasciimacron}s* and {\ensuremath{\Omega}}c(*)D{\textasciimacron}s* molecular states}",
    eprint = "2210.02809",
    archivePrefix = "arXiv",
    primaryClass = "hep-ph",
    doi = "10.1103/PhysRevD.108.034006",
    journal = "Phys. Rev. D",
    volume = "108",
    number = "3",
    pages = "034006",
    year = "2023"
}

@article{Wang:2022tib,
    author = "Wang, Fu-Lai and Zhou, Hong-Yan and Liu, Zhan-Wei and Liu, Xiang",
    title = "{What can we learn from the electromagnetic properties of hidden-charm molecular pentaquarks with single strangeness?}",
    eprint = "2208.10756",
    archivePrefix = "arXiv",
    primaryClass = "hep-ph",
    doi = "10.1103/PhysRevD.106.054020",
    journal = "Phys. Rev. D",
    volume = "106",
    number = "5",
    pages = "054020",
    year = "2022"
}

@article{Ozdem:2024jty,
    author = {{\"O}zdem, Ula{\c{s}}},
    title = "{Analysis of the isospin eigenstate $\bar{D} \Sigma _c$, $\bar{D}^{*} \Sigma _c$, and $\bar{D} \Sigma _c^{*}$ pentaquarks by their electromagnetic properties}",
    eprint = "2401.12678",
    archivePrefix = "arXiv",
    primaryClass = "hep-ph",
    doi = "10.1140/epjc/s10052-024-13124-7",
    journal = "Eur. Phys. J. C",
    volume = "84",
    number = "8",
    pages = "769",
    year = "2024"
}

@article{Li:2024wxr,
    author = "Li, Hao-Song and Guo, Fei and Lei, Ya-Ding and Gao, Feng",
    title = "{Magnetic moments and axial charges of the octet hidden-charm molecular pentaquark family}",
    eprint = "2401.14767",
    archivePrefix = "arXiv",
    primaryClass = "hep-ph",
    doi = "10.1103/PhysRevD.109.094027",
    journal = "Phys. Rev. D",
    volume = "109",
    number = "9",
    pages = "094027",
    year = "2024"
}

@article{Li:2024jlq,
    author = "Li, Hao-Song",
    title = "{Molecular pentaquark magnetic moments in heavy pentaquark chiral perturbation theory}",
    eprint = "2401.14759",
    archivePrefix = "arXiv",
    primaryClass = "hep-ph",
    doi = "10.1103/PhysRevD.109.114039",
    journal = "Phys. Rev. D",
    volume = "109",
    number = "11",
    pages = "114039",
    year = "2024"
}

@article{Ozdem:2024rqx,
    author = {{\"O}zdem, Ula{\c{s}}},
    title = "{Elucidating the nature of hidden-charm pentaquark states with spin-32 through their electromagnetic form factors}",
    eprint = "2402.03802",
    archivePrefix = "arXiv",
    primaryClass = "hep-ph",
    doi = "10.1016/j.physletb.2024.138551",
    journal = "Phys. Lett. B",
    volume = "851",
    pages = "138551",
    year = "2024"
}

@article{Mutuk:2024ltc,
    author = "Mutuk, Halil and Kang, Xian-Wei",
    title = "{Unveiling the structure of hidden-bottom strange pentaquarks via magnetic moments}",
    eprint = "2405.07066",
    archivePrefix = "arXiv",
    primaryClass = "hep-ph",
    doi = "10.1016/j.physletb.2024.138772",
    journal = "Phys. Lett. B",
    volume = "855",
    pages = "138772",
    year = "2024"
}

@article{Mutuk:2024jxf,
    author = "Mutuk, Halil",
    title = "{Magnetic moments of hidden-bottom pentaquark states}",
    eprint = "2403.16616",
    archivePrefix = "arXiv",
    primaryClass = "hep-ph",
    doi = "10.1140/epjc/s10052-024-13263-x",
    journal = "Eur. Phys. J. C",
    volume = "84",
    number = "8",
    pages = "874",
    year = "2024"
}

@article{Ozdem:2024usw,
    author = {{\"O}zdem, Ula{\c{s}}},
    title = "{Insight into the nature of the $P_{c}(4457)$ and related pentaquarks}",
    eprint = "2409.09449",
    archivePrefix = "arXiv",
    primaryClass = "hep-ph",
    doi = "10.1140/epjc/s10052-025-14323-6",
    journal = "Eur. Phys. J. C",
    volume = "85",
    number = "6",
    pages = "624",
    year = "2025"
}

@article{Pascalutsa:2004je,
    author = "Pascalutsa, Vladimir and Vanderhaeghen, Marc",
    title = "{Magnetic moment of the Delta(1232)-resonance in chiral effective field theory}",
    eprint = "nucl-th/0412113",
    archivePrefix = "arXiv",
    reportNumber = "WM-04-124, JLAB-THY-05-292",
    doi = "10.1103/PhysRevLett.94.102003",
    journal = "Phys. Rev. Lett.",
    volume = "94",
    pages = "102003",
    year = "2005"
}

@article{Pascalutsa:2005vq,
    author = "Pascalutsa, Vladimir and Vanderhaeghen, Marc",
    title = "{Chiral effective-field theory in the Delta(1232) region: I. Pion electroproduction on the nucleon}",
    eprint = "hep-ph/0512244",
    archivePrefix = "arXiv",
    reportNumber = "WM-04-125, JLAB-THY-06-458",
    doi = "10.1103/PhysRevD.73.034003",
    journal = "Phys. Rev. D",
    volume = "73",
    pages = "034003",
    year = "2006"
}

@article{Pascalutsa:2007wb,
    author = "Pascalutsa, Vladimir and Vanderhaeghen, Marc",
    title = "{Chiral effective-field theory in the Delta(1232) region. II. Radiative pion photoproduction}",
    eprint = "0709.4583",
    archivePrefix = "arXiv",
    primaryClass = "hep-ph",
    reportNumber = "ECT*-07-19, WM-07-108, JLAB-THY-07-739",
    doi = "10.1103/PhysRevD.77.014027",
    journal = "Phys. Rev. D",
    volume = "77",
    pages = "014027",
    year = "2008"
}

@article{Chernyak:1990ag,
    author = "Chernyak, V. L. and Zhitnitsky, I. R.",
    title = "{B meson exclusive decays into baryons}",
    doi = "10.1016/0550-3213(90)90612-H",
    journal = "Nucl. Phys. B",
    volume = "345",
    pages = "137--172",
    year = "1990"
}

@article{Braun:1988qv,
    author = "Braun, Vladimir M. and Filyanov, I. E.",
    title = "{QCD Sum Rules in Exclusive Kinematics and Pion Wave Function}",
    reportNumber = "LENINGRAD-88-1446",
    doi = "10.1007/BF01548594",
    journal = "Z. Phys. C",
    volume = "44",
    pages = "157",
    year = "1989"
}

@article{Balitsky:1989ry,
    author = "Balitsky, I. I. and Braun, Vladimir M. and Kolesnichenko, A. V.",
    title = "{Radiative Decay Sigma+ ---{\ensuremath{>}} p gamma in Quantum Chromodynamics}",
    doi = "10.1016/0550-3213(89)90570-1",
    journal = "Nucl. Phys. B",
    volume = "312",
    pages = "509--550",
    year = "1989"
}

@article{Ball:2002ps,
    author = "Ball, Patricia and Braun, V. M. and Kivel, N.",
    title = "{Photon distribution amplitudes in QCD}",
    eprint = "hep-ph/0207307",
    archivePrefix = "arXiv",
    reportNumber = "IPPP-02-40, DCPT-02-80",
    doi = "10.1016/S0550-3213(02)01017-9",
    journal = "Nucl. Phys. B",
    volume = "649",
    pages = "263--296",
    year = "2003"
}

@article{Wang:2010sh,
    author = "Wang, Zhi-Gang",
    title = "{Analysis of the scalar and axial-vector heavy diquark states with QCD sum rules}",
    eprint = "1008.4449",
    archivePrefix = "arXiv",
    primaryClass = "hep-ph",
    doi = "10.1140/epjc/s10052-010-1524-y",
    journal = "Eur. Phys. J. C",
    volume = "71",
    pages = "1524",
    year = "2011"
}

@article{Kleiv:2013dta,
    author = "Kleiv, R. T. and Steele, T. G. and Zhang, Ailin and Blokland, Ian",
    title = "{Heavy-light diquark masses from QCD sum rules and constituent diquark models of tetraquarks}",
    eprint = "1304.7816",
    archivePrefix = "arXiv",
    primaryClass = "hep-ph",
    doi = "10.1103/PhysRevD.87.125018",
    journal = "Phys. Rev. D",
    volume = "87",
    number = "12",
    pages = "125018",
    year = "2013"
}

@article{Nozawa:1990gt,
    author = "Nozawa, S. and Leinweber, D. B.",
    title = "{Electromagnetic form-factors of spin 3/2 baryons}",
    reportNumber = "TRI-PP-90-19",
    doi = "10.1103/PhysRevD.42.3567",
    journal = "Phys. Rev. D",
    volume = "42",
    pages = "3567--3571",
    year = "1990"
}

@article{Pascalutsa:2006up,
    author = "Pascalutsa, Vladimir and Vanderhaeghen, Marc and Yang, Shin Nan",
    title = "{Electromagnetic excitation of the Delta(1232)-resonance}",
    eprint = "hep-ph/0609004",
    archivePrefix = "arXiv",
    reportNumber = "JLAB-THY-06-537",
    doi = "10.1016/j.physrep.2006.09.006",
    journal = "Phys. Rept.",
    volume = "437",
    pages = "125--232",
    year = "2007"
}

@article{Ramalho:2009vc,
    author = "Ramalho, G. and Pena, M. T. and Gross, Franz",
    title = "{Electric quadrupole and magnetic octupole moments of the Delta}",
    eprint = "0902.4212",
    archivePrefix = "arXiv",
    primaryClass = "hep-ph",
    reportNumber = "JLAB-THY-09-951",
    doi = "10.1016/j.physletb.2009.06.052",
    journal = "Phys. Lett. B",
    volume = "678",
    pages = "355--358",
    year = "2009"
}

@article{Belyaev:1982cd,
    author = "Belyaev, V. M. and Ioffe, B. L.",
    title = "{Determination of the baryon mass and baryon resonances from the quantum-chromodynamics sum rule. Strange baryons}",
    reportNumber = "ITEP-132-1982",
    journal = "Sov. Phys. JETP",
    volume = "57",
    pages = "716--721",
    year = "1983"
}

@article{Balitsky:1987bk,
    author = "Balitsky, I. I. and Braun, Vladimir M.",
    title = "{Evolution Equations for QCD String Operators}",
    reportNumber = "LENINGRAD-87-1351",
    doi = "10.1016/0550-3213(89)90168-5",
    journal = "Nucl. Phys. B",
    volume = "311",
    pages = "541--584",
    year = "1989"
}

@article{Belyaev:1985wza,
    author = "Belyaev, V. M. and Blok, B. Yu.",
    title = "{CHARMED BARYONS IN QUANTUM CHROMODYNAMICS}",
    doi = "10.1007/BF01560689",
    journal = "Z. Phys. C",
    volume = "30",
    pages = "151",
    year = "1986"
}

@article{ParticleDataGroup:2024cfk,
    author = "Navas, S. and others",
    collaboration = "Particle Data Group",
    title = "{Review of particle physics}",
    doi = "10.1103/PhysRevD.110.030001",
    journal = "Phys. Rev. D",
    volume = "110",
    number = "3",
    pages = "030001",
    year = "2024"
}

@article{Ioffe:2005ym,
    author = "Ioffe, B. L.",
    title = "{QCD at low energies}",
    eprint = "hep-ph/0502148",
    archivePrefix = "arXiv",
    doi = "10.1016/j.ppnp.2005.05.001",
    journal = "Prog. Part. Nucl. Phys.",
    volume = "56",
    pages = "232--277",
    year = "2006"
}

@article{Narison:2018nbv,
    author = "Narison, Stephan",
    editor = "Narison, St{\'e}phan",
    title = "{$\overline{\rm m}_{c,b,}<\alpha_sG^2>$ and $\alpha_s$ from Heavy Quarkonia}",
    doi = "10.1016/j.nuclphysbps.2018.12.026",
    journal = "Nucl. Part. Phys. Proc.",
    volume = "300-302",
    pages = "153--164",
    year = "2018"
}

@article{Rohrwild:2007yt,
    author = "Rohrwild, J.",
    title = "{Determination of the magnetic susceptibility of the quark condensate using radiative heavy meson decays}",
    eprint = "0708.1405",
    archivePrefix = "arXiv",
    primaryClass = "hep-ph",
    doi = "10.1088/1126-6708/2007/09/073",
    journal = "JHEP",
    volume = "09",
    pages = "073",
    year = "2007"
}

@article{Ozdem:2025fks,
    author = {{\"O}zdem, Ula{\c{s}}},
    title = "{Probing the electromagnetic structure of the $P_c(4337)^+$ pentaquark: insights from a diquark{\textendash}diquark{\textendash}antiquark picture for $J^P = \frac{1}{2}^-$ and $\frac{3}{2}^-$ states}",
    eprint = "2506.04345",
    archivePrefix = "arXiv",
    primaryClass = "hep-ph",
    doi = "10.1140/epjc/s10052-025-14439-9",
    journal = "Eur. Phys. J. C",
    volume = "85",
    number = "6",
    pages = "704",
    year = "2025"
}

@article{Ozdem:2024rch,
    author = {{\"O}zdem, Ula{\c{s}}},
    title = "{Shedding light on the nature of the Pcs(4459) pentaquark state}",
    eprint = "2411.11442",
    archivePrefix = "arXiv",
    primaryClass = "hep-ph",
    doi = "10.1103/PhysRevD.111.074038",
    journal = "Phys. Rev. D",
    volume = "111",
    number = "7",
    pages = "074038",
    year = "2025"
}

@article{Vujmilovic:2025czt,
    author = "Vujmilovic, Ivan and Collins, Sara and Leskovec, Luka and Prelovsek, Sasa",
    title = "{Electromagnetic form factors and structure of the $T_{bb}$ tetraquark from lattice QCD}",
    eprint = "2510.17549",
    archivePrefix = "arXiv",
    primaryClass = "hep-lat",
    month = "10",
    year = "2025"
}

@article{Zhu:2025abk,
    author = "Zhu, Sheng-He and Wang, Fu-Lai and Liu, Xiang",
    title = "{Electromagnetic characteristics as probes into the inner structures of the predicted $\Xi_c^{(',*)}D^{(*)}_s$ molecular states}",
    eprint = "2510.18492",
    archivePrefix = "arXiv",
    primaryClass = "hep-ph",
    month = "10",
    year = "2025"
}

@article{Ozdem:2025ion,
    author = {{\"O}zdem, U.},
    title = "{Electromagnetic form factors: A window into the $D\Lambda_c$, $D^*\Lambda_c$, and $D\Lambda_c^*$ molecular structure}",
    eprint = "2511.16052",
    archivePrefix = "arXiv",
    primaryClass = "hep-ph",
    month = "11",
    year = "2025"
}

@article{Ozdem:2025jda,
    author = {{\"O}zdem, Ula{\c{s}}},
    title = "{Electromagnetic tomography of spin-$ \frac{3}{2} $ hidden-charm strange pentaquarks}",
    eprint = "2510.26893",
    archivePrefix = "arXiv",
    primaryClass = "hep-ph",
    doi = "10.1007/JHEP02(2026)207",
    journal = "JHEP",
    volume = "02",
    pages = "207",
    year = "2026"
}

@article{Leinweber:1990dv,
    author = "Leinweber, Derek B. and Woloshyn, R. M. and Draper, Terrence",
    title = "{Electromagnetic structure of octet baryons}",
    reportNumber = "TRI-PP-90-52, UK-PP-90-09",
    doi = "10.1103/PhysRevD.43.1659",
    journal = "Phys. Rev. D",
    volume = "43",
    pages = "1659--1678",
    year = "1991"
}

@article{Mutuk:2024ach,
    author = "Mutuk, Halil",
    title = "{Magnetic moments of hidden-charm pentaquarks in the diquark{\textendash}diquark{\textendash}antiquark scheme}",
    eprint = "2411.16486",
    archivePrefix = "arXiv",
    primaryClass = "hep-ph",
    doi = "10.1016/j.cjph.2025.07.030",
    journal = "Chin. J. Phys.",
    volume = "97",
    pages = "1406--1414",
    year = "2025"
}

@article{Li:2025ddx,
    author = "Li, Hao-Song",
    title = "{Axial charges and magnetic moments of the decuplet pentaquark family}",
    eprint = "2511.12858",
    archivePrefix = "arXiv",
    primaryClass = "hep-ph",
    doi = "10.1103/32n9-j3pp",
    journal = "Phys. Rev. D",
    volume = "113",
    number = "5",
    pages = "056017",
    year = "2026"
}

@article{Wang:2026dqi,
    author = "Wang, Zhi-Gang and Liu, Yang",
    title = "{Analysis of the hidden-charm pentaquark candidates in the $J/\psi \Sigma$ mass spectrum via the QCD sum rules}",
    eprint = "2603.10774",
    archivePrefix = "arXiv",
    primaryClass = "hep-ph",
    month = "3",
    year = "2026"
}

@article{LHCb:2018roe,
    author = "Aaij, Roel and others",
    collaboration = "LHCb",
    title = "{Physics case for an LHCb Upgrade II - Opportunities in flavour physics, and beyond, in the HL-LHC era}",
    eprint = "1808.08865",
    archivePrefix = "arXiv",
    primaryClass = "hep-ex",
    reportNumber = "LHCB-PUB-2018-009, CERN-LHCC-2018-027",
    month = "8",
    year = "2018"
}

@article{Buchmann:2001gj,
    author = "Buchmann, A. J. and Henley, E. M.",
    title = "{Intrinsic quadrupole moment of the nucleon}",
    eprint = "hep-ph/0101027",
    archivePrefix = "arXiv",
    doi = "10.1103/PhysRevC.63.015202",
    journal = "Phys. Rev. C",
    volume = "63",
    pages = "015202",
    year = "2001"
}

@article{Belyaev:1993ss,
    author = "Belyaev, V. M.",
    title = "{Delta - isobar magnetic form-factor in QCD}",
    eprint = "hep-ph/9301257",
    archivePrefix = "arXiv",
    reportNumber = "CEBAF-TH-93-02",
    month = "1",
    year = "1993"
}

@article{Novikov:1983gd,
    author = "Novikov, V. A. and Shifman, Mikhail A. and Vainshtein, A. I. and Zakharov, Valentin I.",
    title = "{Calculations in External Fields in Quantum Chromodynamics. Technical Review}",
    reportNumber = "ITEP-140-1983",
    journal = "Fortsch. Phys.",
    volume = "32",
    pages = "585",
    year = "1984"
}

@article{Ioffe:1983ju,
    author = "Ioffe, B. L. and Smilga, Andrei V.",
    title = "{Nucleon Magnetic Moments and Magnetic Properties of Vacuum in QCD}",
    reportNumber = "ITEP-60-1983",
    doi = "10.1016/0550-3213(84)90364-X",
    journal = "Nucl. Phys. B",
    volume = "232",
    pages = "109--142",
    year = "1984"
}

@article{Zakharov:1968fb,
    author = "Zakharov, V. I. and Kondratyuk, L. A. and Ponomarev, L. A.",
    title = "{Bremsstrahlung and determination of electromagnetic parameters of particles}",
    journal = "Yad. Fiz.",
    volume = "8",
    pages = "783--792",
    year = "1968"
}

@article{LopezCastro:1997dg,
    author = "Lopez Castro, G. and Toledo Sanchez, G.",
    title = "{Effects of the magnetic dipole moment of charged vector mesons in their radiative decay distribution}",
    eprint = "hep-ph/9707202",
    archivePrefix = "arXiv",
    reportNumber = "UCL-IPT-97-10",
    doi = "10.1103/PhysRevD.56.4408",
    journal = "Phys. Rev. D",
    volume = "56",
    pages = "4408--4411",
    year = "1997"
}

@article{GarciaGudino:2013alv,
    author = "Garc{\'\i}a Gudi{\~n}o, D. and Toledo S{\'a}nchez, G.",
    editor = "Gauzzi, Paolo and Venanzoni, Graziano",
    title = "{Determination of the magnetic dipole moment of the rho meson using 4 pion electroproduction data}",
    eprint = "1305.6345",
    archivePrefix = "arXiv",
    primaryClass = "hep-ph",
    doi = "10.1142/S2010194514604633",
    journal = "Int. J. Mod. Phys. Conf. Ser.",
    volume = "35",
    pages = "1460463",
    year = "2014"
}

@article{Ozdem:2026gmn,
    author = {{\"O}zdem, Ula{\c{s}}},
    title = "{Hidden-charm pentaquarks: electromagnetic structure in a diquark{\textendash}diquark{\textendash}antiquark model}",
    eprint = "2603.19151",
    archivePrefix = "arXiv",
    primaryClass = "hep-ph",
    doi = "10.1140/epjc/s10052-026-15591-6",
    journal = "Eur. Phys. J. C",
    volume = "86",
    number = "4",
    pages = "359",
    year = "2026"
}

@article{Mutuk:2026zxp,
    author = "Mutuk, Halil and Kang, Xian-Wei",
    title = "{Magnetic moments of open bottom--charm molecular pentaquark octets}",
    eprint = "2603.27657",
    archivePrefix = "arXiv",
    primaryClass = "hep-ph",
    month = "3",
    year = "2026"
}
\end{document}